\documentclass[preprint,aip,jcp,citeautoscript,nofootinbib,superscriptaddress,amsmath,amssymb]{revtex4-1}
\usepackage{subfigure} 
\usepackage{graphicx} 
\usepackage[mathlines]{lineno}
\usepackage[utf8]{inputenc} 
\usepackage[T1]{fontenc} 
\usepackage{mathptmx} 
\usepackage{etoolbox} 
\usepackage{color} 
\usepackage{newunicodechar} 
\newunicodechar{−}{\ensuremath{-}} 
\newunicodechar{₂}{$_2$} 
\usepackage{graphicx}
\usepackage{tikz}
\usetikzlibrary{positioning}
\usepackage{gensymb}
\usepackage{booktabs}
\makeatletter
\def\@email#1#2{%
 \endgroup
 \patchcmd{\titleblock@produce}	
  {\frontmatter@RRAPformat}
  {\frontmatter@RRAPformat{\produce@RRAP{*#1\href{mailto:#2}{#2}}}\frontmatter@RRAPformat}
  {}{}
}

\makeatother
\begin{document}

\preprint{AIP/123-QED}

\title{Effect of interfacial water structure on the adsorption of weakly
hydrated anions}
\author{Swaminath Bharadwaj}
\affiliation{Department of Chemical Engineering, Shiv Nadar Institution of Eminence, Gautam Buddh Nagar, Uttar Pradesh 201314, India}
\email{swaminath.bharadwaj@snu.edu.in}
\date{\today}

\begin{abstract}
Weakly hydrated anions tend to partition to macromolecular surfaces, leading to
salting-in at low salt concentrations. Recent studies have linked this
partitioning to disruption of the interfacial water network. Here, we
investigate the role of interfacial-water-network disruption as an indirect
driving force for the adsorption of weakly hydrated anions and its interplay
with direct solute–anion and anion-cation interactions using computer
simulations. We study the adsorption of SCN$^{-}$ and I$^{-}$ ions on a model
extended solute that interacts with water and ions only via van der Waals
interactions. To probe the indirect contribution, we increase the effective
solute size while simultaneously decreasing the attractive solute–water and
solute–ion interactions, resulting in progressively greater disorder in the
interfacial water network in pure aqueous solutions. Our results show that
iodide adsorption increases despite weaker direct solute–iodide interactions and
correlates with greater disruption of the interfacial water network, although
iodide remains depleted overall. In contrast, preferential adsorption of
SCN$^{-}$ decreases, but the rate of decrease becomes smaller at higher levels
of interfacial-water-network disorder. This suggests that the indirect
contribution associated with interfacial-water-network disruption partly
compensates for the decrease in SCN$^{-}$ adsorption caused by weaker direct
solute–SCN$^{-}$ interactions. Our results highlight that the indirect
contribution associated with interfacial-water-network disruption promotes the
adsorption of weakly hydrated anions on macromolecular surfaces, while the
extent of adsorption is also influenced by direct solute–ion interactions and
anion–cation pairing.  This interplay may be particularly important in
mixed-salt solutions, where the presence of strongly hydrated anions can also
disrupt the interfacial water network.
\end{abstract}

\maketitle
\section{Introduction}
The anionic Hofmeister series ranks the relative ability of anions to salt-out
proteins in aqueous solutions.\cite{Hofmeister1888,Kunz2004} The salting-out
order is given by $\mathrm{CO_3^{2-}} > \mathrm{SO_4^{2-}} >
\mathrm{S_2O_3^{2-}} > \mathrm{H_2PO_4^{-}} > \mathrm{F^{-}} > \mathrm{Cl^{-}} >
\mathrm{Br^{-}} > \mathrm{NO_3^{-}} > \mathrm{I^{-}} > \mathrm{ClO_4^{-}} >
\mathrm{SCN^{-}}$, indicating a decreasing tendency to precipitate
macromolecules. Anions on the left end of the series are strongly hydrated and
tend to remain in the bulk. These anions are usually depleted from the solute
surface and lead to salting-out of polymers, with the extent increasing
monotonically with salt concentration.\cite{Zhang2005} On the other hand, the
partitioning of weakly hydrated anions (right end of the series) to solute
surfaces and the associated solvation effects depend on factors such solute
size,\cite{Rogers2022} salt
concentration,\cite{Zhang2005,cremert:2020,Moghaddam2015,Jungwirth2001} anion-cation
pairing\cite{Bruce2021,Bui2021} and the presence of other
salts.\cite{cremert:2020, Zhao2022,Goel2026}  Interestingly, preferential
adsorption or depletion is not consistent across weakly hydrated anions, unlike
strongly hydrated anions. In single salt solutions, both I$^{-}$ and SCN$^{-}$
ions have been observed to cause salting-in of the polymer at low salt
concentrations.\cite{Zhang2005,cremert:2020}  However, SCN$^{-}$ ions are
preferentially adsorbed, whereas I$^{-}$ ions remain
depleted.\cite{Rogers2022,Zhao2022,Goel2026} In salt solutions containing a
mixture of weakly and strongly hydrated anions, the partitioning of the weakly
hydrated anion at the polymer interface is enhanced relative to that in
single-salt solutions, contributing significantly to the non-additive effects
observed in such mixtures.\cite{cremert:2020,Zhao2022,Goel2026}  Interestingly,
depending on the polymer, these non-additive effects may be accompanied by a
weak  depletion or preferential adsorption of weakly hydrated anions.\cite{Zhao2022,Goel2026} Additionally, change in the counter cation leads
to qualitative changes in these non-additive effects, which may be due to a
reduction in the adsorption of weakly hydrated anions on account of cation-anion
pairing.\cite{Bui2021} Therefore, it is important to understand the direct and
indirect contributions to the preferential adsorption or depletion of weakly
hydrated anions.

In a recent study, Rogers et al. investigated the relationship between the
solute length scale and the preferential binding of thiocyanate ions in aqueous
solutions using a combination of experiments and molecular
simulations.\cite{Rogers2022} Interestingly, they observed that SCN$^{-}$ ions
preferentially adsorbed to polymer interfaces while remaining depleted from
their monomer counterparts. Even in polymers, ion adsorption was observed mainly
in the center of the chain rather than at its termini. The authors attributed
these trends to variations in surface curvature and their impact on the
interfacial water structure. The center of the chain exhibited a flatter surface
curvature, leading to greater disruption of the interfacial water network,
whereas the termini offered a more curved surface, leading to a relatively
ordered water structure. SCN$^{-}$ preferentially partitioned into regions with
greater disruption of the interfacial water network. Some important questions
arising from this study are: (a) Does disorder in the interfacial water network
contribute indirectly to anion adsorption (an interfacial-water-mediated
contribution), or does it merely provide suitable binding sites for anions
partitioning due to direct solute–anion interactions? (b) How do direct solute–ion
interactions and indirect interfacial-water-mediated contributions jointly
influence ion adsorption? (c) What happens in the case of anions such as iodide,
which salt in polymers despite being depleted from the polymer surface?
Understanding this indirect contribution is important as the interfacial water
network can also be disrupted by other factors, such as the presence of a
strongly hydrated anion.

In this work, we investigate the adsorption of two weakly hydrated anions,
SCN$^{-}$ and I$^{-}$, on a model extended hydrophobic solute using molecular
dynamics simulations. Note that this model solute does not have termini, as our
aim is to understand the interfacial water network and ion adsorption on
surfaces with low curvatures. The solute consists of 80 uncharged beads that
interact with ions and water molecules only via Lennard-Jones (LJ) interactions.
The contributions to ion adsorption from the direct solute-ion van der Waals
interactions  and the indirect interfacial-water-mediated effects are decoupled
by increasing the effective solute size while simultaneously decreasing in the
direct solute bead-water and solute bead-ion attractive interactions (coupled
$\sigma_{\rm bx}-\epsilon_{\rm bx}$ variation). In pure aqueous solutions, such
a variation leads to a decrease in the solute-water affinity and the number of
interfacial water-water hydrogen bonds per water molecule, indicating a
monotonic increase in the disorder of the interfacial water network.
Interestingly, in aqueous alkali iodide solutions, we observe that the
solute-iodide affinity increases and iodide depletion decreases in the coupled
$\sigma_{\rm bx}-\epsilon_{\rm bx}$ variation.  Note that this increase in
iodide adsorption occurs despite a decrease in the direct solute-I$^{-}$ van der
Waals interactions and correlates with the increasing disruption of the
interfacial water network. In contrast, for SCN$^{-}$ ions, the solute-SCN$^{-}$
affinity decreases and the binding shifts from preferential adsorption to
depletion.  However, we observe that the rate of decrease in solute-SCN$^{-}$
affinity is reduced at higher disruptions of the interfacial water network. This
indicates that the reduction in the adsorption of SCN$^{-}$ ions, due to the
decrease in solute-SCN$^{-}$ attractive interactions, is partly compensated by
the indirect interfacial-water-mediated contribution. These results show that an
increase in the disorder of the interfacial water network provides an indirect
driving force for the adsorption of weakly hydrated anions. Note that the extent
adsorption is also influenced by direct solute-anion interactions and
anion-cation pairing. Such
an interplay also becomes very important when the disruption in the  interfacial
water network occurs due to other factors, such as the presence of strongly
hydrated anions. To show that the trends are not a consequence of the effective
solute size, we also performed simulations where the effective solute size is
increased without changing the solute-ion and solute-water attractive
interactions ($\sigma_{\rm bx}$-only variation).  In pure aqueous solutions, the
interfacial water network did not get disrupted as the number of interfacial
water-water hydrogen bonds per water molecule remained nearly unchanged. In
aqueous alkali iodide solutions, this variation led to a decrease in the
solute-I$^{-}$ affinity, and an increase in iodide depletion. In contrast, in
aqueous alkali thiocyanate solutions, both the solute-SCN$^{-}$ affinity and the
preferential adsorption of SCN$^{-}$ ions decreases. Given that the interfacial
water network remains unchanged, these trends can be attributed to the relative
preference of the solute for ion versus water molecules for SCN$^{-}$ and
vice-versa for I$^{-}$.

\section{Methods}
In this study, we employ the model extended hydrophobic solute proposed by
Tripathy et al.\cite{Tripathy2020} which has been created by periodically
stretching a 80 bead polymer in the $Z$-direction. The beads were uncharged and
interacted with the ion-water mixture via LJ interactions with self parameters
$\sigma_{\rm bb}=1$~nm and $\epsilon_{\rm bb}=0.4$~kJ/mol.\cite{Zangi2009} This
extended solute was solvated in aqeuous salt solutions of  LiI, CsI, NaI, NaSCN
and KSCN at a concentration of 1~m.  To study the effect of salt concentration,
simulations were also performed at salt concentrations of 0.2 m and 0.6 m in
aqueous LiI and CsI solutions.  The SPC/E model  was used for
water.\cite{Berendsen1987} Non-polarizable models were employed for all the
ions. For alkali iodide salts, the forcefield parameters for iodide, cesium and
sodium ions were taken from from the work of Fyta and Netz.\cite{Netz:2012}
Models I(4) and Cs(6) were used for iodide and cesium ions,
respectively.\cite{Netz:2012} For lithium ions, the modified model developed by
Bruce et al. was used.\cite{Bruce2021} For NaSCN and KSCN, the forcefield
parameters for thiocyanate, sodium and potassium ions were taken from the work
of Tesei et al.\cite{Tesei2018} The partial charges and, LJ self- and
cross-interaction parameters for the ion-water mixture have been listed in
Tables S1 and S2 in Sec. S1.1 of the Supporting Information.  The LJ
parameters for interactions between the solute beads and the ion-water mixture
were computed using the Lorentz-Bertholet (LB) mixing rules.  To decouple the
effects of direct solute–ion van der Waals interactions from the indirect
interfacial-water-mediated effects on ion adsorption, the repulsive part of the
solute bead (b)–water/ion (x) LJ interactions was scaled by a factor $\alpha$.
\begin{equation}
U_{\rm bx}=\alpha\frac{C_{\rm 12,LB}}{r^{12}}-\frac{C_{\rm
6,LB}}{r^{6}},
\end{equation}
where
\begin{equation}
\begin{split}
C_{\rm 12,LB}&=4\epsilon_{\rm LB}\sigma_{\rm LB}^{12},~~C_{\rm 6,LB}=4\epsilon_{\rm LB}\sigma_{\rm LB}^{6},\\
\epsilon_{\rm LB}&=\sqrt{\epsilon_{\rm bb}\epsilon_{\rm xx}},~~\sigma_{\rm
LB}=\left(\sigma_{\rm bb}+\sigma_{\rm xx}\right)/2,
\end{split}
\end{equation}
where $\epsilon_{\rm xx}$ and $\sigma_{\rm xx}$ are the LJ self parameters for water
and ions. The above potential can be converted to a standard LJ potential in the
following way,
\begin{equation}
\begin{split}
U_{\rm bx}&=U_{\rm LJ,bx}=4\epsilon_{\rm bx}\left(\frac{\sigma_{\rm bx}^{12}}{r_{\rm
bx}^{12}}-\frac{\sigma_{\rm bx}^{6}}{r_{\rm bx}^{6}}\right),\\
\epsilon_{\rm bx}&=\epsilon_{\rm LB}/\alpha,~~\sigma_{\rm
bx}=\alpha^{1/6}\sigma_{\rm LB}.
\end{split}
\end{equation}
This increase in $\alpha$ leads to a decrease in the bead-water/ion attraction
(decreasing $\epsilon_{\rm bx}$) and an increase in the effective bead size
(increasing $\sigma_{\rm bx}$). This methodology has been utilized in previous
works to investigate the binding of different cosolutes/cosolvents on extended
hydrophobic solutes.\cite{Tripathy2022,Bharadwaj2024} We also performed
simulations where only the effective bead size is increased while keeping
$\epsilon_{\rm bx}$ fixed,
\begin{equation}
\begin{split}
U_{\rm LJ,bx}&=4\epsilon_{\rm bx}\left(\frac{\sigma_{\rm bx}^{12}}{r_{\rm
bx}^{12}}-\frac{\sigma_{\rm bx}^{6}}{r_{\rm bx}^{6}}\right),\\
\epsilon_{\rm bx}&=\epsilon_{\rm LB},~~\sigma_{\rm
bx}=\alpha^{1/6}\sigma_{\rm LB}.
\end{split}
\end{equation}
An increase in \(\alpha\) in this case only increases $\sigma_{\rm bx}$, while
$\epsilon_{\rm bx}$ remains fixed. Thus, increasing $\alpha$ increases the
effective bead size and, consequently, the number of solute–water/ion contacts
at a fixed interaction strength. This provides a reference system for assessing
the effect of increasing effective solute size on ion adsorption. Hereafter, the
two cases will be referred to as the coupled $\sigma_{\rm bx}-\epsilon_{\rm bx}$
variation and the $\sigma_{\rm bx}$-only variation, respectively. For
$\alpha=1$, the solute-water/ion LJ parameters are identical for both
variations, thereby leading to identical results. These results at $\alpha=1$
serve as a common reference for both variations. For each salt solution,
simulations were performed at three different values of $\alpha~(=1,2,4)$ for
both the variations. The variation of the solute-ion and solute-water LJ
parameters with $\alpha$ for the coupled  $\sigma_{\rm bx}-\epsilon_{\rm bx}$
variation has been listed in Table S3 in Sec. S1.2 of the Supporting
Information.

The number of water molecules were fixed at 15000 and 270 for all the
simulations, respectively leading to a 1~m salt solution. Note that the number
of water molecues was fixed at 15000 for other concentrations as well. Molecular
dynamics simulations were performed using the Gromacs 2022.1 package using a
time step of 2 fs.\cite{Abraham2015} The cutoff for the van der Waals and short
range Coulomb interactions was fixed at 1.4 nm. The Particle Mesh Ewald method
was used for calculating the long range electrostatic interactions. All bonds
were constrained using the LINCS algorithm \cite{Hess2008}. All simulations were
performed at 1 bar pressure and 300 K temperature. The time constants for the
thermostats and barostats were kept at 1 and 2 ps, respectively.  The end-to-end
distance, $R_{\rm ee}$, of the extended solute along the $Z$-direction was 12.24
nm and its conformation was kept fixed by applying position restraints on all
beads (force constant $\kappa=10^{5}$~kJ~mol$^{-}$~nm$^{2}$).  To maintain a
fixed $R_{\rm ee}$, all NPT runs were performed with semi-isotropic pressure
coupling where the simulation box was uniformly scaled only in the $x-y$ plane.
The system was first energy minimized and subsequently a 1 ns NVT run was
perfomed using the v-rescale thremostat.\cite{Bussi2007} The NPT equilibration
was performed in two stages. In the first stage, a 2 ns NPT run was performed
using the v-rescale thermostat and the Berendsen barostat.\cite{Berendson1984}
The second stage comprised of a 5 ns NPT run using the v-rescale thermostat and
the Parrinello-Rahman barotat.\cite{Parrinello1981}  The average box dimensions
computed from this run were used for the subsequent 800 ns NVT-production run.
The v-rescale thermostat was utilized for this run.  The production run was
perfomed in the NVT ensemble so that solvation shell partial molar properties
could be computed through the small system method.\cite{Tripathy2022,
Bharadwaj2024} System snapshots and thermodynamic data were collected at every
40 ps and 20 ps, respectively.

The densities of water and ion molecules around the extended solute were
quantified using the proximal radial distribution function, $g_{\rm p}$, in
which particles are binned in cylindrical shells. To ensure that the shells are
cylindrical, the proximal distance is computed from the line connecting the
centers of all the solute beads. The preferential accumulation of any ion $i$
around the solute, or depletion from its vicinity, was quantified using the
preferential binding coefficient,\cite{Pierce2008}

\begin{equation}
\Gamma_{\rm s-i}(r_{\rm p})=\left< n_{\rm i}(r_{\rm p}) - \frac{N_{\rm i}-n_{\rm i}(r_{\rm p})}{N_{\rm w}-n_{\rm w}(r_{\rm p})}n_{\rm w}(r_{\rm p})\right>,
\end{equation}
where $n_{\rm i}$ ($r_{\rm p}$) and $n_{\rm w}$ ($r_{\rm p}$) are the number of
ions of type $i$ and water molecules within a proximal distance $r_{\rm p}$ from
the solute, respectively.  $N_{\rm i}$ and $N_{\rm w}$ are the total number of
ions of type $i$ and water molecules in the simulation box, respectively. The
preferential binding coefficient is an integrated quantify which plateaus at
large $r_{\rm p}$. This plateau value is the thermodynamic value of the
preferential binding coefficient and quantifies the overall preferential
accumulation or depletion. Hereafter, $\Gamma_{\rm s-i}$ will refer to this
thermodynamic value, and $\Gamma_{\rm s-i}(r_{\rm p})$ will refer to the
distance-dependent profile. For the calculation of $\Gamma_{\rm s-i}(r_{\rm
p})$, the 800 ns trajectory was split into 40 blocks of 20 ns each.  For each
block, the average value of $\Gamma_{\rm s-i}(r_{\rm p})$ and the associated
standard deviation were computed.  Subsequently, the block averages and
deviations were combined to obtain the overall average and standard error.  Note
that the standard error was computed from the block standard deviations using
Gaussian error propagation.

To characterize the interfacial water hydrogen bond network in pure aqueous
solutions, the average numbers of hydrogen bonds formed by first hydration shell
water molecules with other water molecules in the first hydration shell within
(intra) and outside (inter) the first shell were computed using an in-house
code. The quantities $\left<n_{\rm HB,intra}^{\rm ww}\right>$ and $\left<n_{\rm
HB,inter}^{\rm ww}\right>$ are the average numbers of intra-shell and
inter-shell hydrogen bonds per first shell water molecule. Two water molecules
were considered to be hydrogen bonded if their oxygen atoms were within a
distance of 0.35 nm of each other and the angle between the O–O vector and one
of the donor O–H bond axes was 30$\degree$ or less.\cite{Luzar1993} The first
minimum in the solute-water proximal RDF was considered as the boundary of the
first hydration shell. 

\section{Results and Discussion}
\subsection{Extended solute in aqueous solutions}
\begin{figure}[h!]
\includegraphics[width=0.4\textwidth]{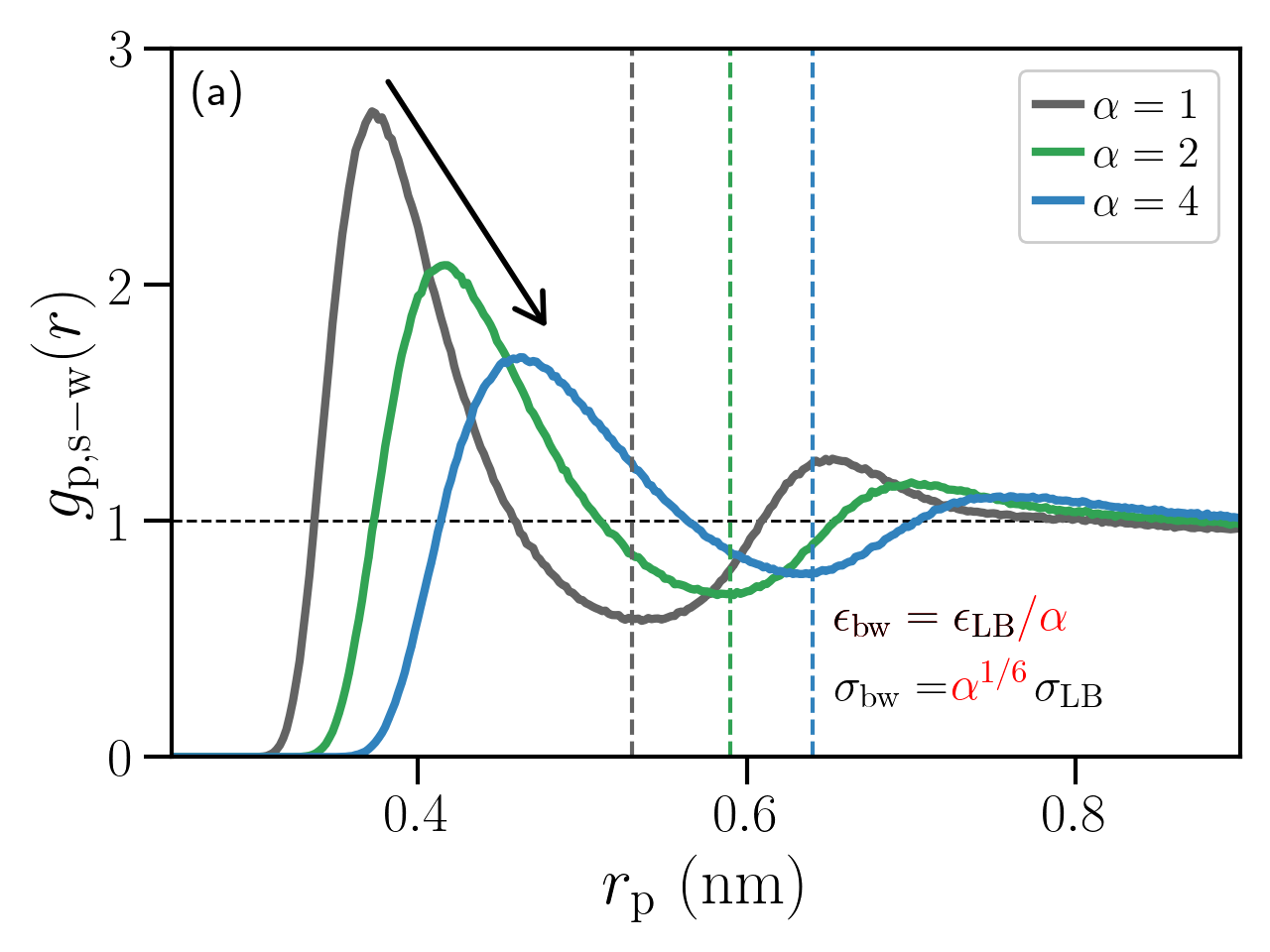}
\includegraphics[width=0.4\textwidth]{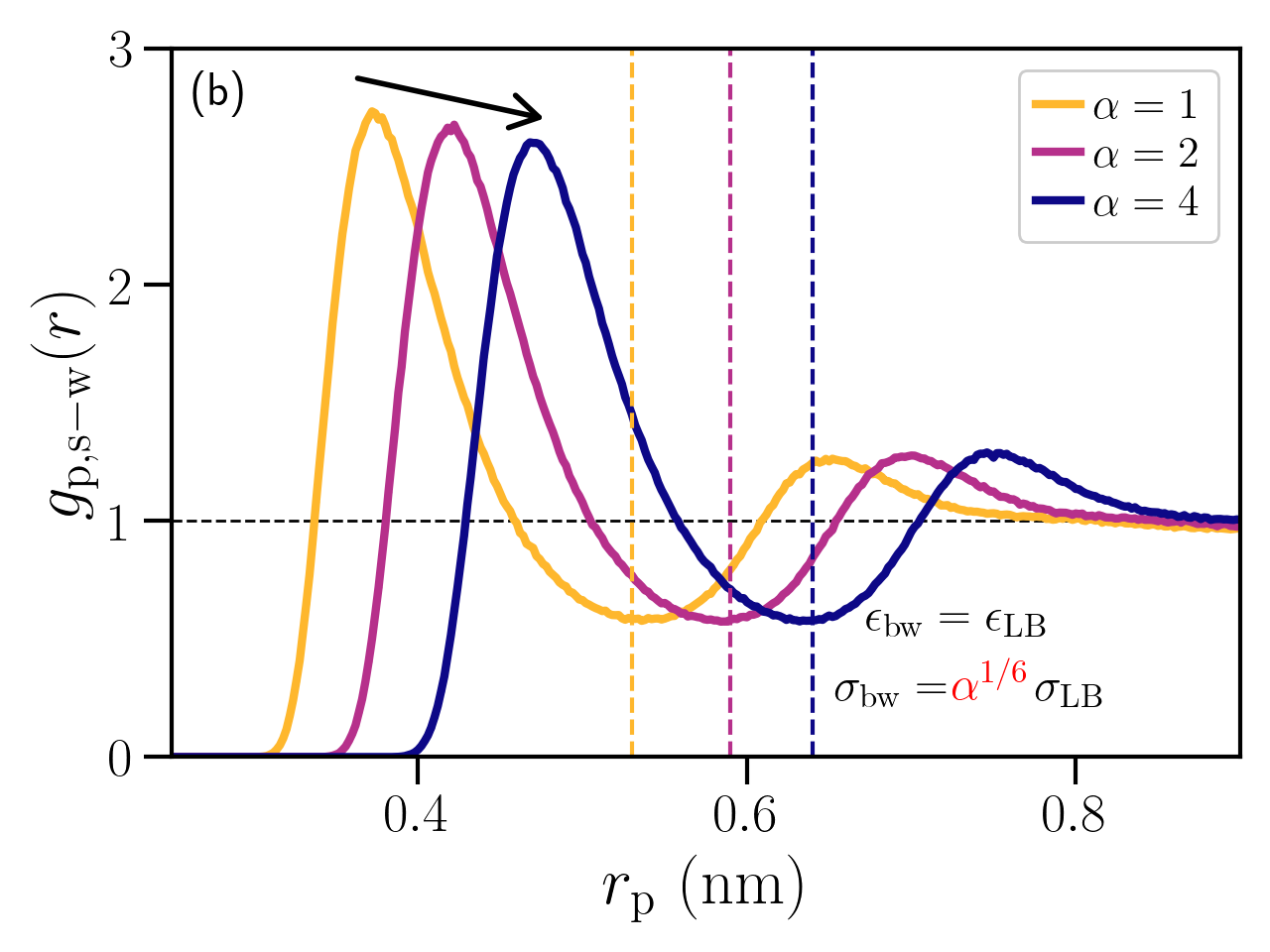}
\caption{Solute-water proximal RDFs at different values of $\alpha$ under (a)
coupled $\sigma_{\rm bw}-\epsilon_{\rm bw}$ variation and (b) $\sigma_{\rm
bw}-$only variation. Vertical dashed lines denote the boundary of the first
hydration shell for each value of $\alpha$. }
\label{fig:neatw_proximal}
\end{figure}

\begin{figure}[h!]
\includegraphics[width=0.4\textwidth]{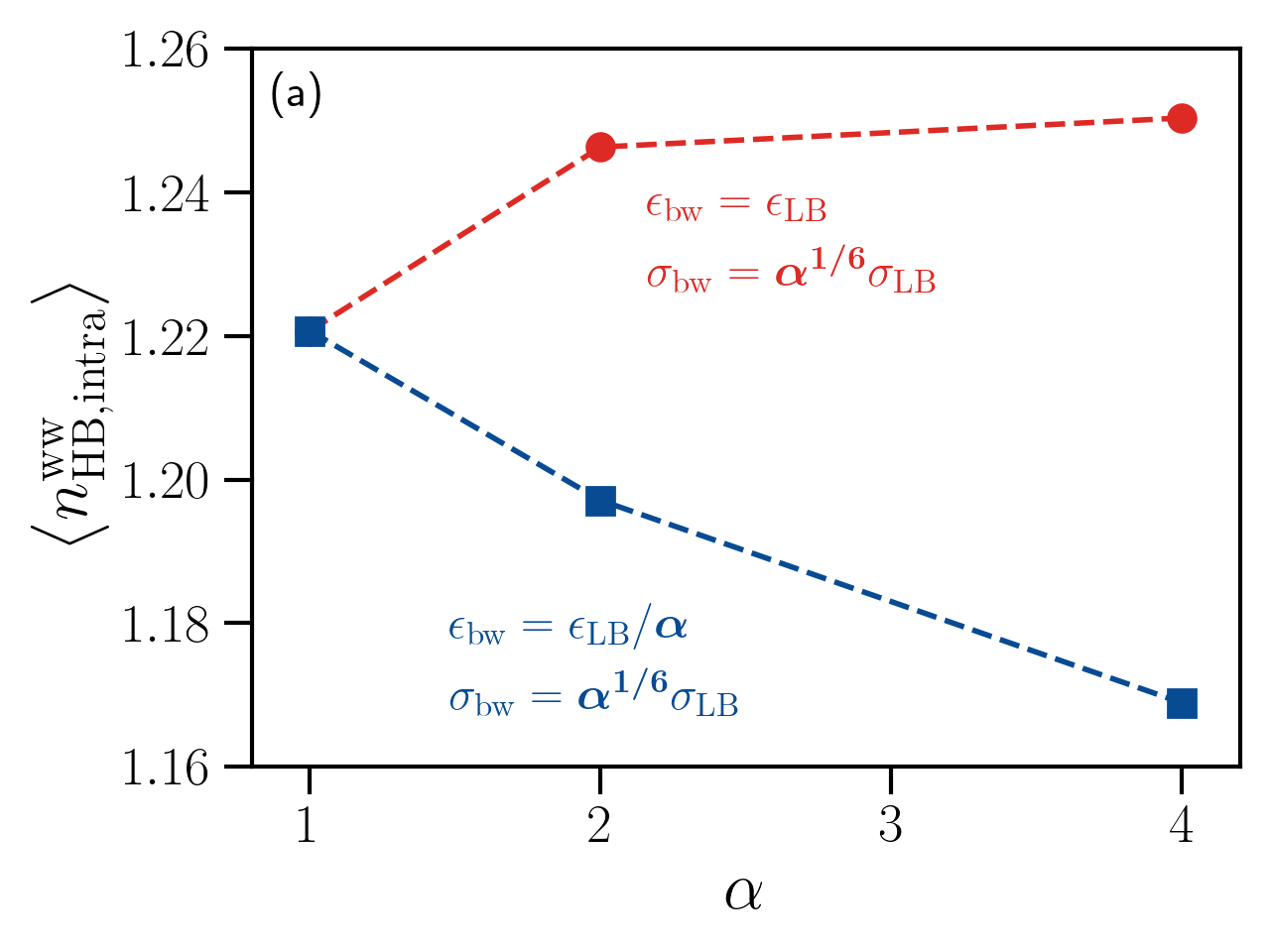}
\includegraphics[width=0.4\textwidth]{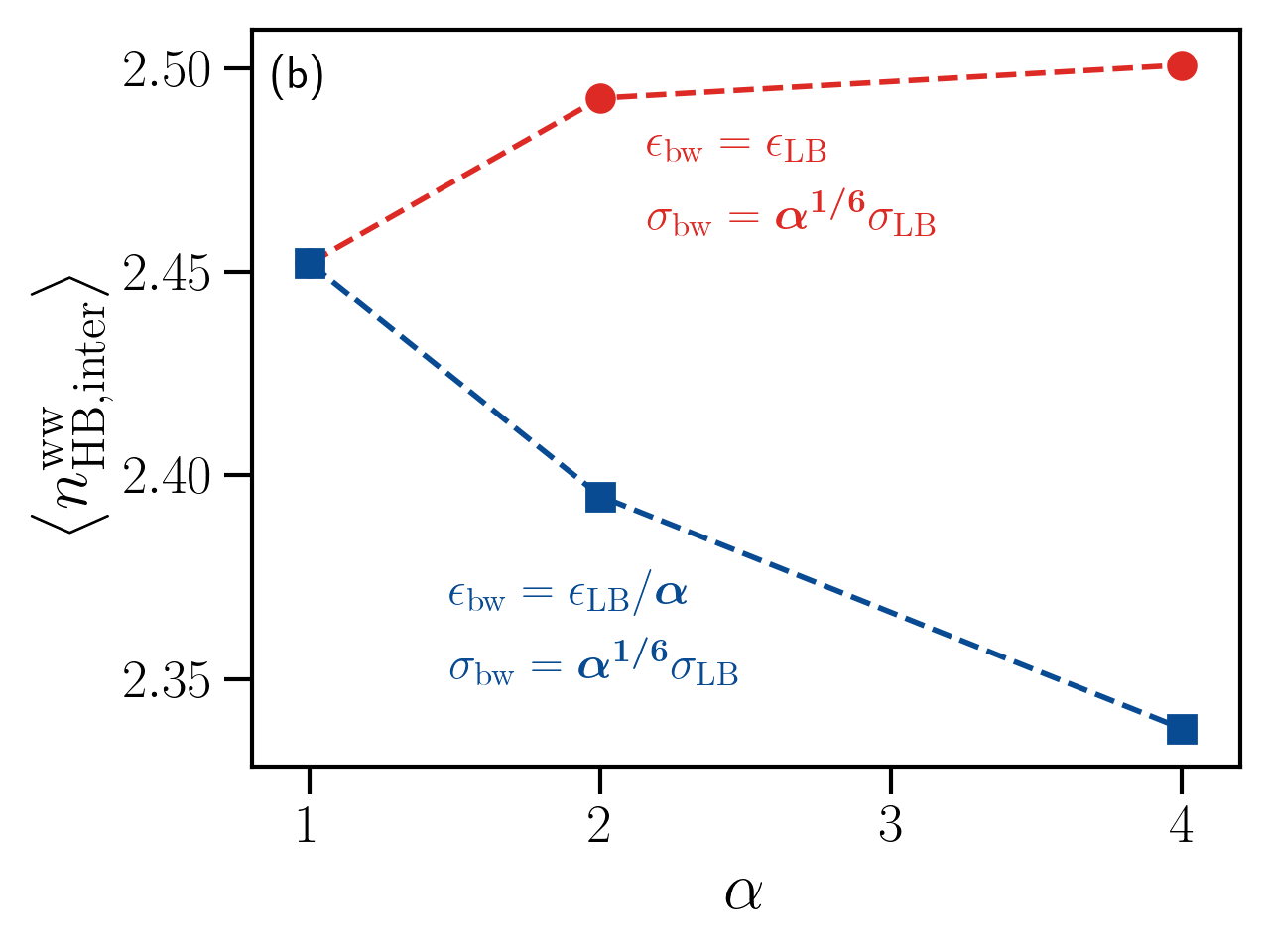}
\caption{Dependence of the average number of water–water hydrogen bonds per
first-shell water molecule on $\alpha$: (a) hydrogen bonds formed with other
first-shell water molecules (intra-shell) and (b) hydrogen bonds formed with
water molecules outside the first hydration shell (inter-shell), for both
coupled $\sigma_{\rm bw}-\epsilon_{\rm bw}$ and $\sigma_{\rm bw}$-only
variations}
\label{fig:neatw_hb}
\end{figure}

Figure~\ref{fig:neatw_proximal}(a) shows the solute-water proximal RDF profiles,
$g_{\rm p,s-w}(r_{\rm p})$, at different values of $\alpha$ for the coupled
$\sigma_{\rm bw}-\epsilon_{\rm bw}$ case. The first peak of $g_{\rm p,s-w}$
decreases significantly and shifts to larger $r_{\rm p}$ with increasing
$\alpha$, indicating a decrease in the solute-water affinity. These trends are
expected, as the solute-water attraction decreases ($\epsilon_{\rm
bw}\downarrow$) and the effective solute size increases ($\sigma_{\rm
bw}\uparrow$) with increasing $\alpha$. For the $\sigma_{\rm bw}-$only case (see
Fig.~\ref{fig:neatw_proximal}(b)), the first peak of $g_{\rm p,s-w}$ shifts to
larger $r_{\rm p}$ with a minor decrease in its magnitude with increasing
$\alpha$, indicating that the solute-water affinity remains largely the same.
Note that an increase in $\sigma_{\rm bw}$, at fixed $\epsilon_{\rm bw}$, leads
to a larger number of favorable solute-water contacts. 
To understand these trends further, we examine the hydrogen bonds formed by the
water molecules in the first hydration shell of the solute (defined by dashed
vertical lines in Fig.~\ref{fig:neatw_proximal}). Figures~\ref{fig:neatw_hb}(a)
and \ref{fig:neatw_hb}(b) show that both $<n_{\rm HB,intra}^{\rm ww}>$ and
$<n_{\rm HB,inter}^{\rm ww}>$ decrease monotonically with increasing $\alpha$
for the coupled $\sigma_{\rm bw}-\epsilon_{\rm bw}$ case, indicating an
increasing disruption of the interfacial water network. On the contrary, for the
$\sigma_{\rm bw}$-only case, both <$n_{\rm HB,intra}^{\rm ww}$> and <$n_{\rm
HB,inter}^{\rm ww}$> initially increase and subsequently remain unchanged with
increasing $\alpha$, indicating that the interfacial water network largely
remains unchanged.
Based on these trends, it can be concluded that the coupled $\sigma_{\rm
bw}-\epsilon_{\rm bw}$ variation leads to monotonic disruption of the
interfacial water network, providing a means to examine the role of interfacial
water structure in ion adsorption. Additionally, for solutes interacting solely
via LJ interactions, increasing only the effective size does not substantially
disrupt the interfacial water network.
\subsection{Extended solutes in aqueous alkali iodide solutions} 
\begin{figure}[h!]
\includegraphics[width=0.32\textwidth]{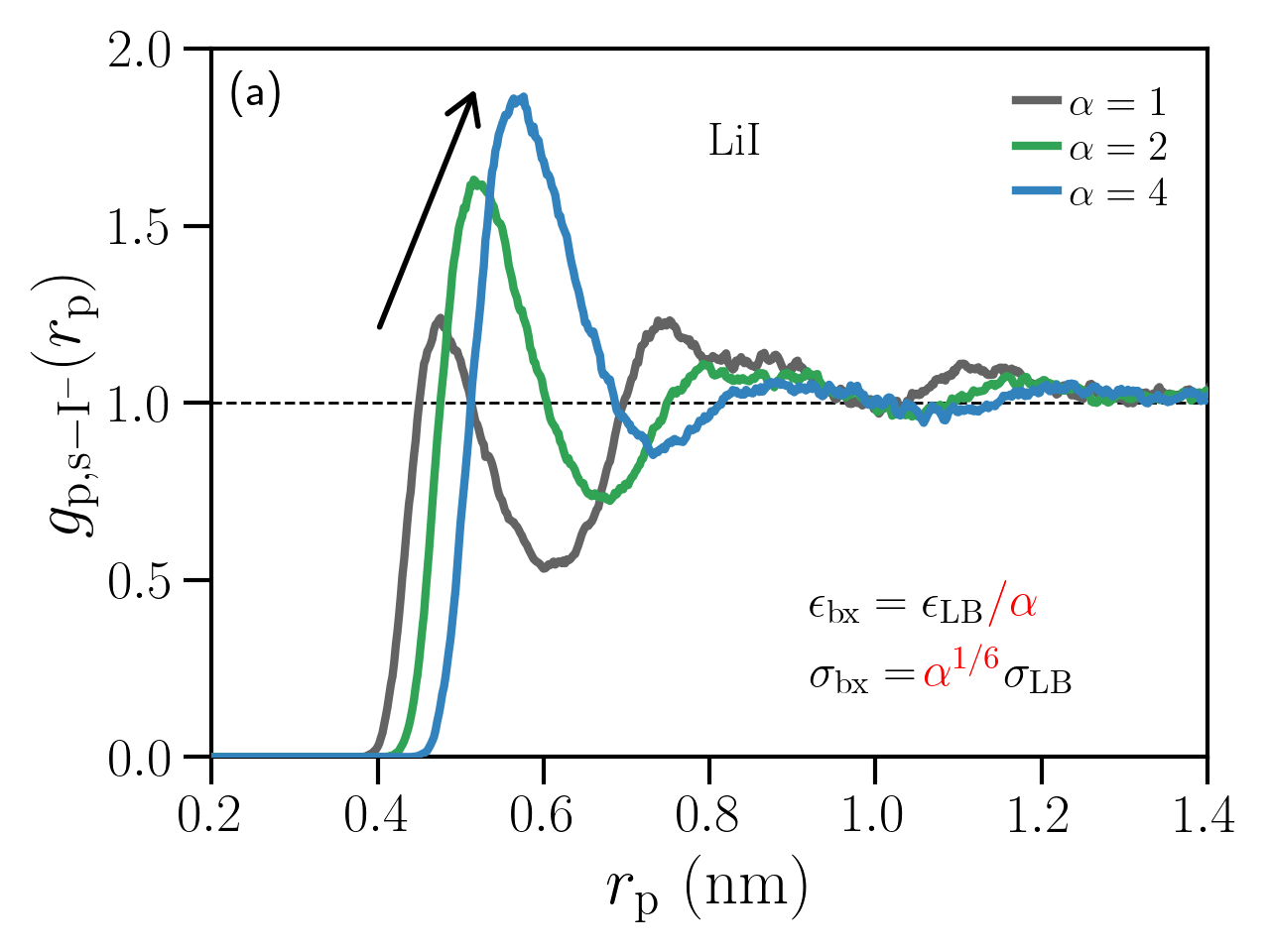}
\includegraphics[width=0.32\textwidth]{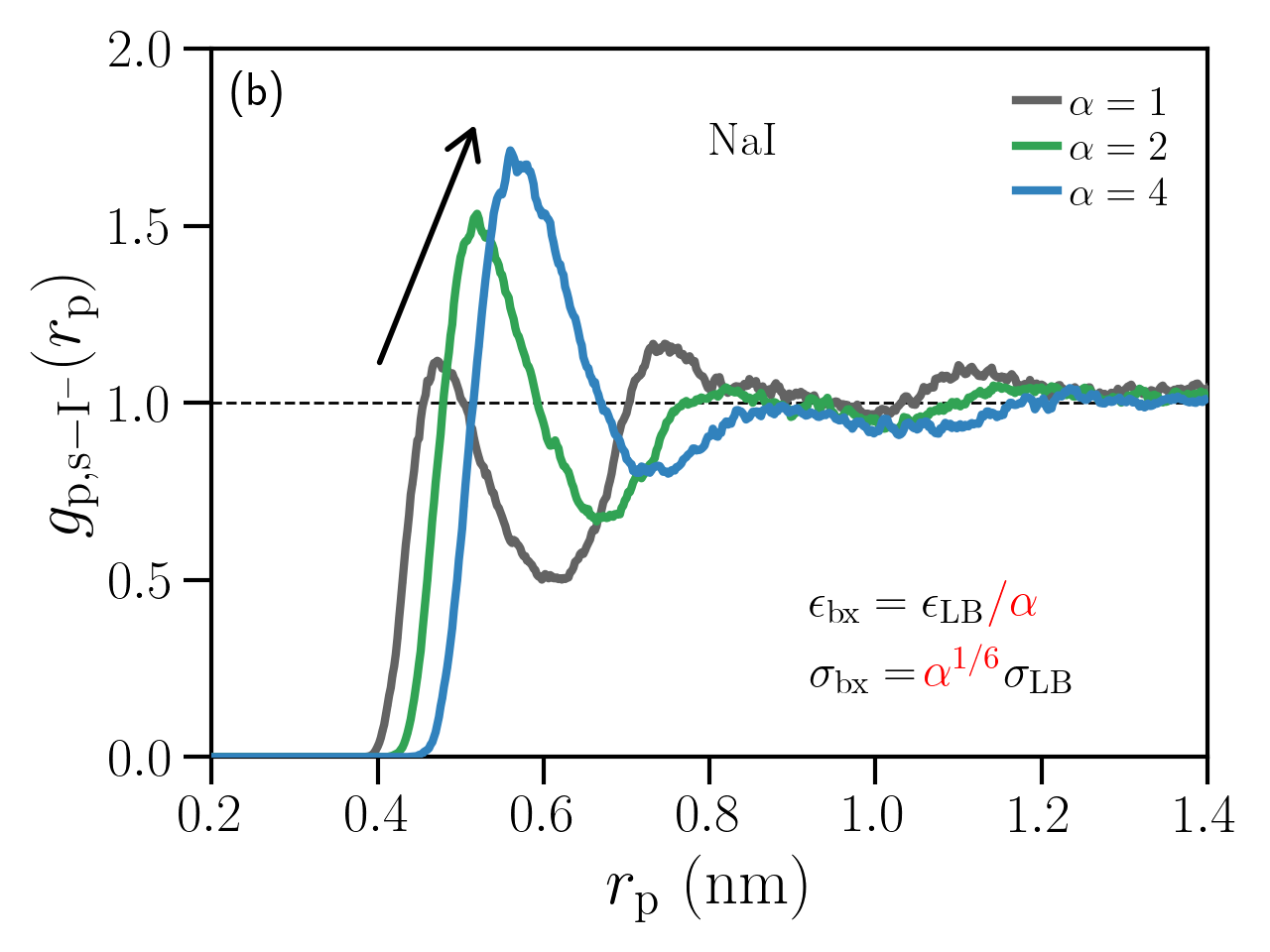}
\includegraphics[width=0.32\textwidth]{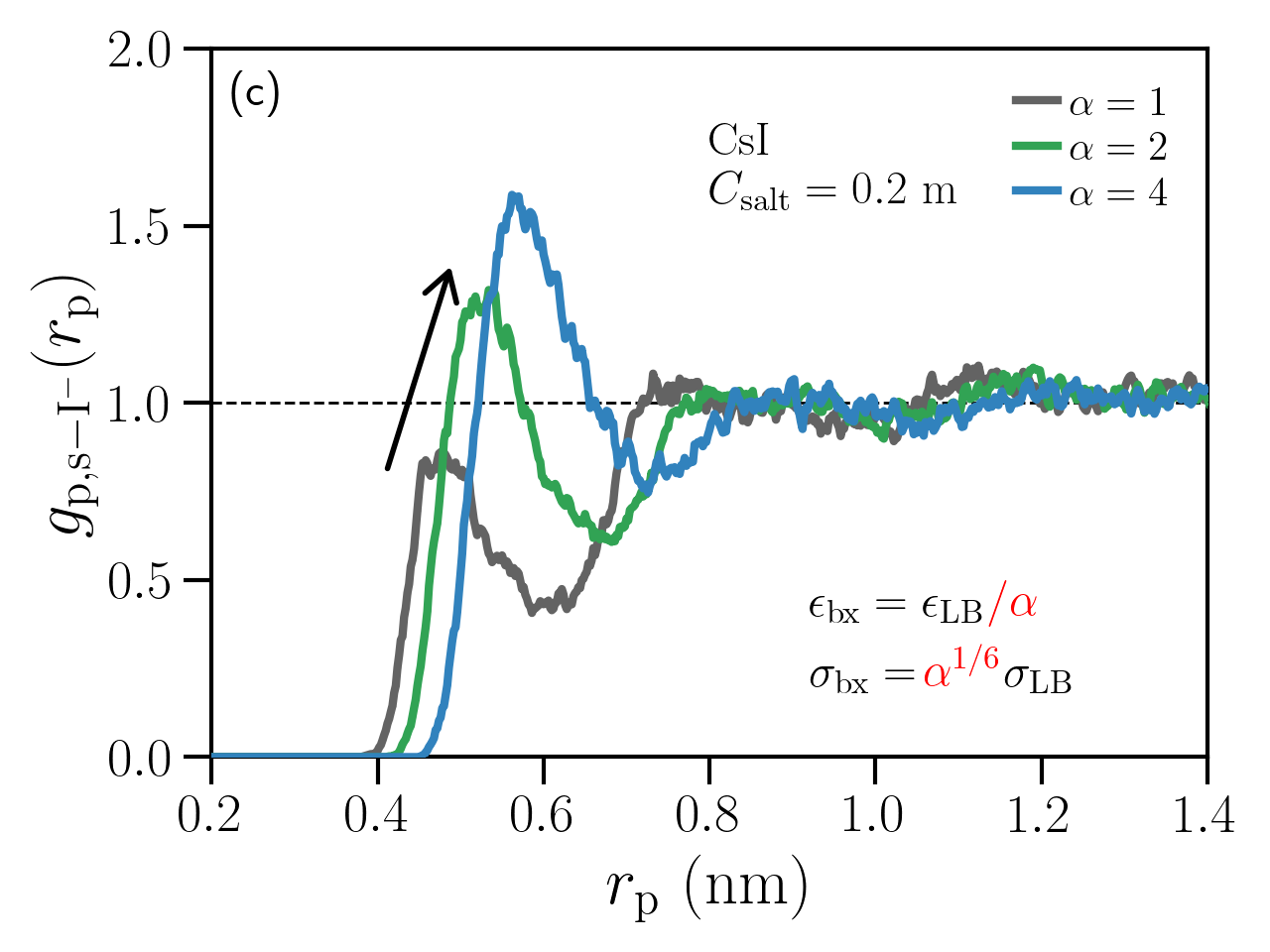}
\includegraphics[width=0.32\textwidth]{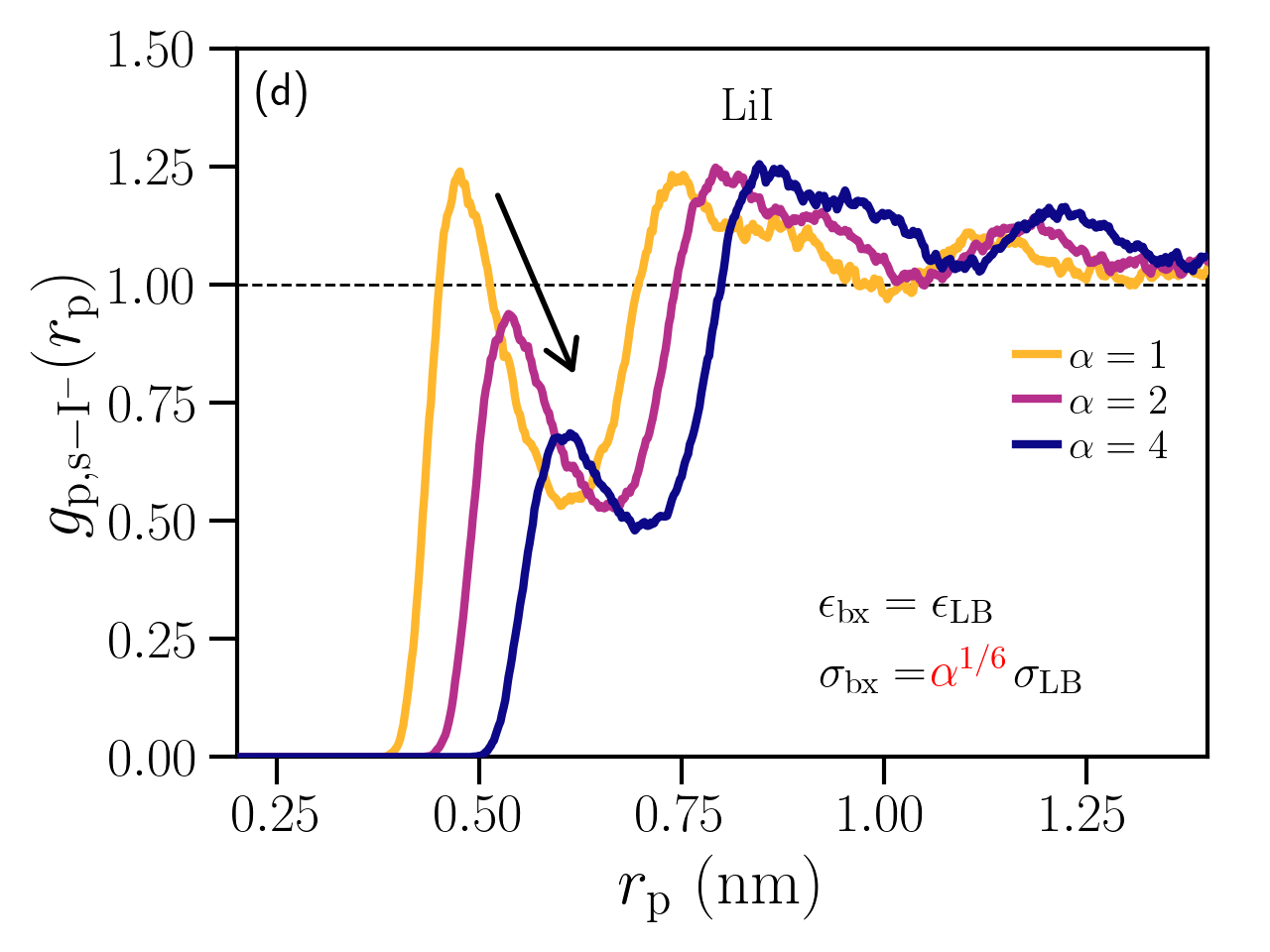}
\includegraphics[width=0.32\textwidth]{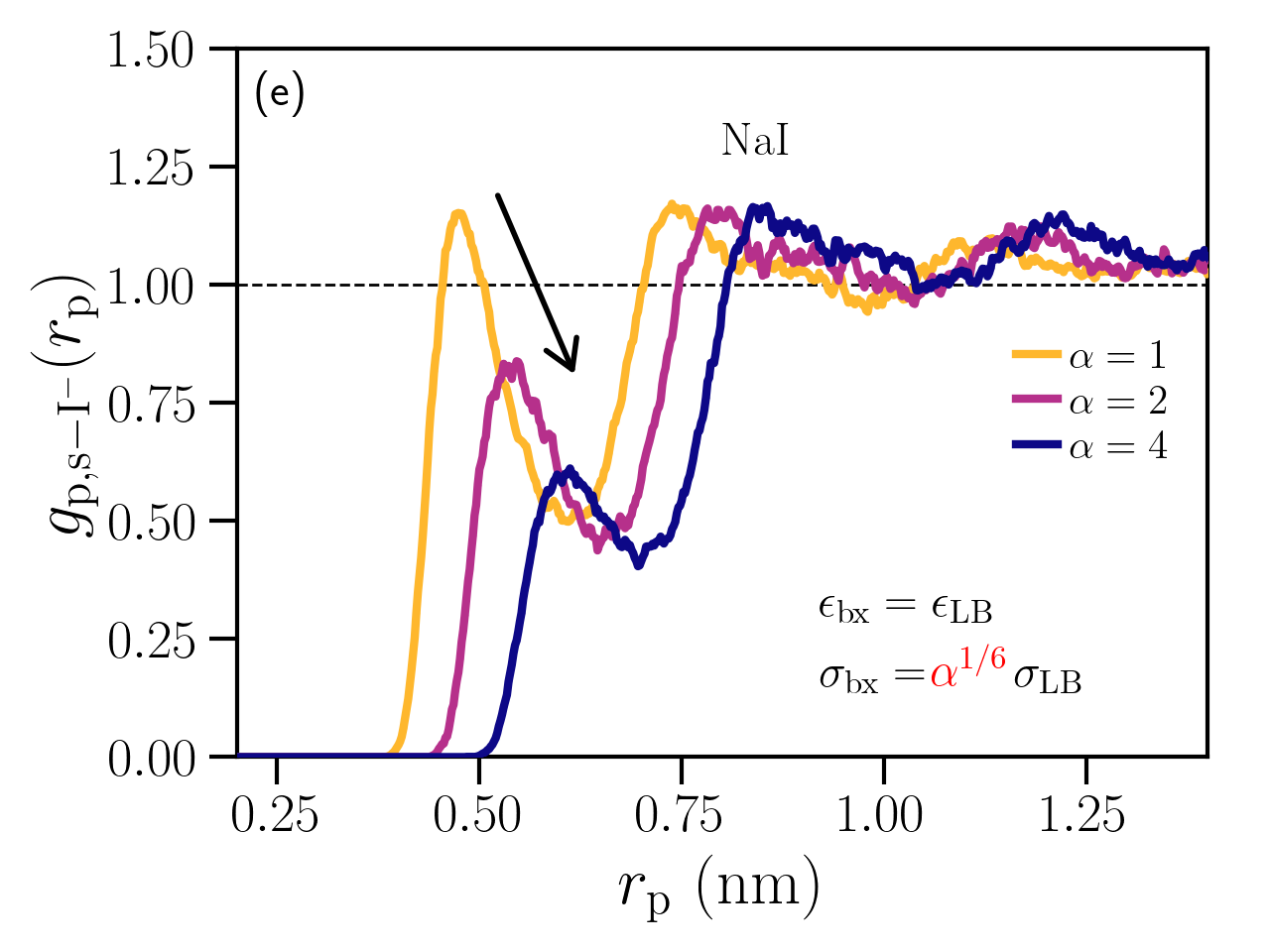}
\includegraphics[width=0.32\textwidth]{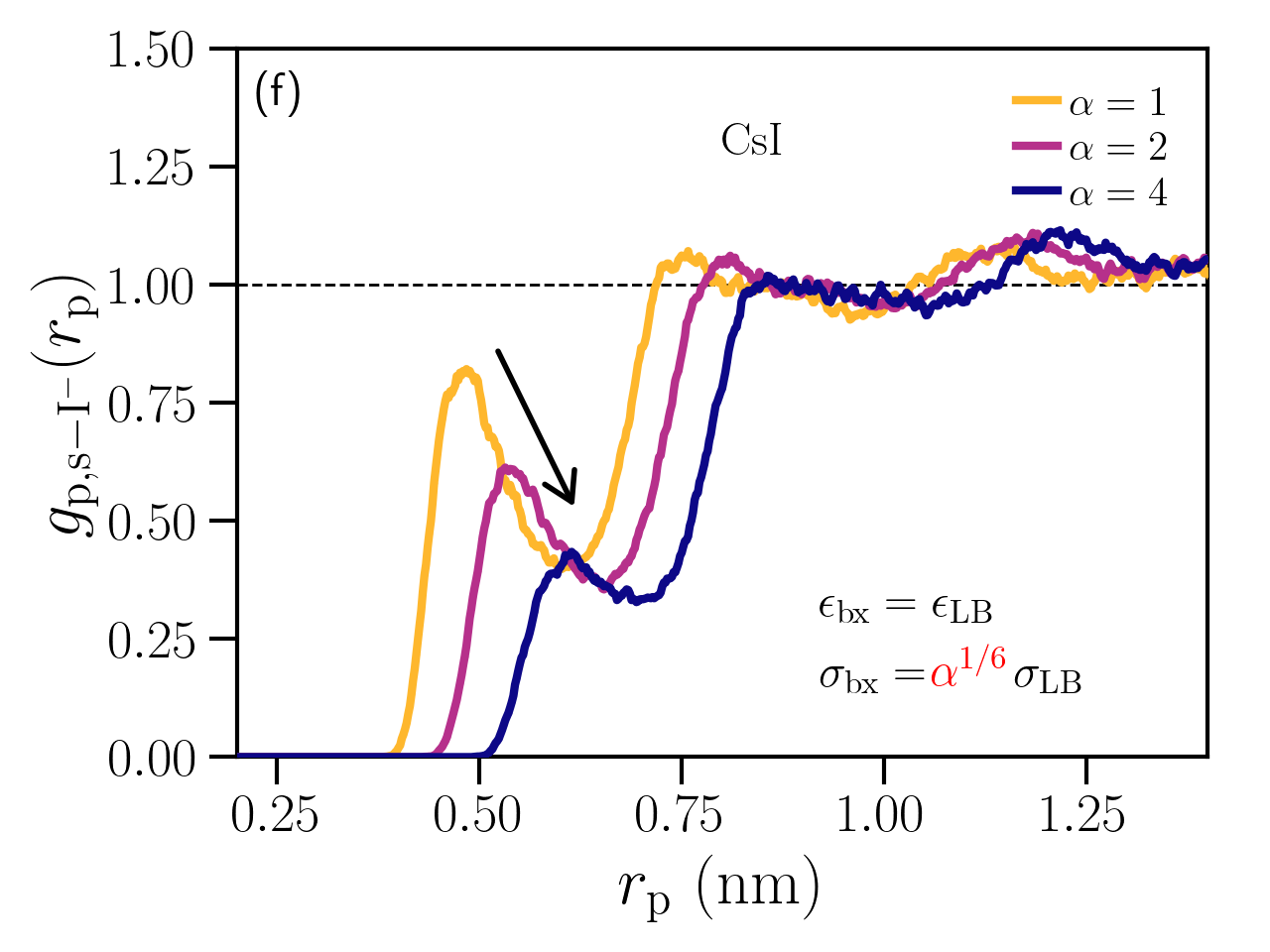}
\caption{Solute-iodide proximal RDF profiles, $g_{\rm p,s-I^{-}}(r_{\rm p})$,
for different values of $\alpha$ in aqueous LiI, NaI and CsI solutions.  The top
and bottom rows correspond to profiles for the coupled $\sigma_{\rm
sx}-\epsilon_{\rm bx}$ and the $\sigma_{\rm bx}$-only variations, respectively.}
\label{fig:prdf_iodide}
\end{figure}

Figures~\ref{fig:prdf_iodide}(a)-(c) show the solute-I$^{-}$ proximal RDFs at
different values of $\alpha$ in aqueous LiI, NaI and CsI solutions for the
coupled $\sigma_{\rm bx}-\epsilon_{\rm bx}$ variation. In all cases, the first
peak of $g_{\rm p,s-I^{-}}$ increases and shifts to larger $r_{\rm p}$ with
increasing $\alpha$, indicating that the solute-I$^{-}$ affinity increases
despite the decrease in the direct solute-I$^{-}$ van der Waals attraction. Note
that the first peak of $g_{\rm p,s-w}$ decreases with increasing $\alpha$
similar to the trend observed in pure aqueous solutions although it remains
larger that that of  $g_{\rm p,s-I^{-}}$ in all cases (see Figs. S1(a)-(c) in
the Supporting Information).  Combining these trends with those in
Fig.~\ref{fig:neatw_hb}, it can be seen that the increase in solute-I$^{-}$
affinity correlates with an increasing disruption of the interfacial water
network, suggesting a interfacial-water-mediated contribution to I$^{-}$
adsorption. Note that an increase in $\alpha$ increases the number solute-water
and solute-ion contacts due to the increase in $\sigma_{\rm bx}$, while
decreasing the solute-water and solute-ion attraction due to the decrease in
$\epsilon_{\rm bx}$.  Given the increasing disruption of the interfacial water
network, replacement of some solute–water contacts by solute–I$^{-}$ contacts
becomes favorable, as iodide is a weakly hydrated anion. 

On the other hand, from Figs.~\ref{fig:prdf_iodide}(d)-(f), it can be observed
that the first peak of $g_{\rm p,s-I^{-}}$ decreases significantly with
increasing $\alpha$ for the $\sigma_{\rm bx}$-only variation in all cases,
indicating a decrease in the solute-I$^{-}$ affinity. The first peak of $g_{\rm
p,s-w}$ shows a minor decrease with increasing $\alpha$ (see Figs. S1(d)-(f) in the
Supporting Information) similar to the trend observed in
Fig.~\ref{fig:neatw_proximal}(b) and remains larger than first peak of $g_{\rm
p,s-I^{-}}$ in all cases.  In the $\sigma_{\rm bx}$-only variation,  an increase
in $\alpha$ results in a larger number of  solute-water and solute-I$^{-}$
contacts at the same $\epsilon_{\rm bx}$. The different responses of $g_{\rm
p,s-I^-}$ and $g_{\rm p,s-w}$ to increasing $\alpha$ indicate a preference for
solute–water contacts over solute-I$^{-}$ contacts.

\begin{figure}[h!]
\includegraphics[width=0.32\textwidth]{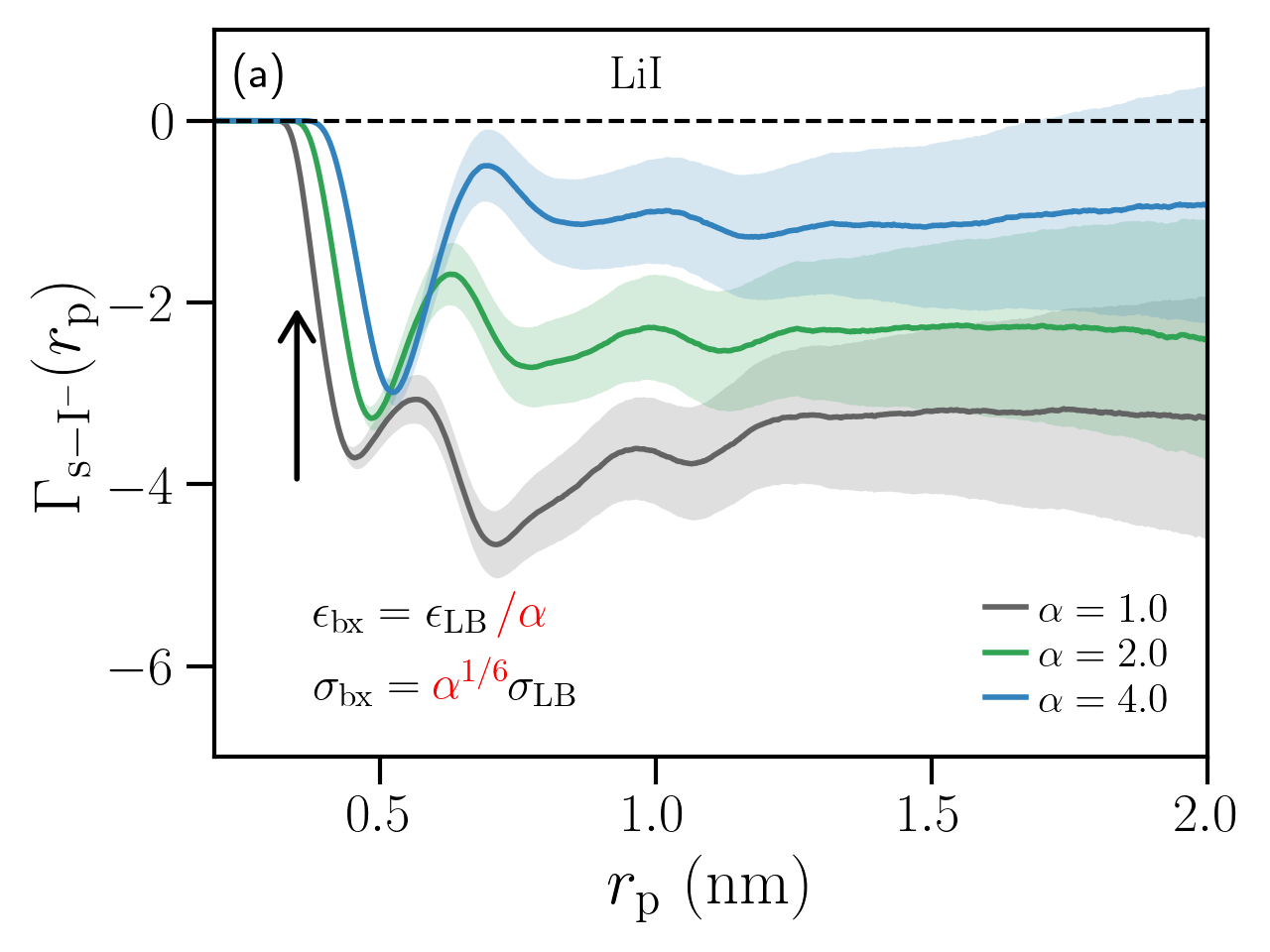}
\includegraphics[width=0.32\textwidth]{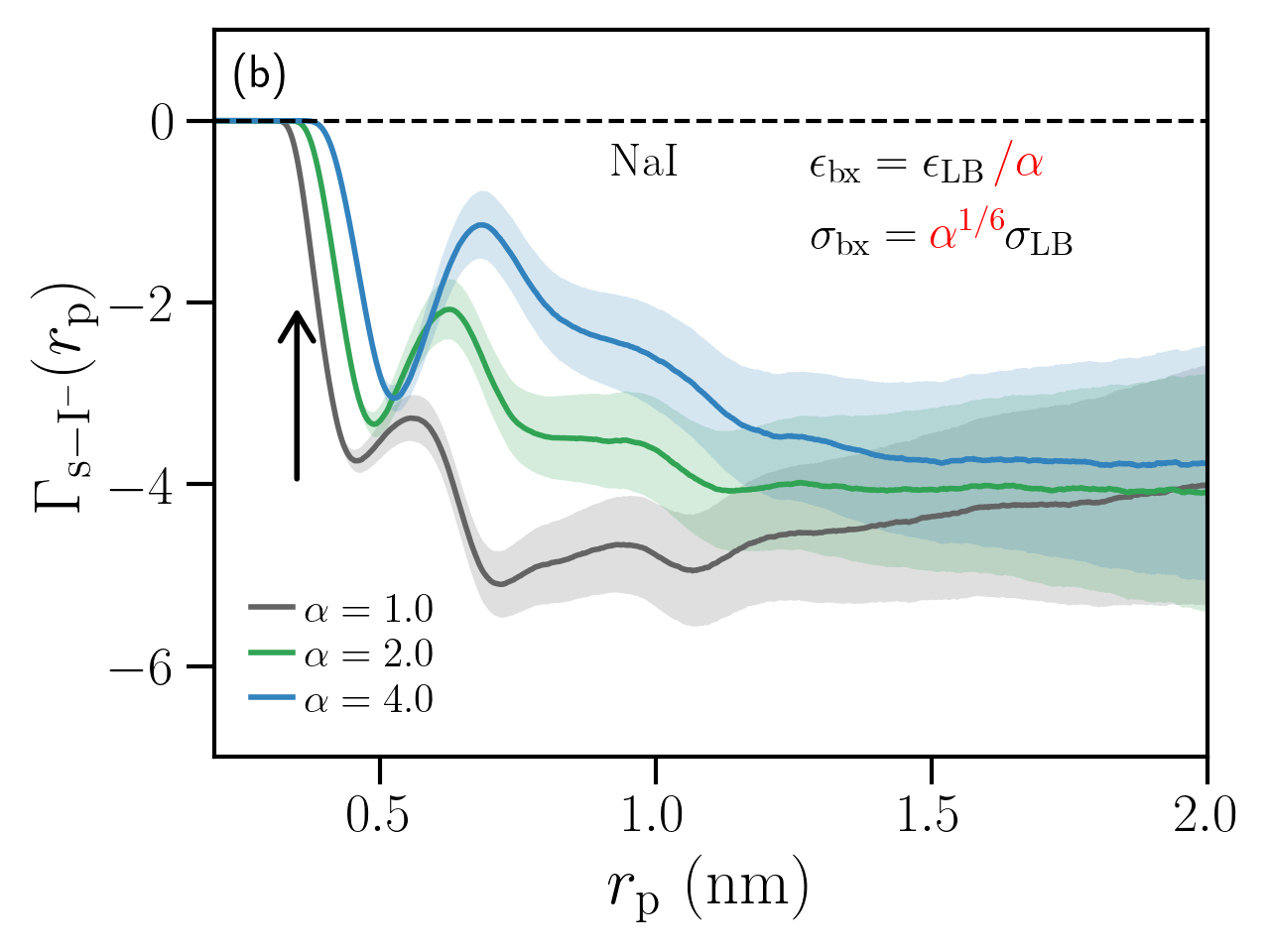}
\includegraphics[width=0.32\textwidth]{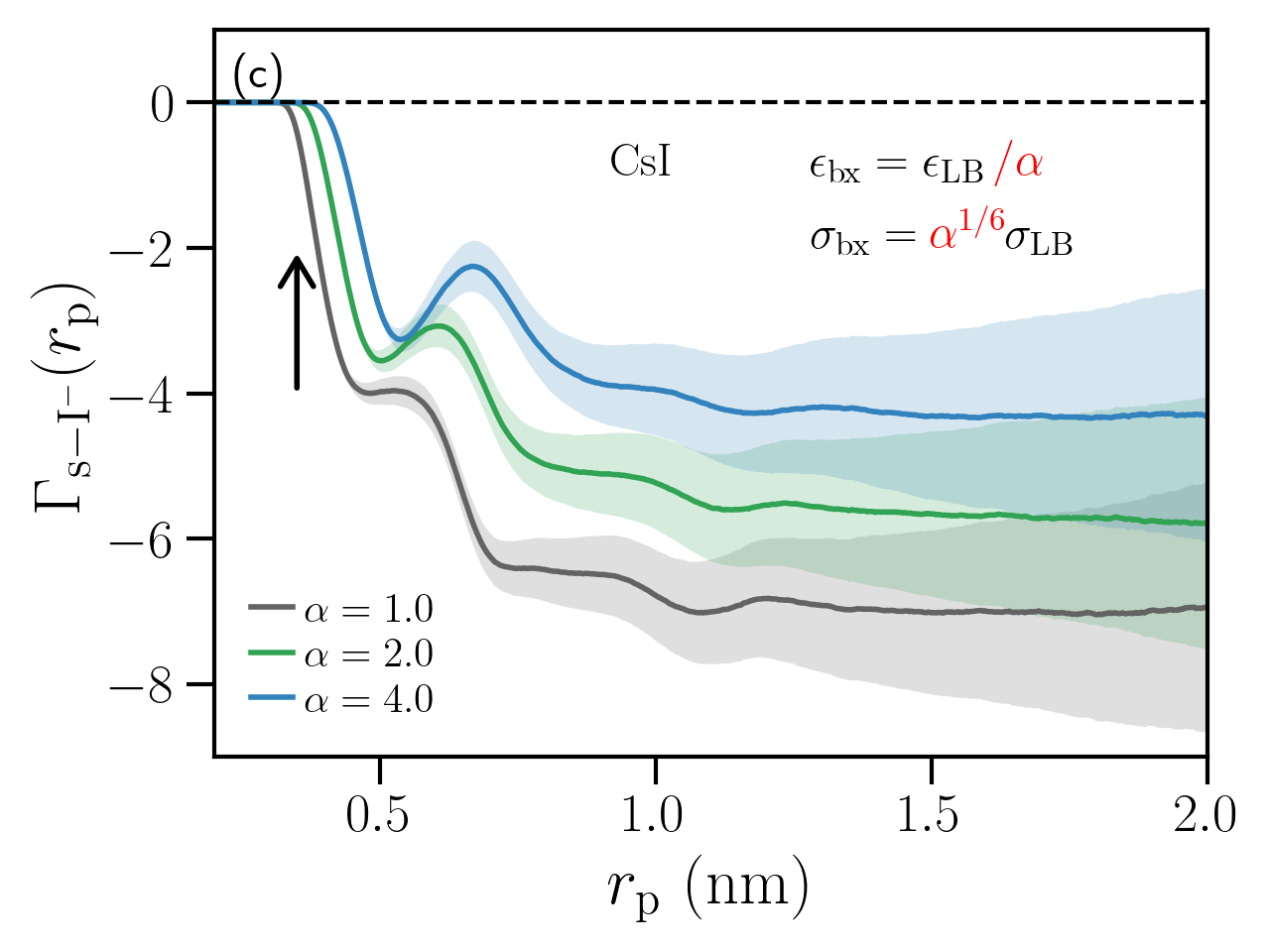}
\includegraphics[width=0.32\textwidth]{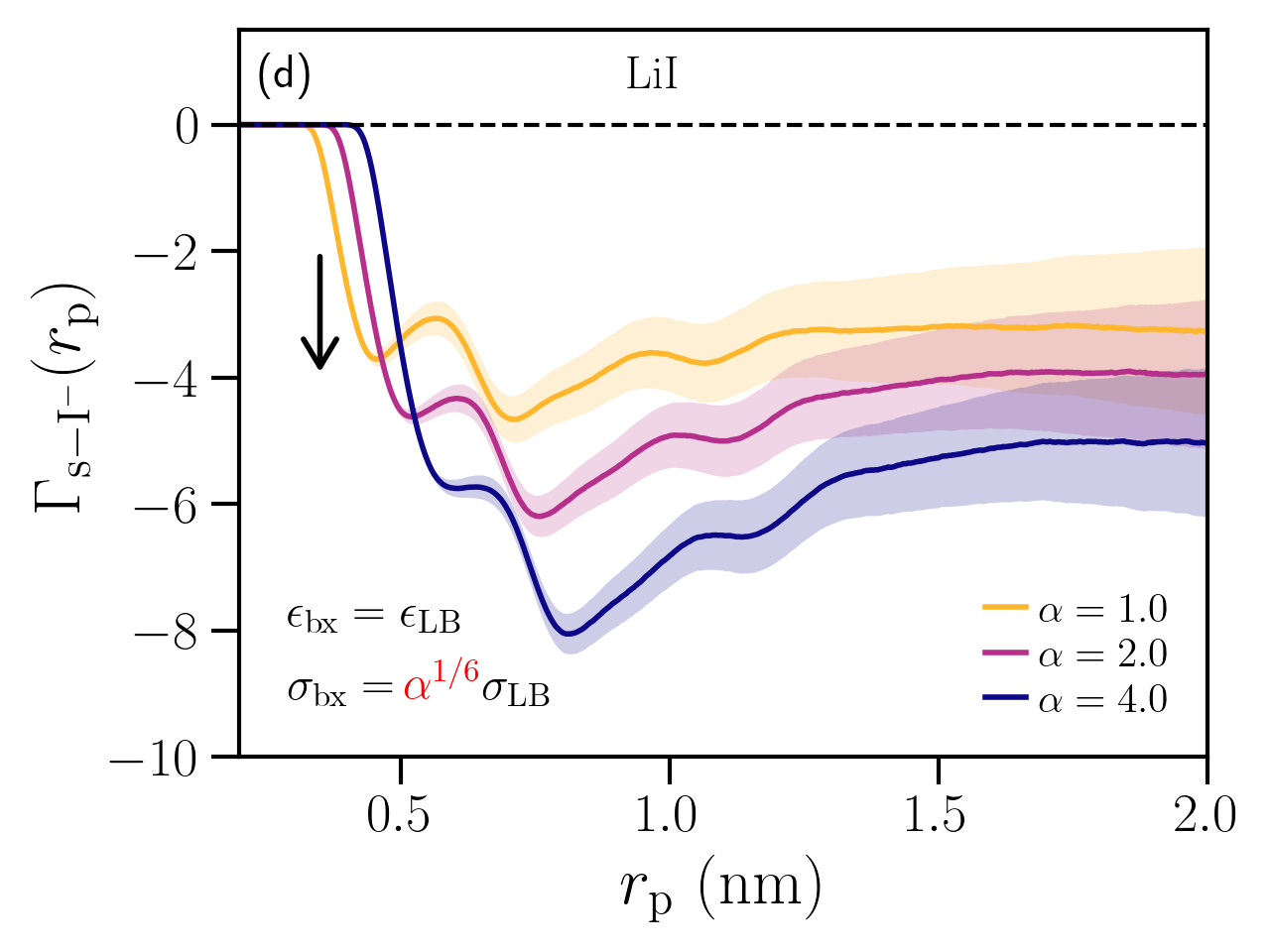}
\includegraphics[width=0.32\textwidth]{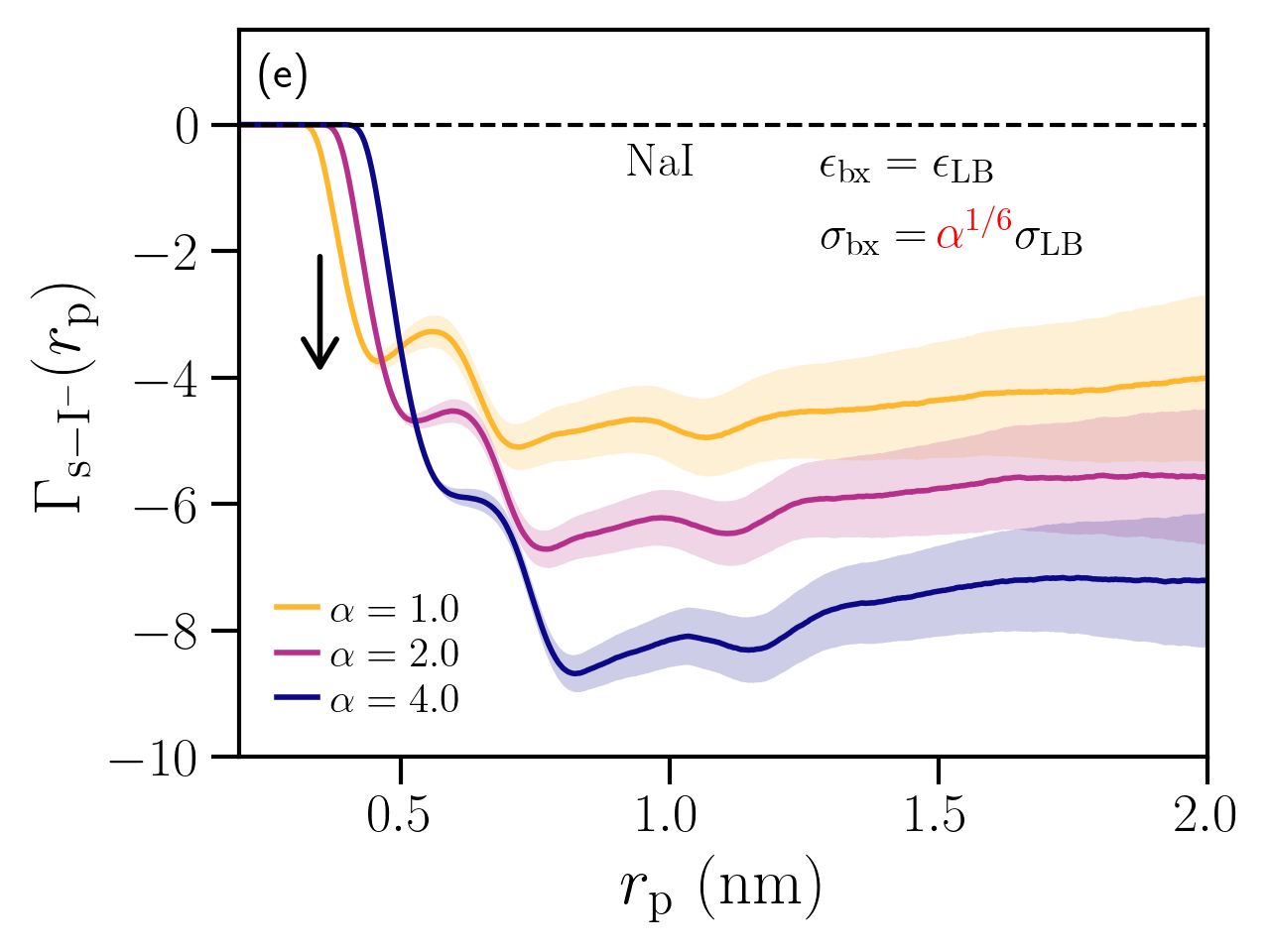}
\includegraphics[width=0.32\textwidth]{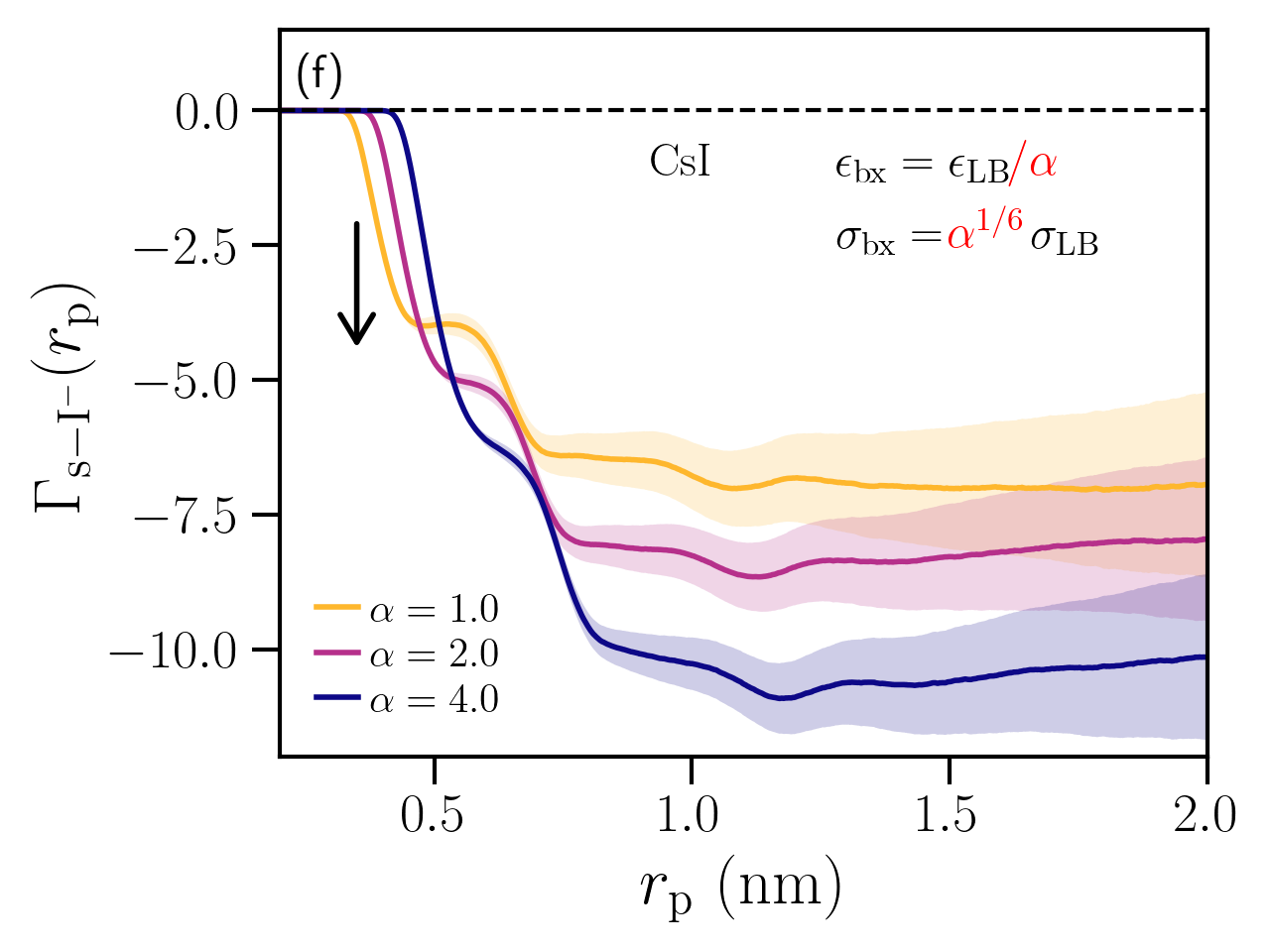}
\caption{Solute-iodide preferential binding coefficient profiles, $\Gamma_{\rm
s-I^{-}}~(r_{\rm p})$, for different values of $\alpha$ in aqueous LiI, NaI and
CsI solutions.  The top and bottom rows correspond to profiles for the coupled
$\sigma_{\rm bx}-\epsilon_{\rm bx}$ and the $\sigma_{\rm bx}$-only variations,
respectively.}
\label{fig:pa_iodide}
\end{figure}

To quantify this relative preference in both the variations, we compute the
preferential binding coefficient, $\Gamma_{\rm s-I^{-}}$. From
Fig.~\ref{fig:pa_iodide}, it can be seen that  $\Gamma_{\rm s-I^{-}}$ is
negative for both kinds of variations, indicating that iodide ions are depleted
from the solute surface. This in agreement with simulation studies on aqueous
single salt solutions of PNIPAM and generic uncharged polymers where it has been
observed that iodide ions are depleted irrespective of whether they salt-in or
salt-out the polymer.\cite{Zhao2022,Bruce2021,Goel2026}
Figs.~\ref{fig:pa_iodide}(a)-(c) show that $\Gamma_{\rm s-I^{-}}$ increases with
increasing $\alpha$ for the coupled $\sigma_{\rm bx}-\epsilon_{\rm bx}$
variation. This increase in $\Gamma_{\rm s-I^{-}}$ occurs despite a decrease in
the direct solute-I$^{-}$ van der Waals attraction and correlates with the
increasing disruption of the interfacial water network. This suggests that an
indirect interfacial-water-mediated effect contributes to the enhancement in ion
adsorption. On the contrary, $\Gamma_{\rm s-I^{-}}$ decreases with increasing
$\alpha$ for the $\sigma_{\rm bx}$-only variation. Together with the observation
that the interfacial water network remains unchanged with increasing $\alpha$,
this suggests that the effective solute-water attraction is stronger than the
effective solute-I$^{-}$ attraction. These trends indicate that
interfacial-water-mediated effects may play a more prominent role in ion
adsorption than direct solute–I$^{-}$ van der Waals interactions in this case.

From Figs.~\ref{fig:pa_iodide}, it can be seen that for a fixed value of
$\alpha$, $\Gamma_{\rm s-I^{-}}$ decreases in the order LiI $>$ NaI $>$ CsI.
This can be attributed to iodide–cation pairing which impedes the partitioning
of iodide ions to the solute interface. This extent of ion-pairing is largest
with Cs$^{+}$, followed by Na$^{+}$ and then Li$^{+}$. This interplay between
iodide accumulation around the solute and iodide-cation pairing is consistent
with the simulation studies of PNIPAM in aqueous alkali iodide
solutions.\cite{Bruce2021} Note that the cations themselves do not partition to
the solute interface. This can be observed from solute-cation proximal rdfs,
$g_{\rm p,s-X^{+}}$, which do not show any solute-cation contact peaks (see
Figs. S2(a)-(f) in the Supporting Information). Note that qualitatively similar
trends were also observed at smaller salt concentrations (see Figs.~S3 and S4 in the
Supporting Information). 

Given these trends in aqueous alkali iodide solutions, a question that arises is
whether similar behavior would arise in other salts containing weakly hydrated
anions.  To understand this, we investigated the anion adsorption in aqueous
alkali thiocyanate solutions for the two kinds of variations. Note that the
thiocyanate ion lies to the right of iodide in the Hofmeister series, indicating
that it is a more weakly hydrated anion.
\subsection{Extended solutes in aqueous alkali thiocyanate solutions} 
\begin{figure}[h!]
\includegraphics[width=0.4\textwidth]{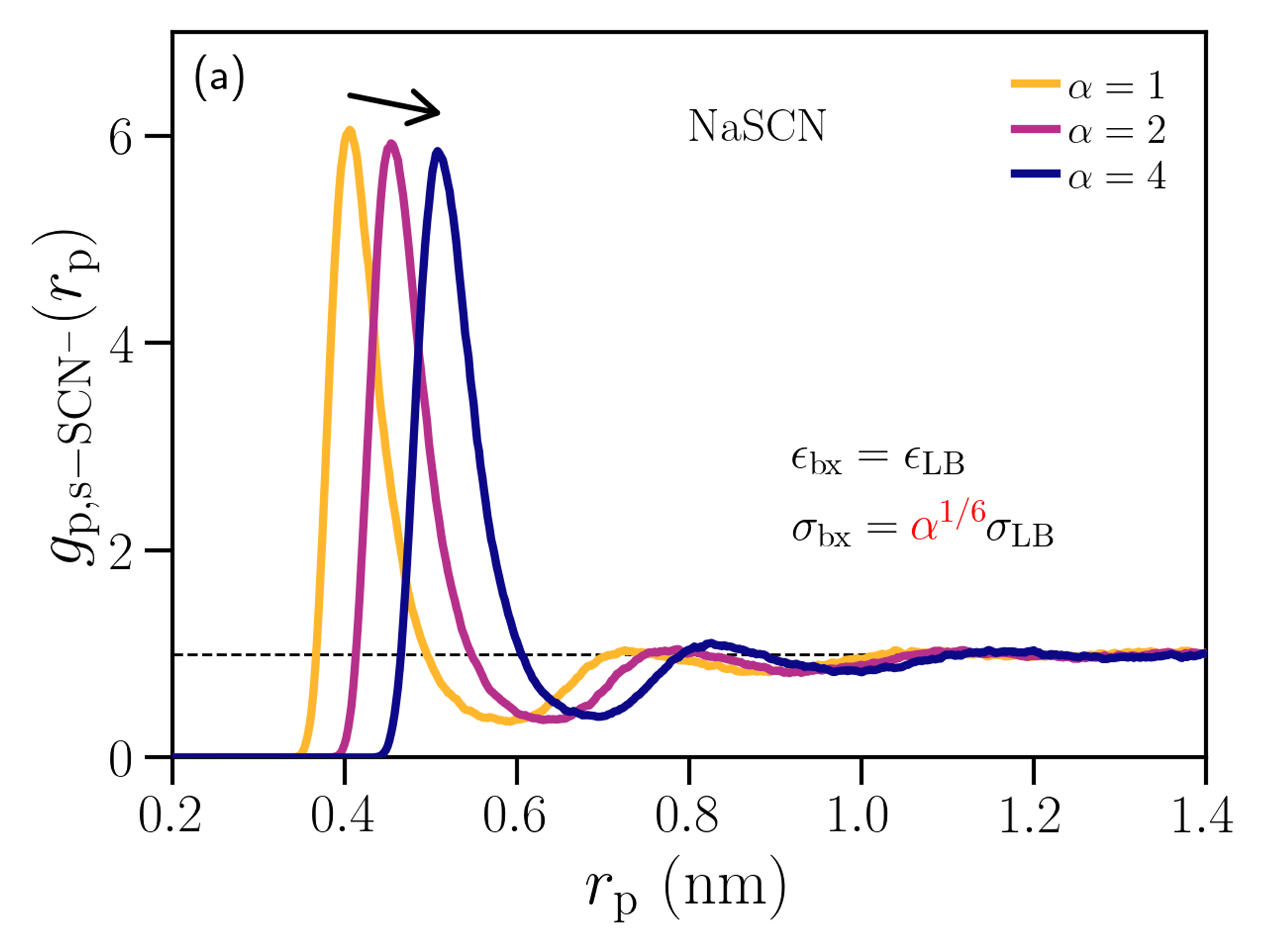}
\includegraphics[width=0.4\textwidth]{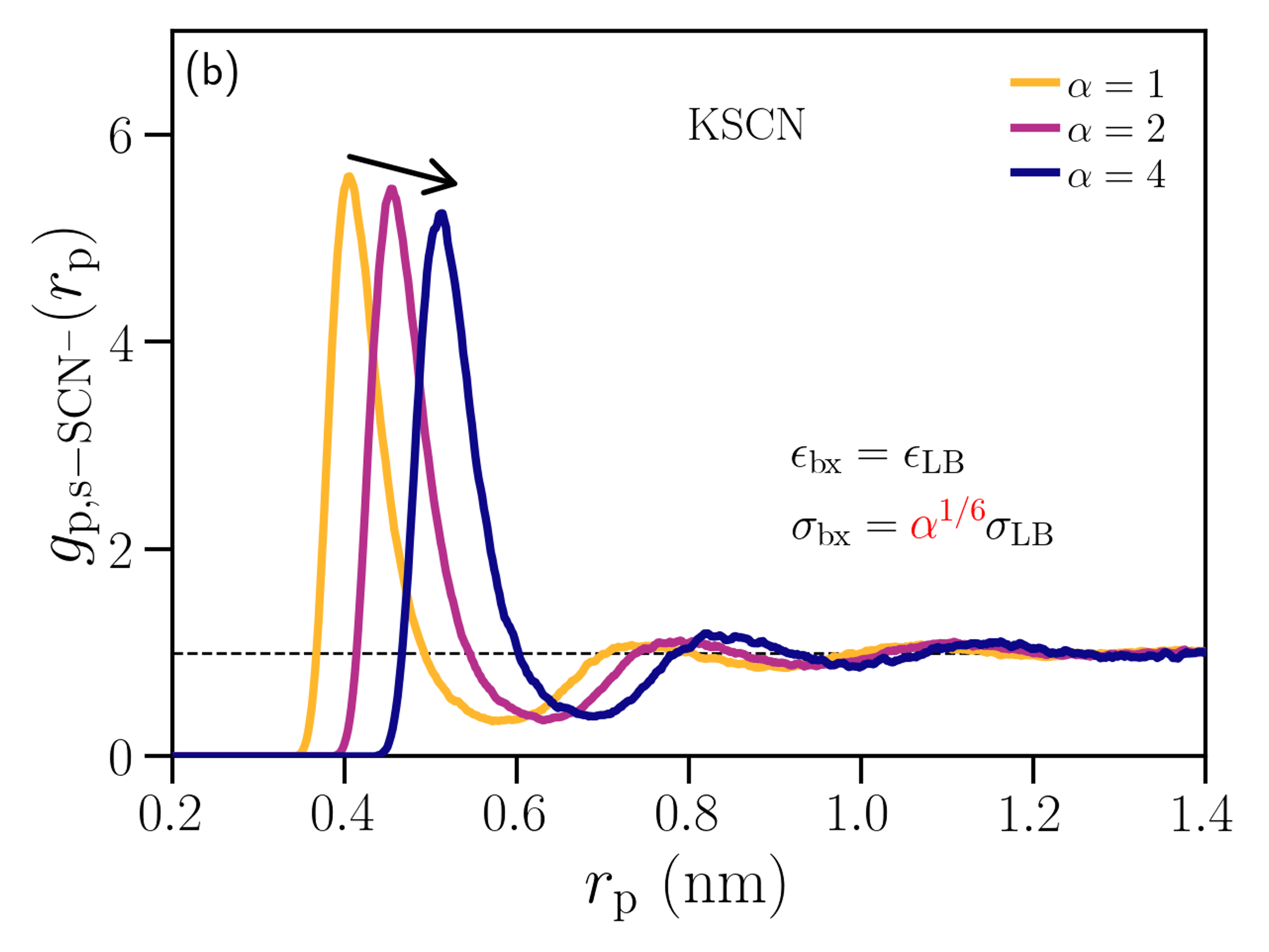}
\includegraphics[width=0.4\textwidth]{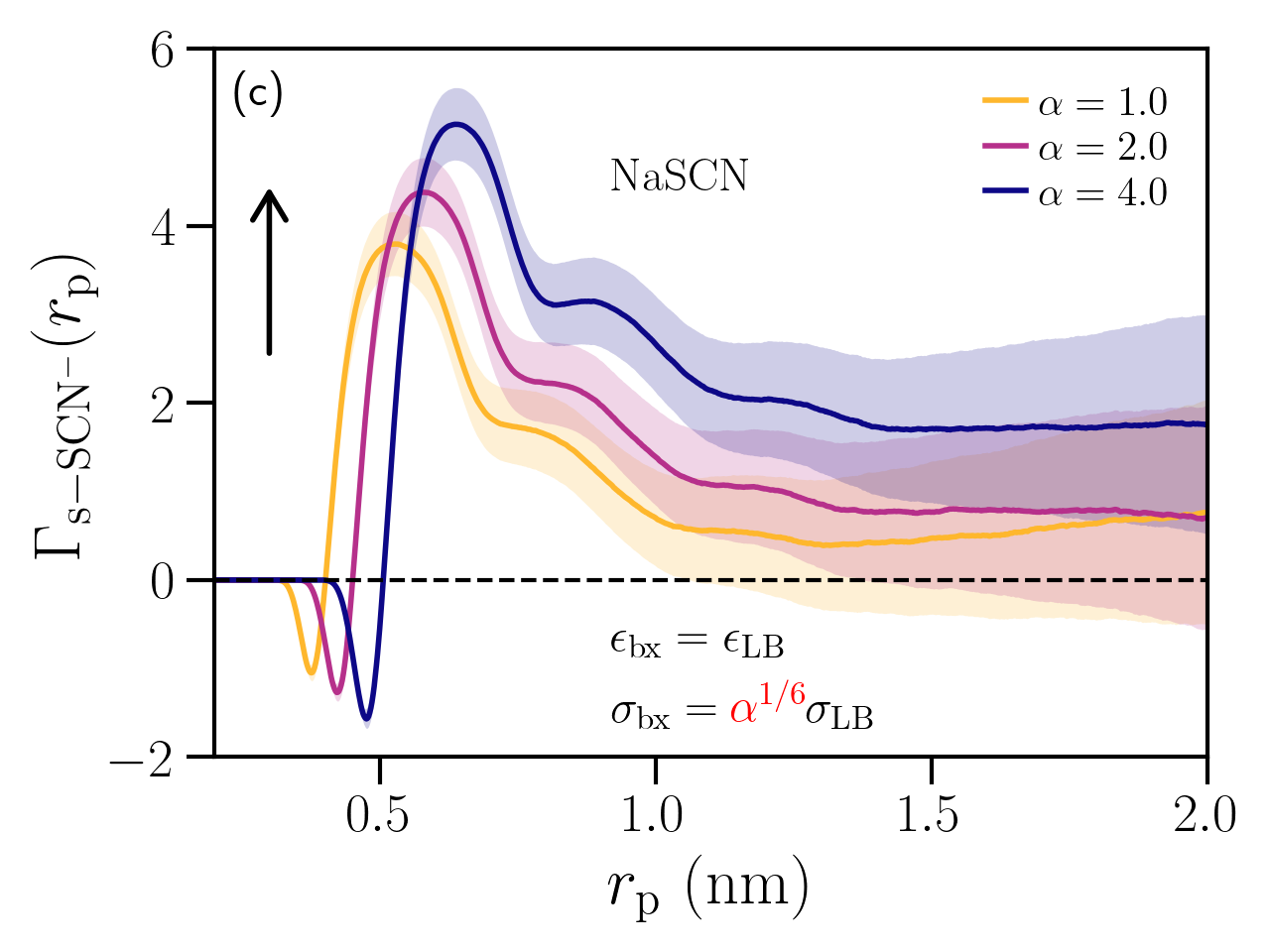}
\includegraphics[width=0.4\textwidth]{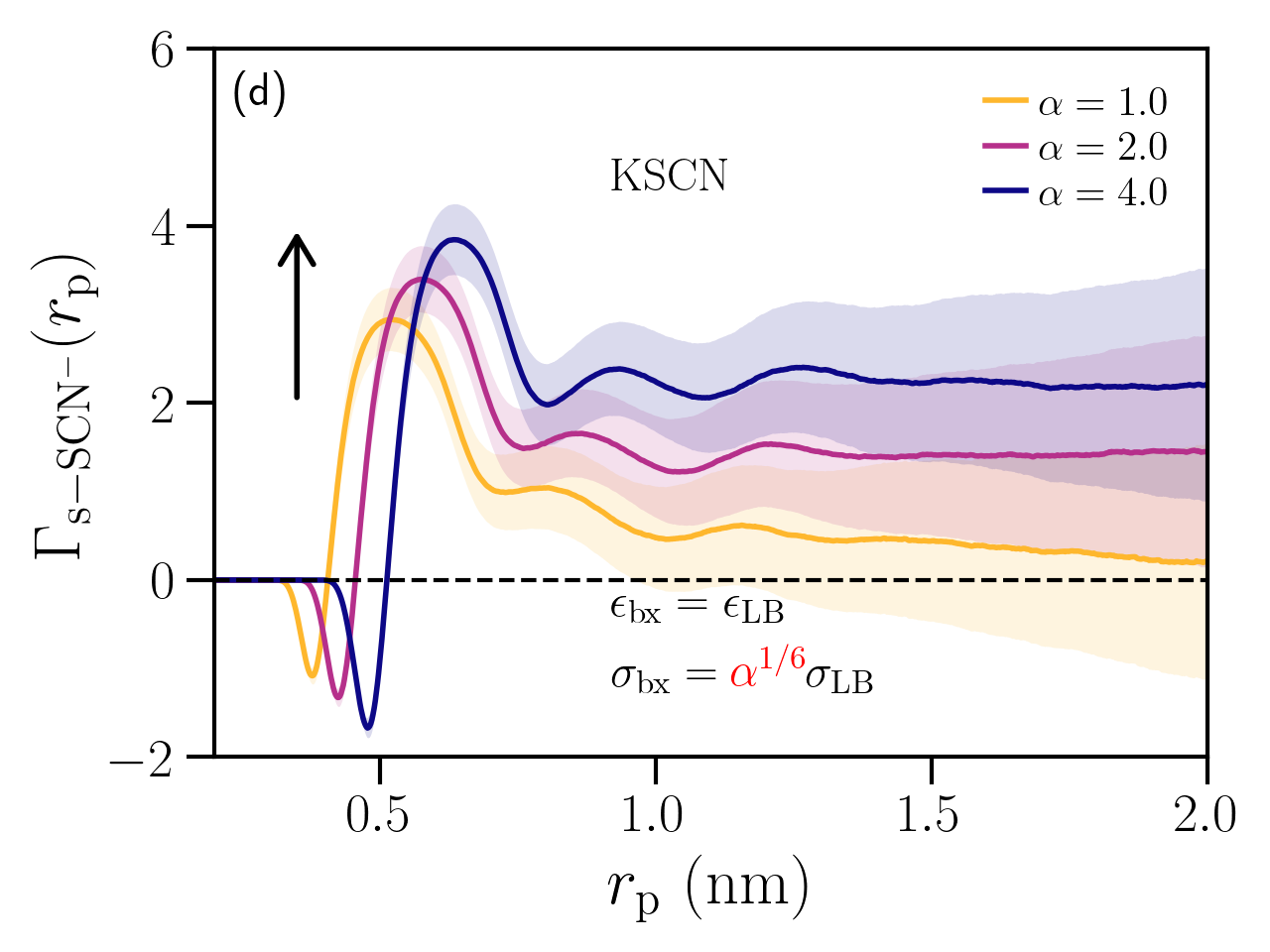}
\caption{Solute-thiocyanate proximal RDF profiles, $g_{\rm p,s-SCN^{-}}(r_{\rm
p})$, and preferential binding coefficients, $\Gamma_{\rm s-SCN^{-}}(r_{\rm
p})$, at different values of $\alpha$ in aqueous NaSCN and KSCN solutions for
the $\sigma_{\rm bx}$-only variation.}
\label{fig:prdf_thiocyanate_sigma}
\end{figure}

Figure~\ref{fig:prdf_thiocyanate_sigma} shows the solute-SCN$^{-}$ proximal RDFs
and the preferential binding coefficients at different values of $\alpha$ for
the $\sigma_{\rm bx}$-only variation in aqueous NaSCN and KSCN solutions. From
Figs.~\ref{fig:prdf_thiocyanate_sigma} (a) and (b), it can be seen that the
first peak of $g_{\rm p,s-SCN^{-}}$, at $\alpha=1$, is significantly larger than
the first peak of $g_{\rm p,s-I^{-}}$ indicating that solute has a larger
affinity for SCN$^{-}$ ions than I$^{-}$ ions. Note that the first peak of
$g_{\rm p,s-SCN^{-}}$ decreases slightly with increasing $\alpha$. A similar
decrease is observed for $g_{\rm p,s-w}$ (see Fig. S5 (a)-(b) in the Supporting
Information), consistent with the trends observed in pure water and aqueous
alkali iodide solutions for the $\sigma_{\rm bx}$-only variation. Thus, in contrast
to aqueous alkali iodide solutions, $g_{\rm p,s-w}$ and $g_{\rm p,s-SCN^{-}}$
show similar responses to increasing $\alpha$.
Figs.~\ref{fig:prdf_thiocyanate_sigma}(c) and
~\ref{fig:prdf_thiocyanate_sigma}(d) show that $\Gamma_{\rm s-SCN^{-}}$ is
positive and increases with increasing $\alpha$.  Since the increase in $\alpha$
increases the number of solute-water and solute-SCN$^{-}$ contacts (same
$\epsilon_{\rm bx}$), this shows that the solute has a preference for
solute-SCN$^{-}$ contacts over solute-water contacts. This suggests that the
effective solute-SCN$^{-}$ attraction is stronger than the effective
solute-water attraction, indicating that direct solute-SCN$^{-}$ interactions
play an important role in ion adsorption. The observed preferential accumulation
of SCN$^{-}$ ions around the solute is consistent with previous simulation
studies of Polyethylene Glycol and generic uncharged polymers in aqueous NaSCN
solutions.\cite{Rogers2022,Goel2026}

\begin{figure}[h!]
\includegraphics[width=0.4\textwidth]{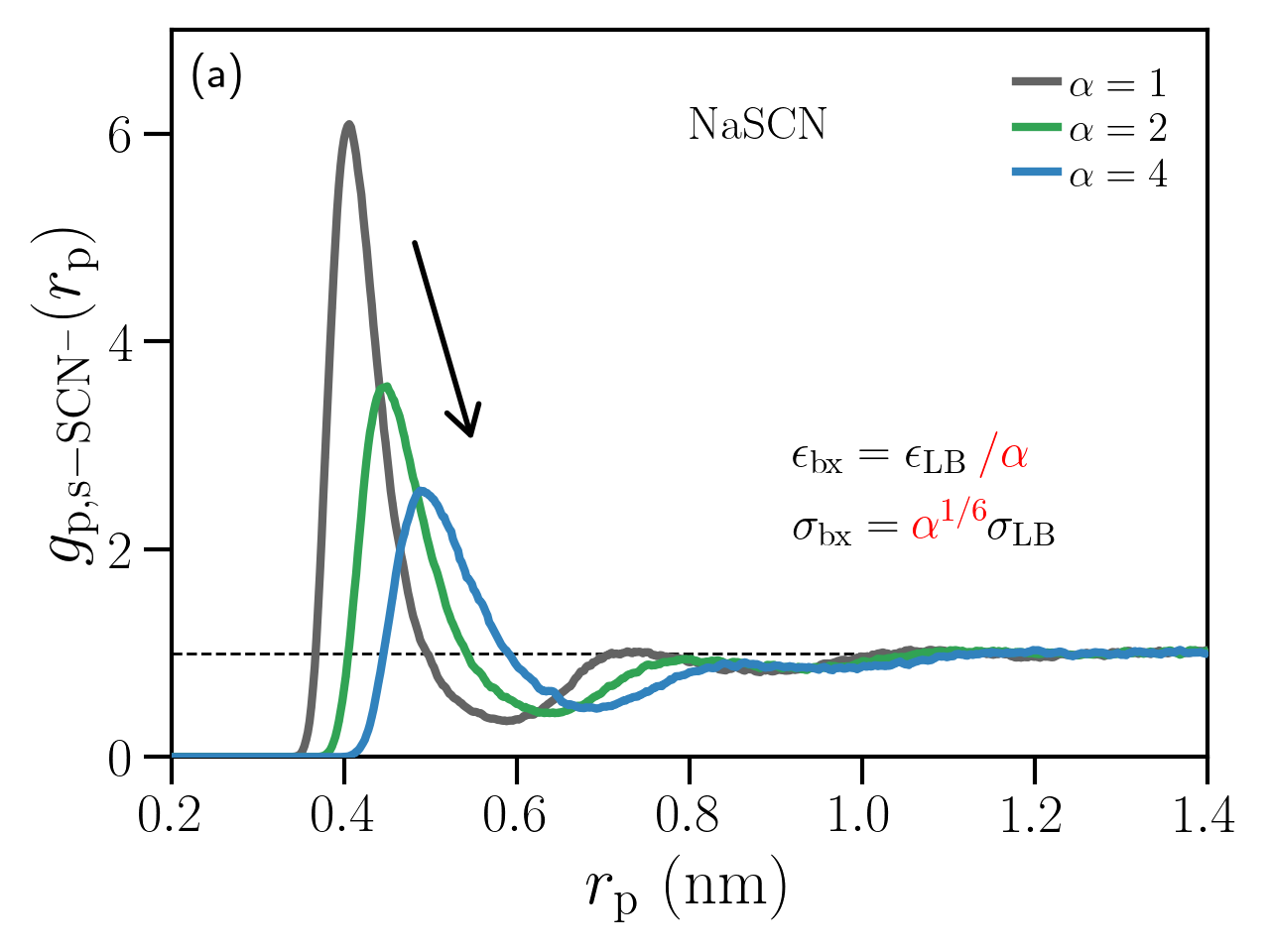}
\includegraphics[width=0.4\textwidth]{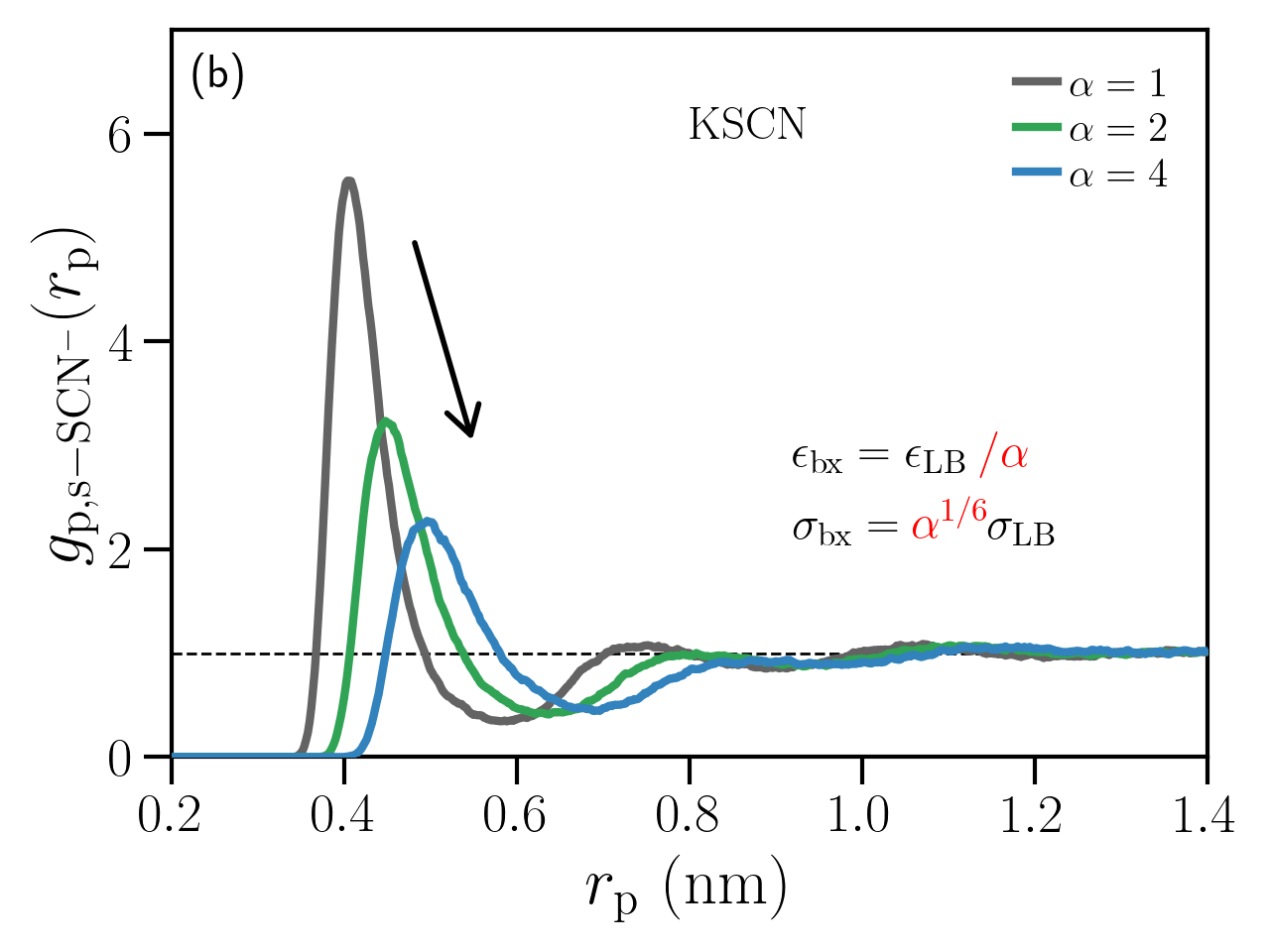}
\includegraphics[width=0.4\textwidth]{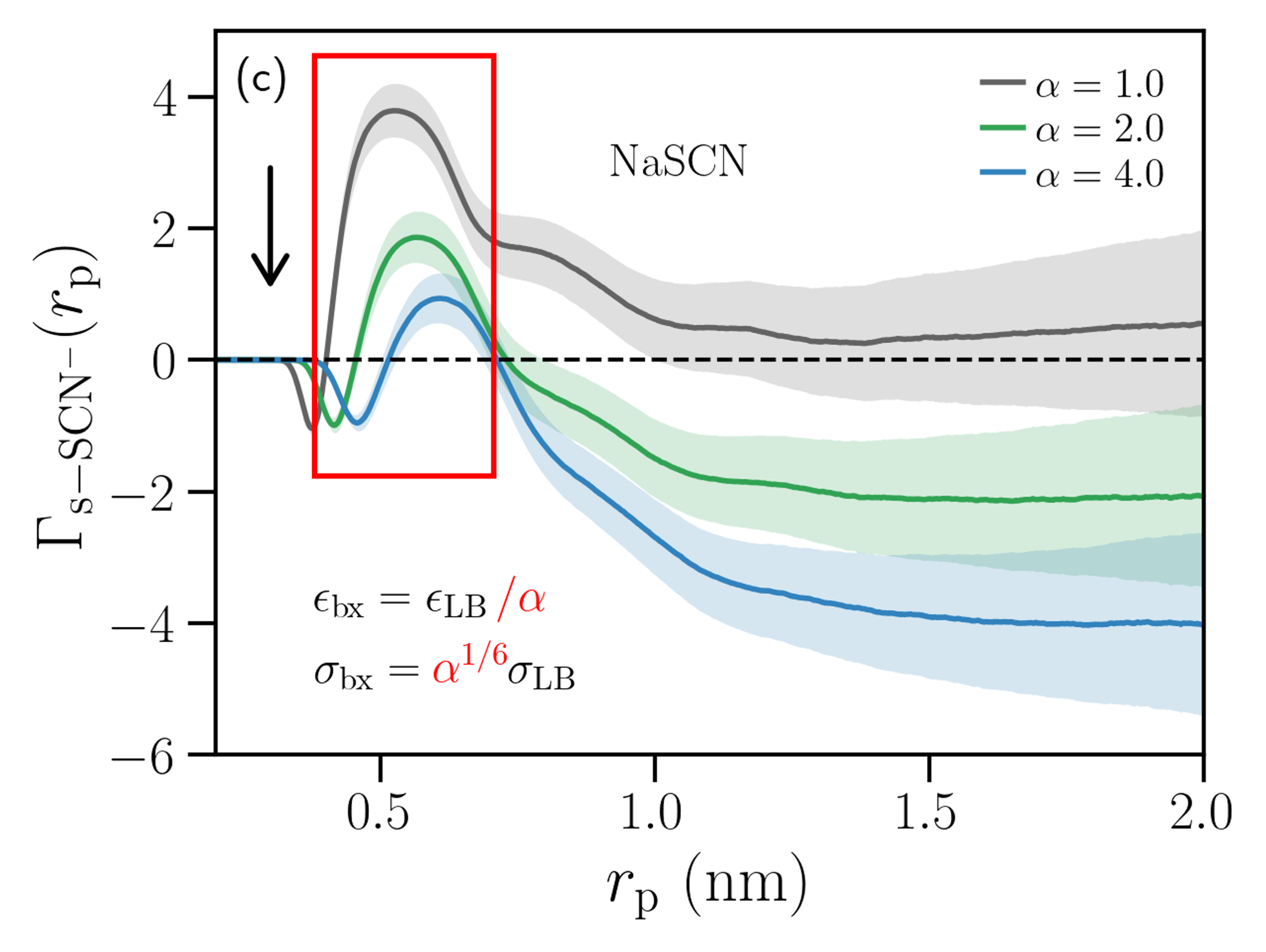}
\includegraphics[width=0.4\textwidth]{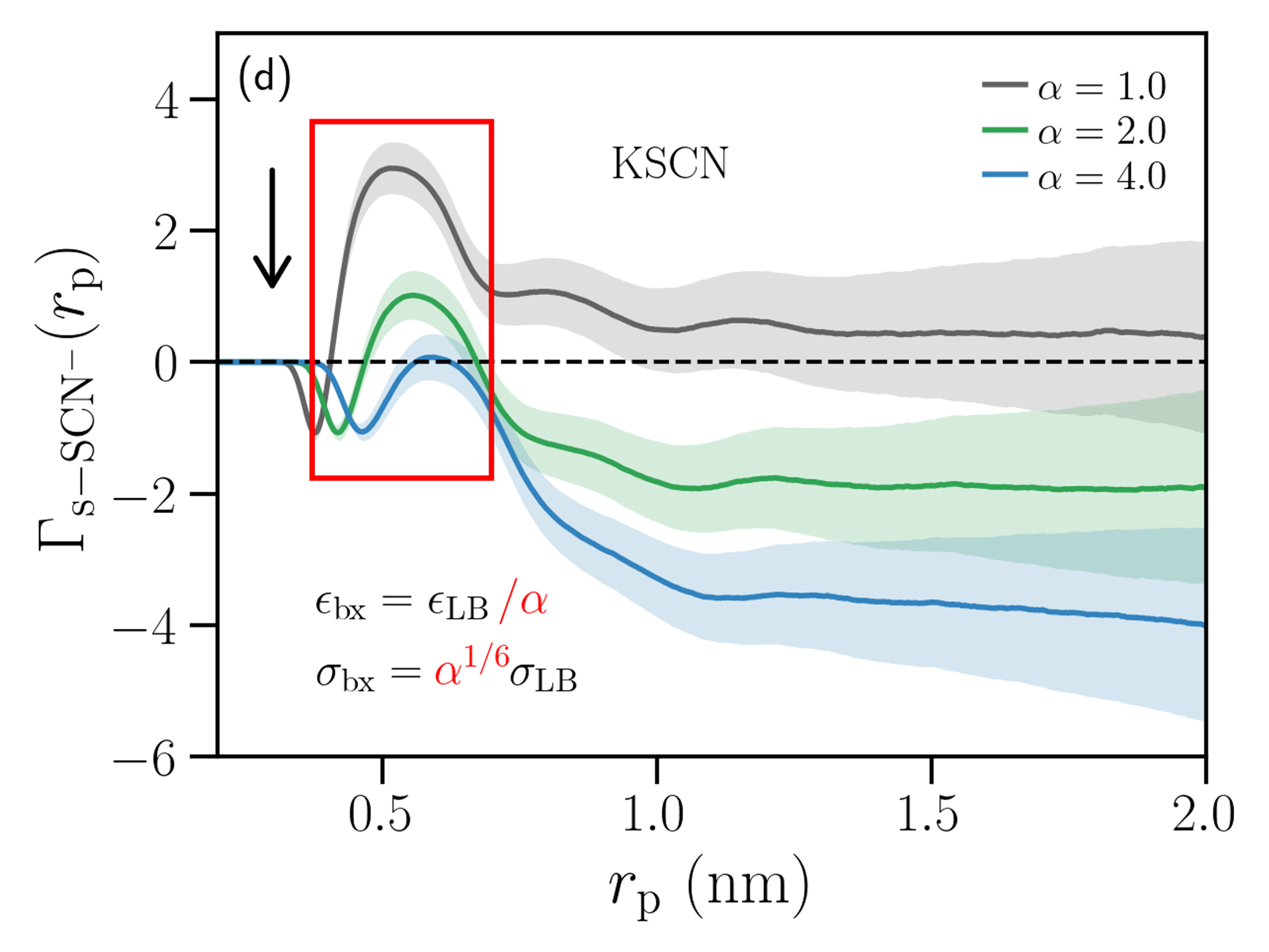}
\caption{Solute-thiocyanate proximal RDF profiles, $g_{\rm p,s-SCN^{-}}(r_{\rm
p})$, and preferential binding coefficients, $\Gamma_{\rm s-SCN^{-}}(r_{\rm
p})$, at different values of $\alpha$ in aqueous NaSCN and KSCN solutions for
the coupled $\sigma_{\rm bx}-\epsilon_{\rm bx}$ variation.}
\label{fig:prdf_thiocyanate_coupled}
\end{figure}

Interestingly, for the coupled $\sigma_{\rm bx}-\epsilon_{\rm bx}$ case,  the
first peak of $g_{\rm p,s-SCN^{-}}$ decreases significantly with increasing
$\alpha$, indicating a decrease in the solute-SCN$^{-}$ affinity (see
Figs.~\ref{fig:prdf_thiocyanate_coupled}(a) and
~\ref{fig:prdf_thiocyanate_coupled}(b)). Similar to the trends observed in pure
aqueous and alkali iodide solutions, $g_{\rm p,s-w}$ shows a significant
decrease with increasing $\alpha$ (See Fig. S5(c)-(d) in the Supporting
Information). Note that both  $g_{\rm p,s-w}$ and $g_{\rm
p,s-SCN^{-}}$ have a similar response to increasing $\alpha$ in this case as
well. From Figs.~\ref{fig:prdf_thiocyanate_coupled}(c) and
~\ref{fig:prdf_thiocyanate_coupled}(d), it can seen that $\Gamma_{\rm
s-SCN^{-}}$ decreases with increasing $\alpha$, indicating that the solute-water
contacts decrease at a smaller rate than solute-SCN$^{-}$ contacts. These trends
indicate that the direct solute-SCN$^{-}$ interactions plays an important role
in the adsorption of SCN$^{-}$ ions. Interestingly, both $g_{\rm s-SCN^-}$ and
$\Gamma_{\rm s-SCN^-}$ exhibit a more pronounced decrease between $\alpha=$1 and
2 than between $\alpha=$2 and 4. This disproportionate decrease can be clearly
observed in the first peaks of $\Gamma_{\rm s-SCN^{-}} (r_{\rm p})$ (see red
rectangles in Figs.~\ref{fig:prdf_thiocyanate_coupled}(c) and
~\ref{fig:prdf_thiocyanate_coupled}(d)) which correspond to the contribution to
$\Gamma_{\rm s-SCN^{-}}$ from the shells in the immediate vicinity of the
solute. Note that such disproportionate changes are not observed in $\Gamma_{\rm
s-SCN^{-}}(r_{\rm p})$ for the $\sigma_{\rm bx}$-only variation and $\Gamma_{\rm
s-I^{-}}(r_{\rm p})$
(see Fig.~\ref{fig:pa_iodide}). In the coupled $\sigma_{\rm bx}-\epsilon_{\rm
bx}$ case, the indirect interfacial-water-mediated contribution to ion
adsorption increases with increasing $\alpha$ due to a larger disruption of the
interfacial water network. This contribution partly compensates for the decrease
in SCN$^{-}$ binding due to weaker solute-SCN$^{-}$ attraction, and may explain
the smaller decrease in $\Gamma_{\rm s-SCN^{-}}$ and $g_{\rm p,s-SCN^{-}}$ with
increasing $\alpha$ at larger values of $\alpha$. These results indicate that
both direct solute–SCN$^{-}$ interactions and indirect
interfacial-water-mediated effects play important roles in SCN$^{-}$ ion
adsorption.

Similar to aqueous alkali iodide solutions, it can be seen that $\Gamma_{\rm
s-SCN^{-}}$ decreases in the order NaSCN$>$KSCN for a fixed value of $\alpha$
(see Figs.~\ref{fig:prdf_thiocyanate_sigma} and
~\ref{fig:prdf_thiocyanate_coupled}). This can be attributed to anion-cation
pairing which impeded the partitioning of the SCN$^{-}$  anions to the solute
interface. In NaSCN solutions, Na$^{+}$ do not partition ot the solute surface
leading to the absence of any solute-cation contact peaks (see Figs. S6(a) and
(c) in the Supporting Information). K$^{+}$ ions do exhibit a mild solute-cation
contact peak though the intensity of the peak is less than 1 (see Figs. S6(b)
and (d) in the Supporting Information).
\section{Conclusions}
We have studied the adsorption of two weakly hydrated anions, I$^{-}$ and
SCN$^{-}$, on a model extended hydrophobic solute. The contributions to ion
adsorption from the direct solute-ion van der Waals attraction and the indirect
interfacial-water-mediated effects were investigated using the coupled
$\sigma_{\rm bx}-\epsilon_{\rm bx}$ and $\sigma_{\rm bx}$-only variations.  In
the coupled $\sigma_{\rm bx}-\epsilon_{\rm bx}$ variation, the effective solute
size was increased with a simultaneous decrease in the direct solute-water and
solute-ion attractive interaction. In pure aqueous solutions, this led to a
decrease in both the solute-water affinity and the number of water-water
hydrogen bonds (intra and inter shell) per first solvation shell water molecule,
indicating a monotonic increase in the disruption of the interfacial water
network. For the $\sigma_{\rm bx}$-only variation, the effective solute size was
increased without changing the direct solute-water and solute-ion attractive
interactions. In pure aqueous solutions, the solute-water affinity and the
number of water-water hydrogen bonds (intra- and inter-shell) per first
solvation shell water molecule remained largely the same under this variation,
indicating that the interfacial water network was not disrupted. These two
variation allow us to isolate the contribution from the indirect
interfacial-water-mediated effects to ion adsorption.

In aqueous alkali iodide solutions, the solute-iodide affinity increases in the
coupled $\sigma_{\rm bx}-\epsilon_{\rm bx}$ variation, indicating that the
increase in ion adsorption correlates with the enhanced disruption of the
interfacial water network. Interestingly, this increase in affinity occurs even
though  the direct solute-iodide attractive interaction decreases.  Note that
iodide ions are depleted from the solute surface, although the extent of
depletion decreases monotonically in this variation. Based on these results, it
can be said that the disruption of the interfacial water network provides a
significant indirect driving force for the adsorption of iodide ions. On the
contrary, in the $\sigma_{\rm bx}$-only variation, the solute-iodide affinity
decreases and the extent of iodide depletion from the solute surface increases.
As the interfacial area increases, the solute favors contacts with water
molecules over iodide ions. This also correlates with the observation that the
interfacial water network remains undisturbed in this variation. These results
show that the adsorption of iodide ions is dependent on both direct
solute-iodide interactions and the indirect interfacial-water-mediated effect,
with the latter contribution playing a prominent role. In contrast to iodide
ions, the adsorption of thiocyanate ions is predominantly dictated by the direct
solute-thiocyanate interactions, with the indirect interfacial-water-mediated
effects becoming significant at higher of disruption of the interfacial
water network. This can be seen from the observations that both the
solute-thiocyanate affinity and preferential adsorption of thiocyanate ions
increase in the $\sigma_{\rm bx}$-only variation. In the coupled $\sigma_{\rm
bx}-\epsilon_{\rm bx}$ variation, both the solute-thiocyanate affinity and the
preferential adsorption of thiocyanate ions decrease. Interestingly, there is a
reduction in the rate of this decrease, which may attributed to the indirect
interfacial-water-mediated effects which become prominent at higher levels of
disruption of the solvation shell water network. For both the weakly hydrated
anions, the choice of cation influences the extent of ion adsorption, although
it does not affect the qualitative trends. It was observed that the adsorption
of the anion decreases with increasing cation-anion pairing, which is consistent
with the simulation studies of PNIPAM in aqueous alkali iodide
solutions.\cite{Bruce2021} 

Overall, our results highlight that the indirect interfacial-water-mediated
contribution promotes the adsorption of weakly hydrated anions on macromolecular
surfaces. This indirect effect may be a significant factor contributing to the
enhanced adsorption of weakly hydrated anions in mixed-salt solutions containing
both weakly and strongly hydrated anions. Given that strongly hydrated anions
are depleted from polymer surfaces and progressively salt out polymers, it is
reasonable to expect that the interfacial water network would be disrupted to a
greater extent in their presence. Consequently, the indirect contribution to
weakly hydrated anion adsorption may be greater in solutions containing strongly
hydrated anions than in solutions without them, and may increase with increasing
concentration of the strongly hydrated anions. These inferences are consistent
with the trends observed in previous simulation studies of
Poly(N-isopropylacryamide) (PNIPAM) and generic uncharged polymers in aqueous
mixed salt solutions.\cite{Zhao2022,Goel2026} Direct solute–anion interactions
can reinforce the adsorption of weakly hydrated anions. This can be seen in the
trends observed in mixed-salt solutions of PNIPAM and the generic uncharged
polymer. For PNIPAM, iodide ions transition from being depleted to being
preferentially adsorbed in the presence of 0.3 m sodium sulfate. However, for
the generic polymer, iodide depletion decreases, although iodide ions remain
depleted even at 0.5 m sodium sulfate. This difference may arise from favorable
PNIPAM–iodide interactions that reinforce iodide adsorption driven by the
interfacial-water-mediated contribution, which may be very weak for the generic
polymer.
\section*{Data Availability Statement}
The data that support the findings of this study are available from the
corresponding author upon reasonable request.
\section{Acknowledgments}
SB acknowledges financial assistance provided by the Ministry of Education,
India through SERB, India (SRG/2023/000202). We acknowledge the super-computing
facility of Shiv Nadar Institution of Eminence, Magus02, for providing the
computational support.

\bibliography{reference,ref}
\section*{Supplementary Information}
\setcounter{section}{0}
\setcounter{figure}{0}
\setcounter{table}{0}
\renewcommand{\thesection}{S\arabic{section}}
\renewcommand{\thefigure}{S\arabic{figure}}
\renewcommand{\thetable}{S\arabic{table}}
\renewcommand{\thesection}{S\arabic{section}}
\renewcommand{\thesubsection}{\thesection.\arabic{subsection}}
\renewcommand{\thesubsubsection}{\thesubsection.\arabic{subsubsection}}
\section{Non-bonded interaction Parameters}
\subsection{Ion-ion and ion-water interactions}
\label{sec:interaction_parameters}
Water is modeled using the SPC/E model. Non-polarizable force fields were
employed for all the ions. For aqueous iodide solutions, the forcefield
parameters for sodium, cesium (model Cs6), potassium and iodide (model I4) ions
were taken from the work of Fyta and Netz was used.\cite{Netz:2012} For LiI, the
model developed by Bruce et al.\cite{} by combining the Li$^{+}$ model from the
work of Hess and van der Vegt\cite{Hess2009} with model I(4) from the work of
Fyta and Netz\cite{Netz:2012} has been used. For NaSCN and KSCN, the force-field
parameters for sodium, potassium and thiocyanate ions were taken from the work
of Tesei et al.\cite{Tesei2018} The partial charges and self-interaction
Lennard-Jones parameters are listed in Table~\ref{tab:SI_nonbonded}.

\begin{table}[ht!]
    \centering
    \caption{Partial charges and self-interaction Lennard-Jones parameters. S-P,
CSP and NSP respresent the sulphur, carbon and nitrogen atoms of the thiocyanate
anion.}
    \label{tab:SI_nonbonded}
    \begin{tabular}{l  c c c }
        \toprule
        \textbf{Name}  & \textbf{Charge}& \textbf{$\sigma_{\rm ii}$ (nm)} &
\textbf{$\epsilon_{\rm ii}$ (kJ/mol)} \\\hline
        OW  & -0.8476& 0.31656 & 0.65019 \\
        HW  & 0.4238 & 0.00000 & 0.00000 \\\hline
        I$^{-}$   & -1.0000& 0.53310 & 0.15700 \\
        Na$^{+}$   & 1.0000 & 0.25830 & 0.41860\\ 
        Li$^{+}$   & 1.0000 & 0.2 & 0.5  \\
        Cs$^{+}$   & 1.0000 & 0.3491 & 0.325  \\
        \bottomrule
    \end{tabular}
\hspace{2ex}
    \begin{tabular}{l  c c c }
        \toprule
        \textbf{Name}  & \textbf{Charge}& \textbf{$\sigma_{\rm ii}$ (nm)} &
\textbf{$\epsilon_{\rm ii}$ (kJ/mol)} \\\hline
        S-P  & -0.573 & 0.38308 & 1.52256 \\
        CSP &0.483& 0.334977  & 0.425093 \\
        NSP &-0.910& 0.369722  & 0.310034\\\
        Na$^{+}$   & 1.0000 & 0.255 & 0.28\\
        K$^{+}$   & 1.0000 & 0.403& 0.85  \\
        \bottomrule
    \end{tabular}
\end{table}

 \begin{table}[h!]
    \centering
    \caption{$\lambda_{\rm \epsilon,ij}$ and $\lambda_{\rm,\sigma,ij}$ values
for cross-interactions utilizing LB rule.}
    \label{tab:SI_LB}
    \begin{tabular}{l l c c}
        \toprule
        i & j & $\lambda_{\rm \epsilon,ij}$ & $\lambda_{\rm \sigma,ij}$ \\
        \midrule
        OW  & Na$^{+}$ &1 & 1 \\
		OW& Li$^{+}$ & 0.4&1\\\hline  		 
        \bottomrule
    \end{tabular}
\hspace{2ex}
    \begin{tabular}{l l c c}
        \toprule
        i & j & $\lambda_{\rm \epsilon,ij}$ & $\lambda_{\rm \sigma,ij}$ \\
        \midrule
        Na$^{+}$  & I$^{-}$  &0.9 &1  \\
		Li$^{+}$ & I$^{-}$ &1&0.93\\\hline
        \bottomrule
    \end{tabular}
\end{table}

For certain cross-interactions in the alkali iodide solutions, the
Lorentz-Berthelot (LB) rule along with scaling factors was employed,
\begin{equation}
\begin{split}
\epsilon_{\rm ij}=\lambda_{\rm \epsilon,ij}\sqrt{\epsilon_{\rm ii}\epsilon_{\rm
jj}},~~\sigma_{\rm ij}=\lambda_{\rm \sigma,ij}\left(\frac{\sigma_{\rm ii}+\sigma_{\rm
jj}}{2}\right).
\end{split}
\end{equation}
Table~\ref{tab:SI_LB} lists the $\lambda_{\rm \epsilon,ij}$ and
$\lambda_{\rm,\sigma ij}$ values for the cross-interactions utilizing LB rule.
All other cross-interaction parameters were computed using standard LB rule.

\subsection{Solute-water and solute-ion interactions}
The extended solute consists of 80 uncharged beads (CT1) which interact with
water molecules and ions via LJ interactions. The self interaction LJ parameters
for the beads are $\epsilon_{\rm bb}=1$ kJ/kmol and $\sigma_{\rm bb}=0.4$ nm.
The LJ parameters for solute-water and solute-ion interactions at different values of
$\alpha$ for the coupled $\sigma_{\rm bx}-\epsilon_{\rm bx}$ variation are
listed in Table~\ref{tab:SI_alpha}. For the $\sigma_{\rm bx}$-only variation,
$\epsilon_{\rm bx}$ is equal to the corresponding $\epsilon_{\rm LB}$, and
$\sigma_{\rm bx}$ varies with $\alpha$ as given in Table~\ref{tab:SI_alpha}.
 \begin{table}[h!]
    \centering
    \caption{Lennard-Jones interaction parameters for bead-ion and
bead-water combinations at different $\alpha$ values for the coupled
$\sigma_{\rm bx}-\epsilon_{\rm bx}$ variation. The beads in the solute are
represented by CT1.}
    \label{tab:SI_alpha}
    \begin{tabular}{p{2.8cm} p{1cm}||p{1.5cm} p{1.5cm}|| p{1.5cm} p{1.5cm}||  p{1.5cm}
p{1.5cm}|| p{1.5cm} p{1.5cm}}
        \toprule
$\sigma_{\rm bx}=\alpha^{1/6}\sigma_{\rm LB}$ $\epsilon_{\rm bx}=\epsilon_{\rm
LB}/\alpha$&&&&$\alpha=1$&&$\alpha=2$&&$\alpha=4$&\\\hline
        b & x &$\sigma_{\rm LB}$&$\epsilon_{\rm LB}$&$\sigma_{\rm bx}$&$\epsilon_{\rm bx}$&$\sigma_{\rm bx}$&$\epsilon_{\rm bx}$&$\sigma_{\rm bx}$&$\epsilon_{\rm bx}$\\\hline
Iodide salts&&&&&&&&&\\\hline
CT1&Li&  	0.3	        &0.70710	&0.3	    &0.70710	&0.3367    &0.3535	&0.3779	&0.1767\\
CT1&Cs&		0.37455	    &0.57008	&0.3745     &0.57008    &0.4204    &0.2850	&0.4719	&0.1425\\
CT1&Na&		0.32915	    &0.64699	&0.3291     &0.64699    &0.3694	   &0.3234	&0.4147	&0.1617\\
CT1&I &  	0.46655	    &0.39623	&0.4665     &0.39623    &0.5236	   &0.1981	&0.5878	&0.0990\\
CT1&OW&		0.35827	    &0.80634    &0.3582	    &0.80634    &0.4021    &0.4031
&0.4514	&0.2015\\\hline\hline
Thiocyanate salts&&&&&&&&&\\\hline
CT1&	K	&	0.4015&	0.9219&0.4015&	0.9219&	0.4507&	0.4610&0.5058&	0.2305\\
CT1&	NA	&	0.3275&	0.5292&0.3275&	0.5292&	0.3676&0.2646&0.41262&	0.1323\\
CT1&	S-P&	0.3915&	1.2341&0.3915&	1.2341&	0.4394&	0.6170&0.49325&0.3085\\
CT1&	CSP&	0.3675&	0.6519&0.3675&	0.6519&	0.41250&0.3260 &0.4630&0.1629\\
CT1&	NSP&	0.3850&	0.5568&0.3850&	0.5568&	0.43215&0.27848&0.4851&
0.1392\\\hline
\bottomrule 
\end{tabular}
\end{table}
\section{Extended solute in aqueous alkali iodide solutions}
\subsection{Solute-water and solute-cation RDFs at 1 m salt concentration}
\begin{figure}[h!]
\includegraphics[width=0.32\textwidth]{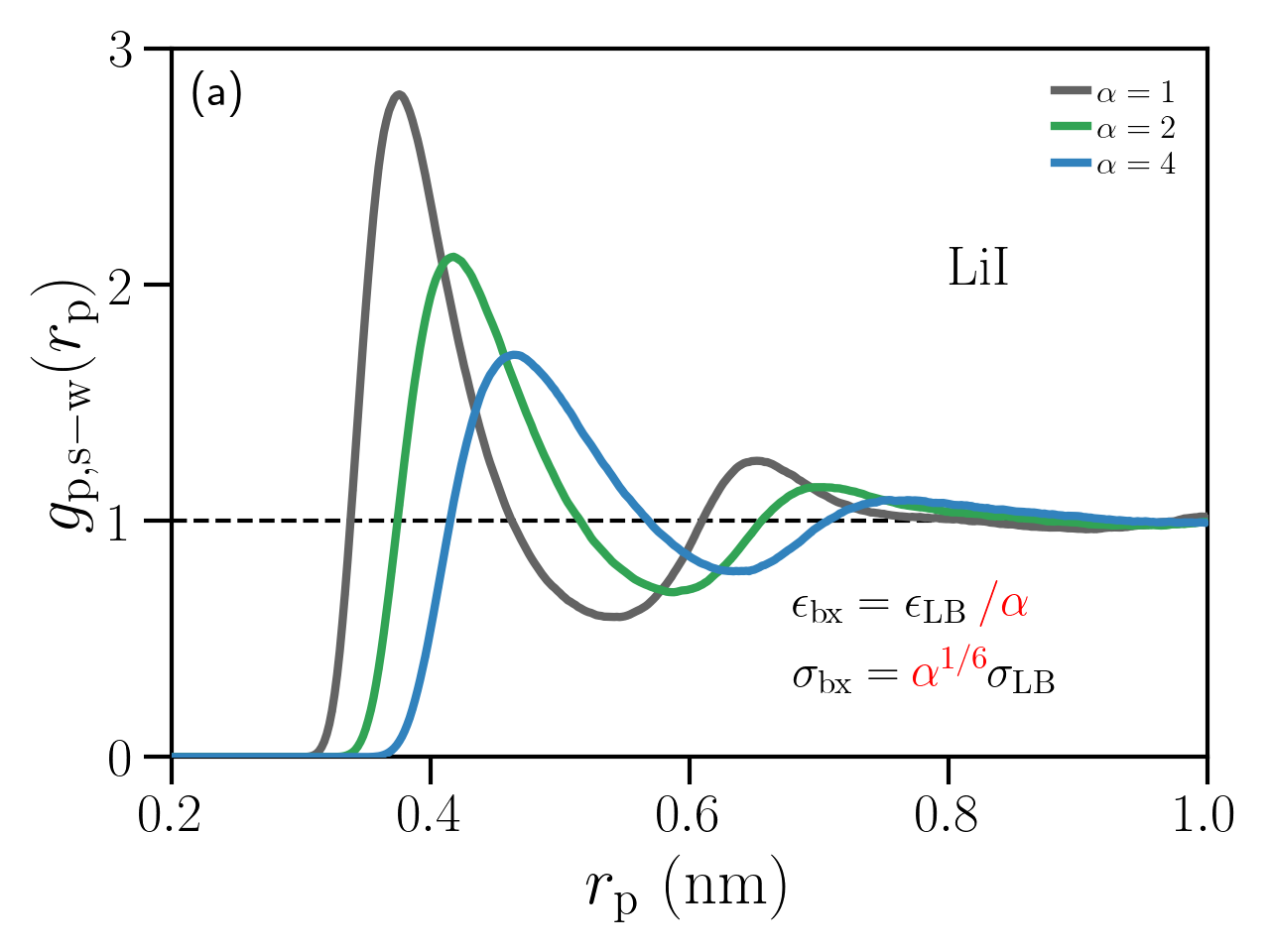}
\includegraphics[width=0.32\textwidth]{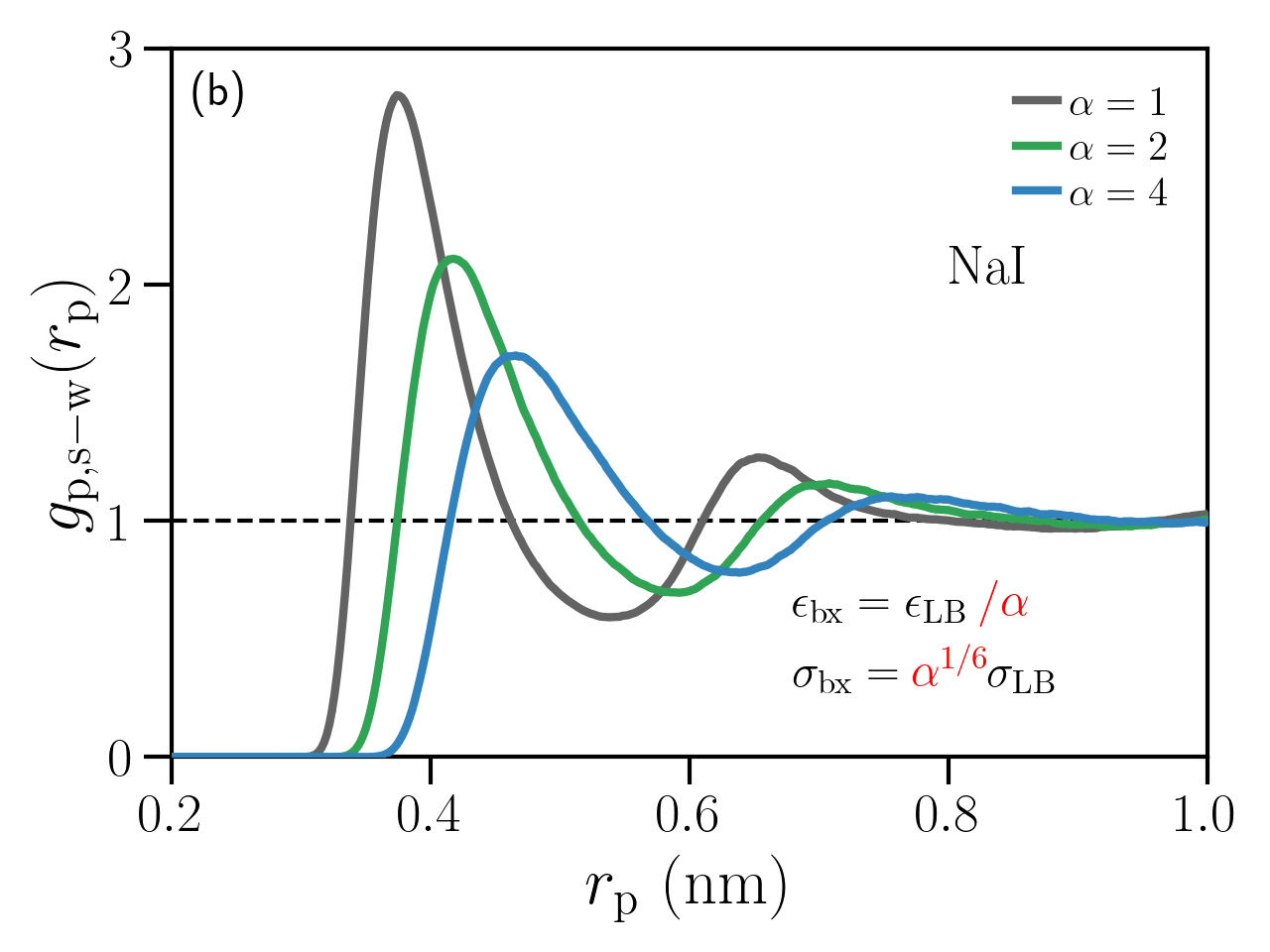}
\includegraphics[width=0.32\textwidth]{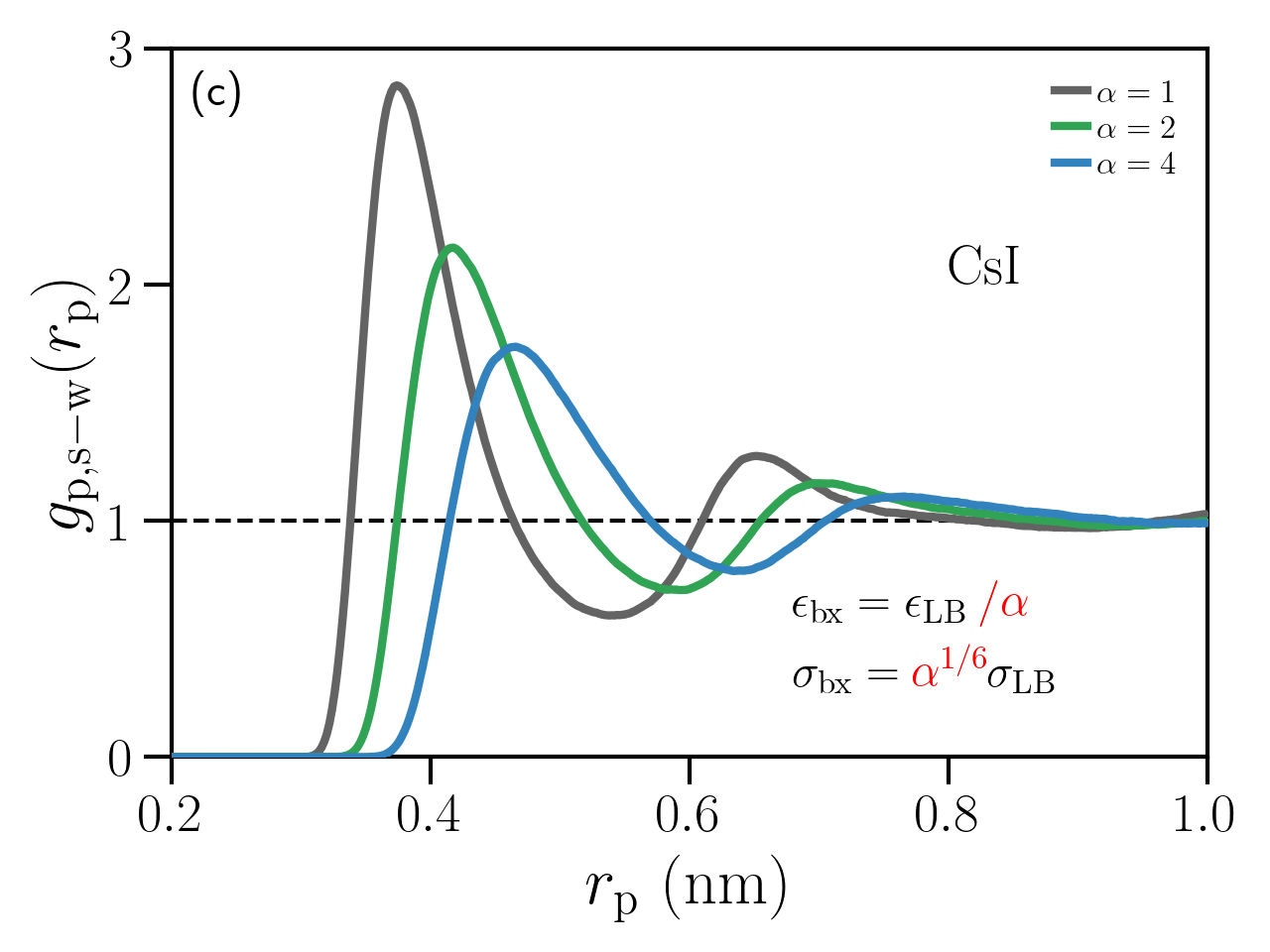}
\includegraphics[width=0.32\textwidth]{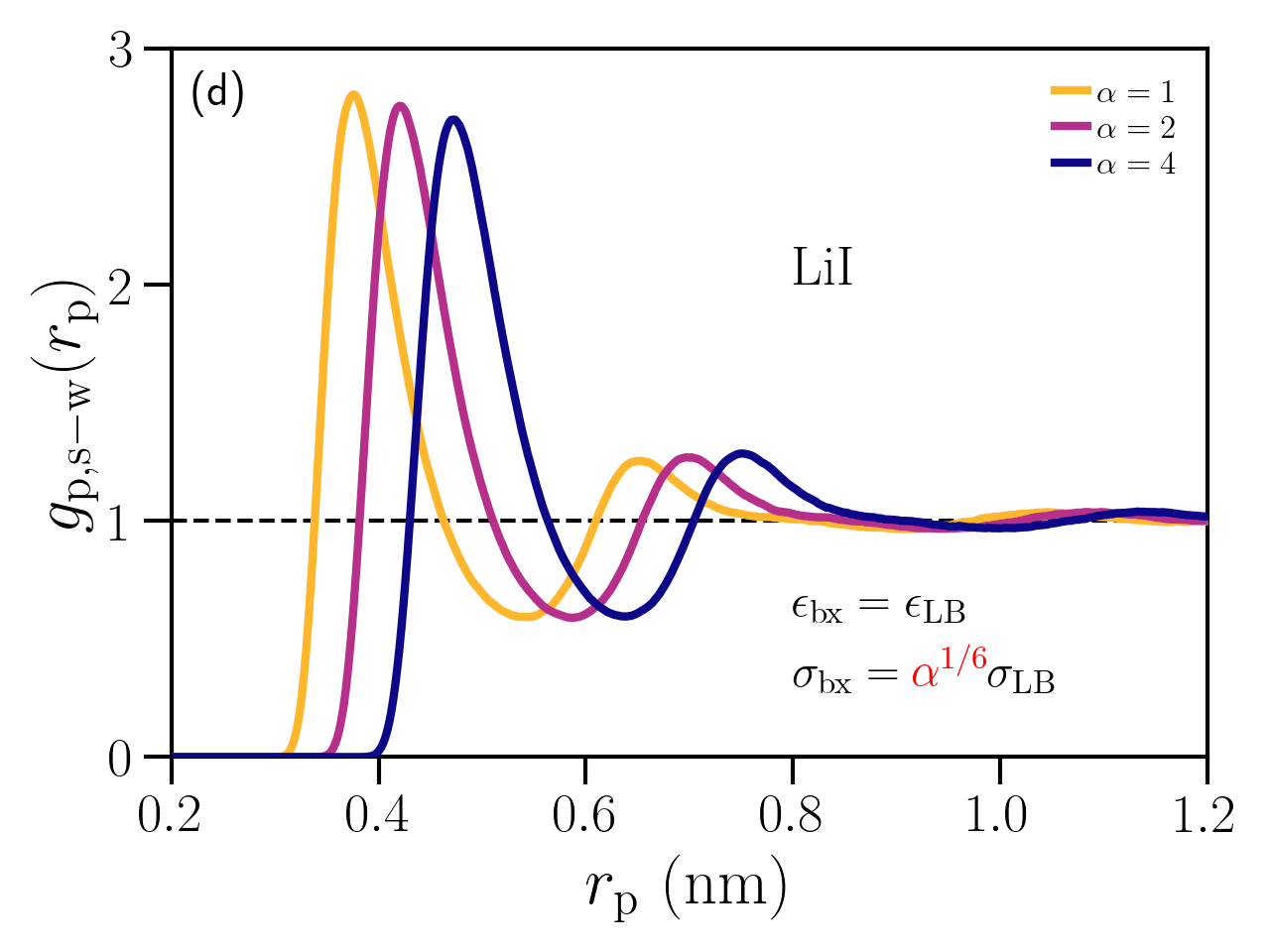}
\includegraphics[width=0.32\textwidth]{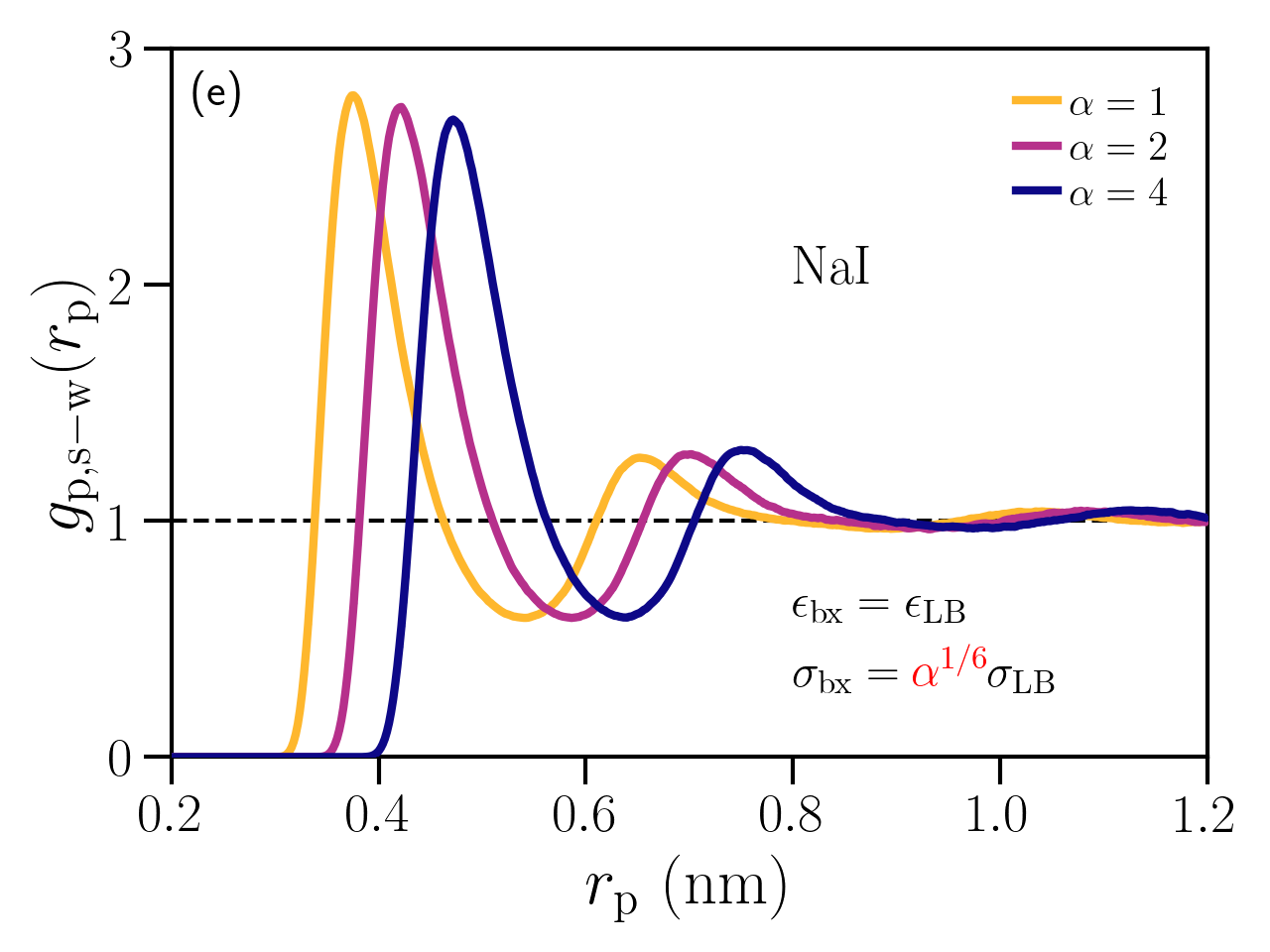}
\includegraphics[width=0.32\textwidth]{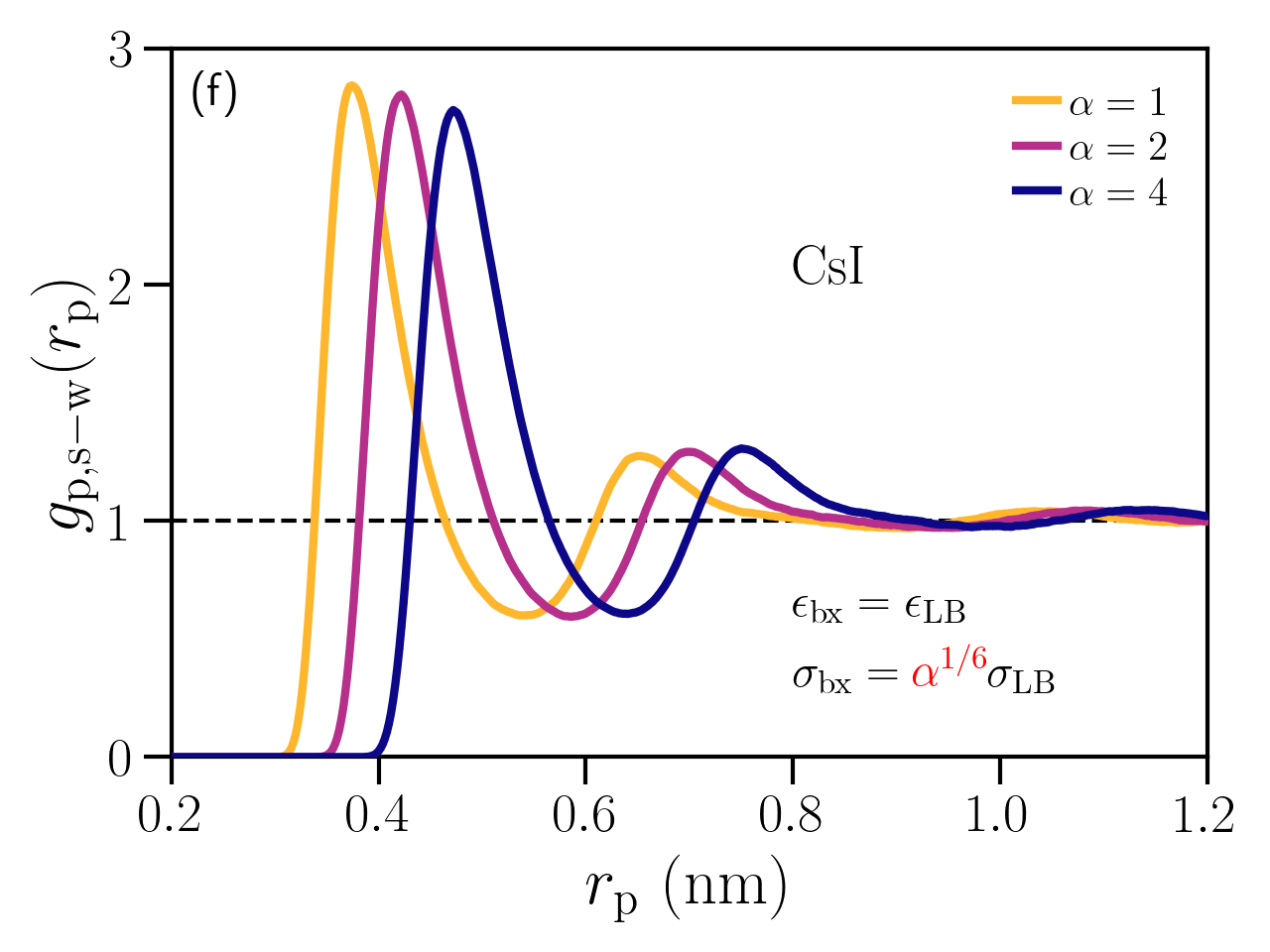}
\caption{Solute-water proximal RDF profiles, $g_{\rm p,s-w}(r_{\rm p})$,
for different values of $\alpha$ in aqueous LiI, NaI and CsI solutions.  The top
and bottom rows correspond to profiles for the coupled $\sigma_{\rm
sx}-\epsilon_{\rm bx}$ and the $\sigma_{\rm bx}$-only variations, respectively.}
\label{fig:prdf_iodide_water}
\end{figure}

\begin{figure}[h!]
\includegraphics[width=0.32\textwidth]{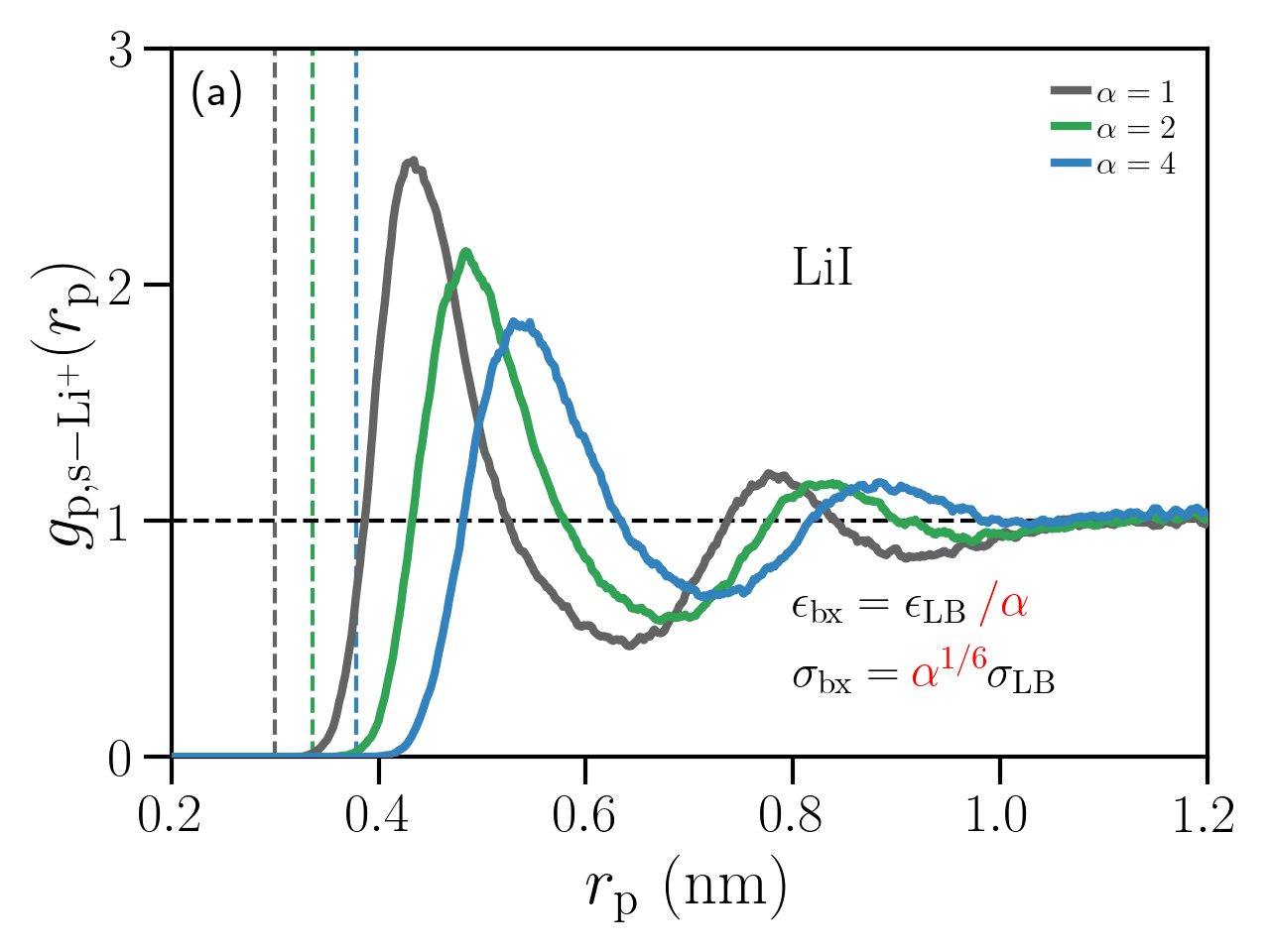}
\includegraphics[width=0.32\textwidth]{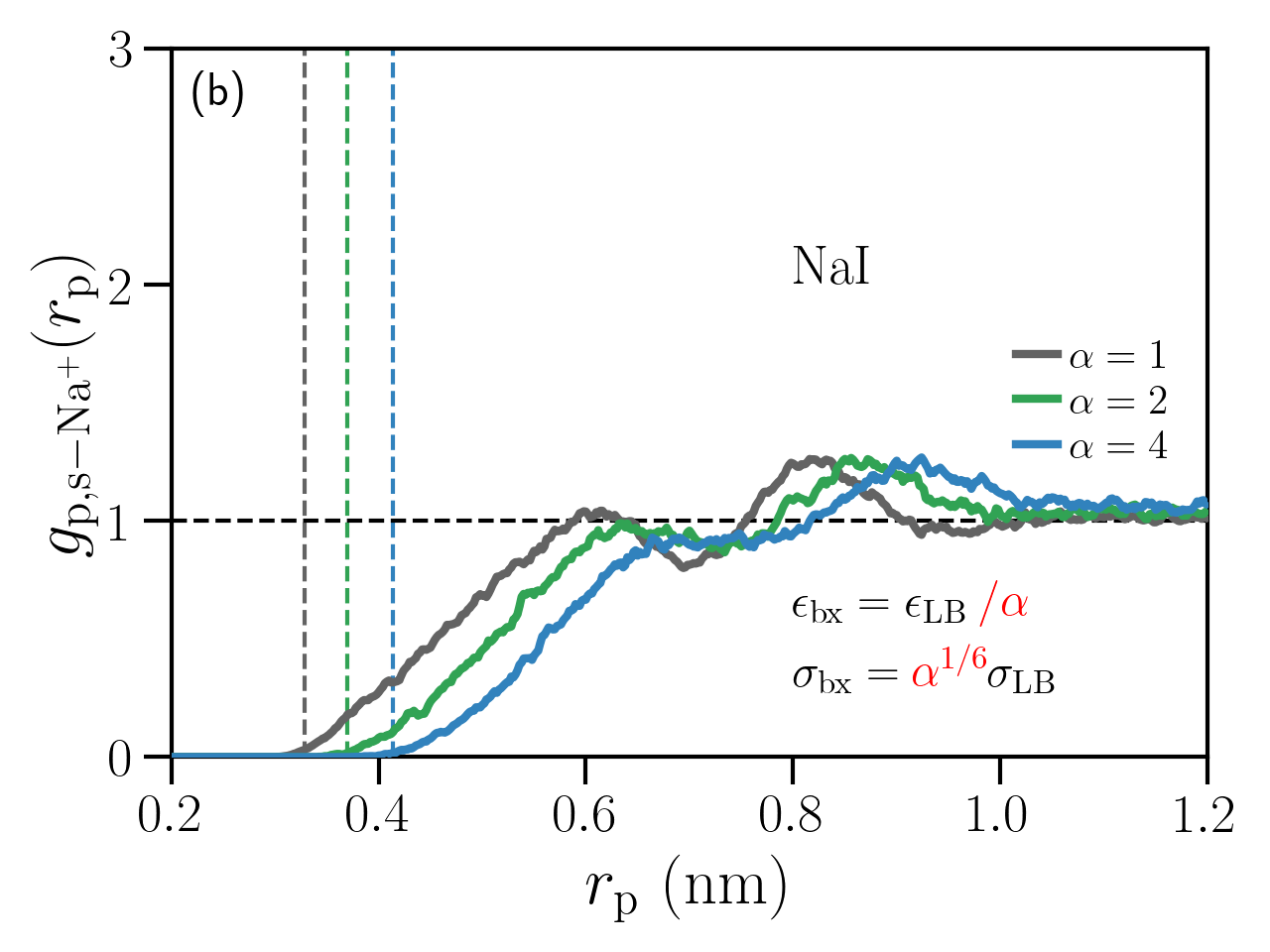}
\includegraphics[width=0.32\textwidth]{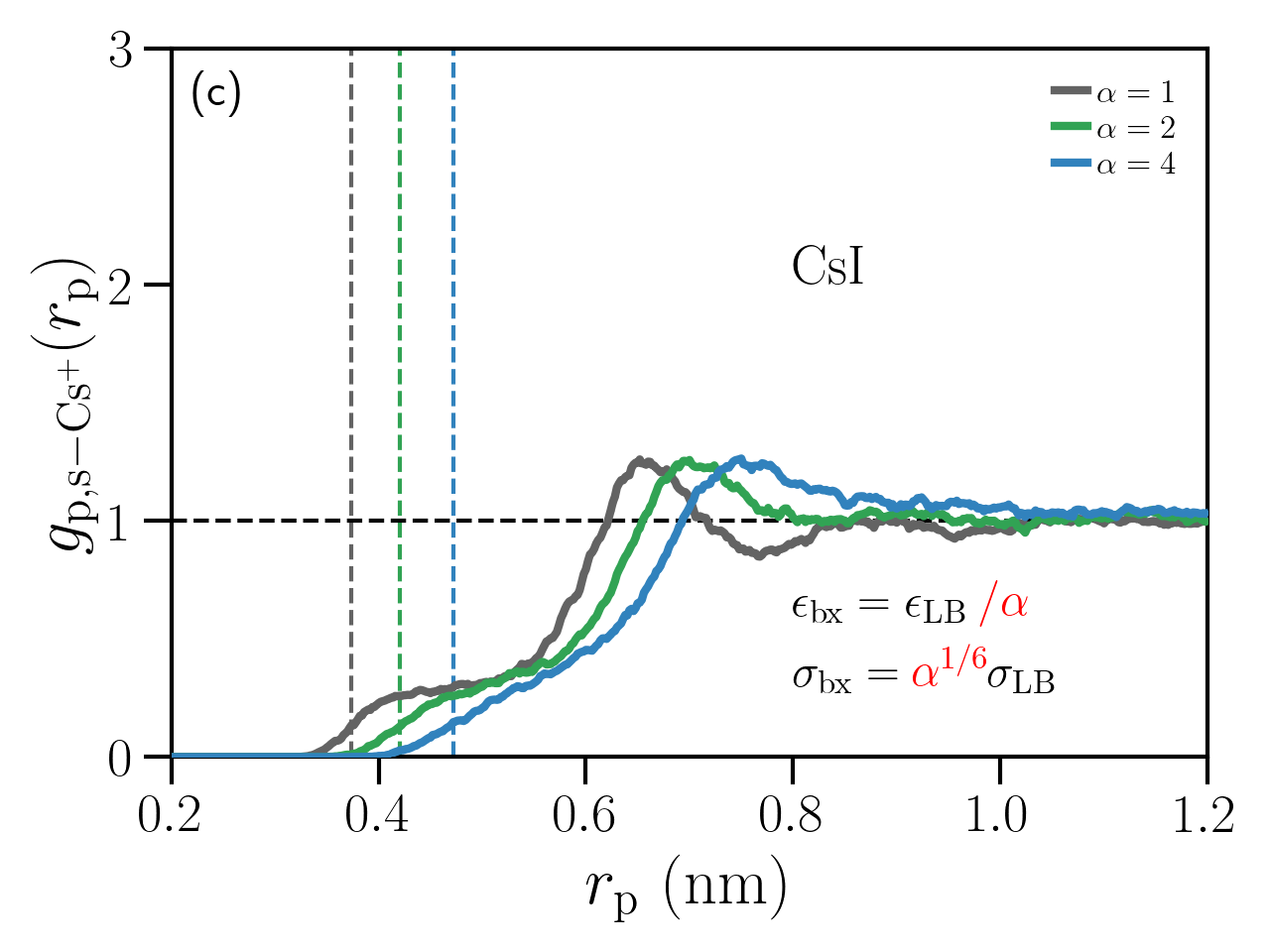}
\includegraphics[width=0.32\textwidth]{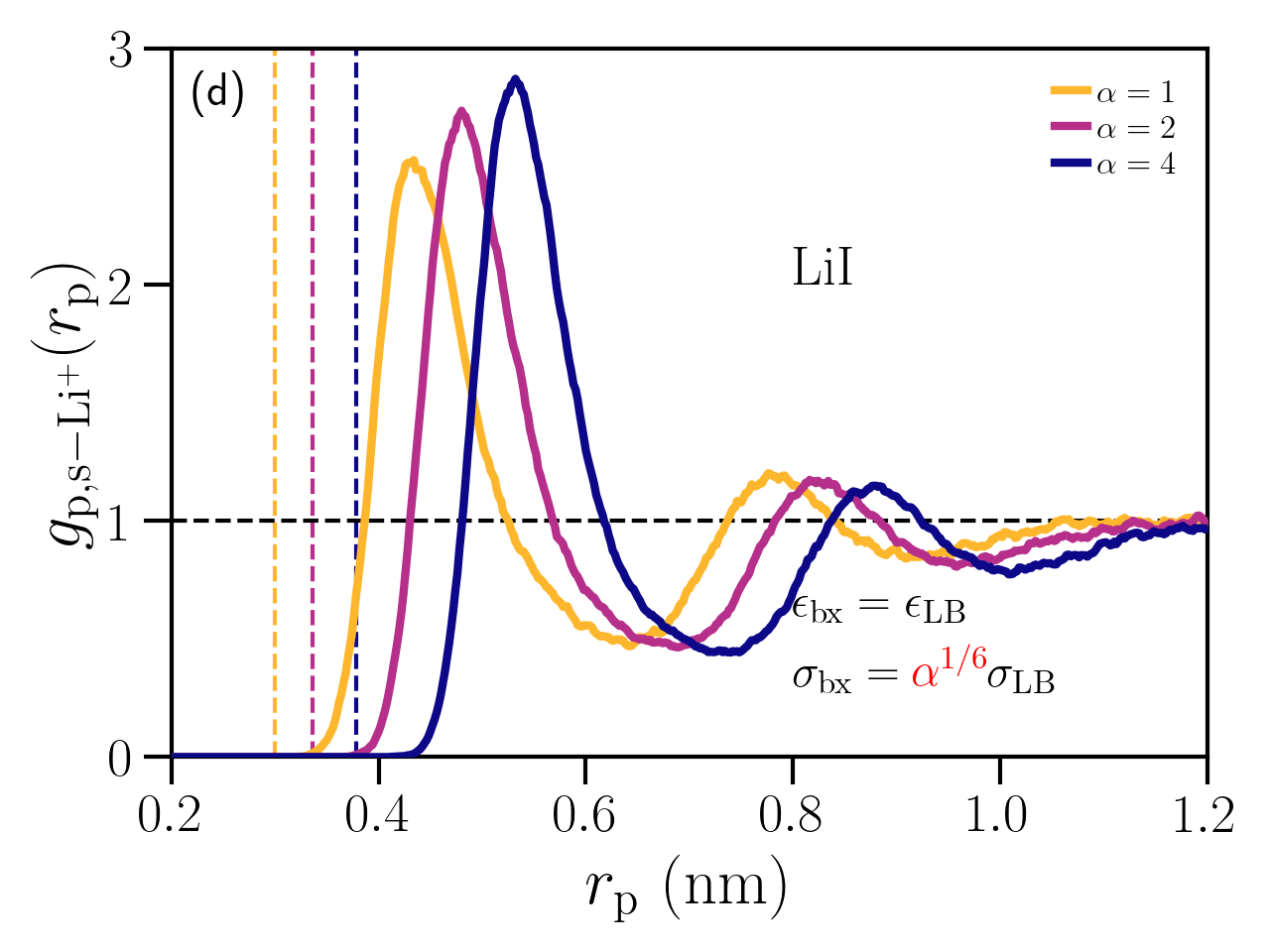}
\includegraphics[width=0.32\textwidth]{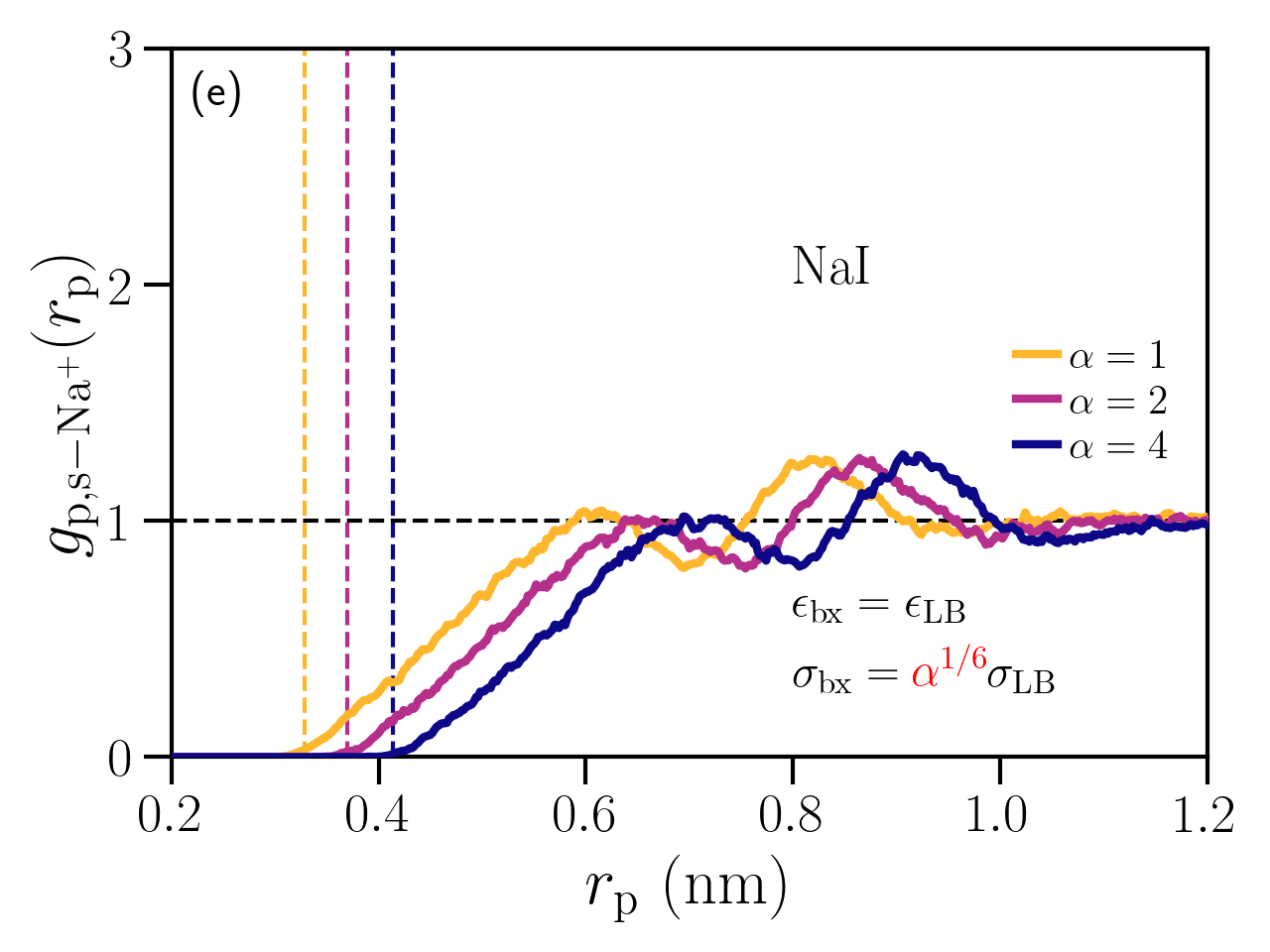}
\includegraphics[width=0.32\textwidth]{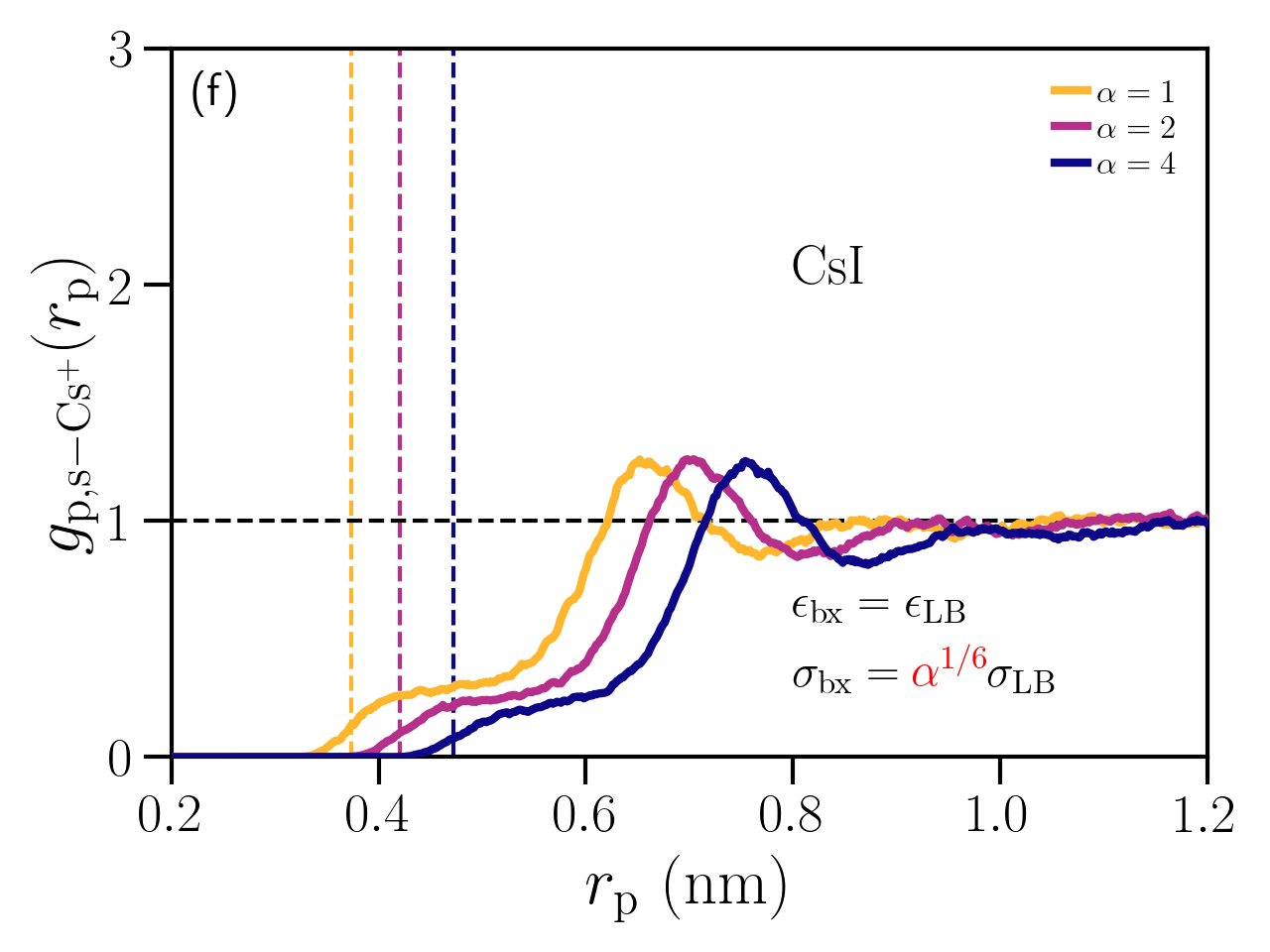}
\caption{Solute-cation proximal RDF profiles, $g_{\rm p,s-X^{+}}(r_{\rm p})$,
at different values of $\alpha$ in aqueous LiI, NaI and CsI solutions.  The top
and bottom rows correspond to profiles for the coupled $\sigma_{\rm
sx}-\epsilon_{\rm bx}$ and the $\sigma_{\rm bx}$-only variations, respectively.}
\label{fig:prdf_iodide_cation}
\end{figure}

Figure~\ref{fig:prdf_iodide_water} (a)-(c) show the solute-water proximal RDF
profiles at different values of $\alpha$ for the coupled $\sigma_{\rm
bx}-\epsilon_{\rm bx}$ variation in aqueous LiI, NaI and CsI solutions with 1 m
salt concentration. Figure~\ref{fig:prdf_iodide_water} (d)-(f) show the
solute-water proximal RDF profiles at different values of $\alpha$ for the
$\sigma_{\rm bx}$-only variation in aqueous LiI, NaI and
CsI solutions with 1 m salt concentration. 
Figure~\ref{fig:prdf_iodide_cation} (a)-(c) show the solute-cation proximal RDF
profiles at different values of $\alpha$ for the coupled $\sigma_{\rm
bx}-\epsilon_{\rm bx}$ variation in aqueous LiI, NaI and CsI solutions with 1 m
salt concentration. Figure~\ref{fig:prdf_iodide_cation} (d)-(f) show the
solute-cation proximal RDF profiles at different values of $\alpha$ for the
$\sigma_{\rm bx}$-only variation in aqueous LiI, NaI and CsI solutions with a
salt concentration of 1 m. The vertical lines indicate the corresponding
$\sigma_{\rm bx}$ for solute-cation combination. Note that the peaks in $g_{\rm
p, s-X^{+}}$ occur at much larger distances indicating the absence of
solute-cation contact and solvent-shared peaks for Li$^{+}$ and Na$^{+}$. For
Cs$^{+}$, there is slight shoulder region at small $r_{\rm p}$, although the
peak appears beyond the solvent-separated region. Note that, at small $r_{\rm
p}$, $g_{\rm p, s-Cs^{+}}$ becomes non-zero at distances slightly smaller than
the corresponding $\sigma_{\rm bx}$. This occurs as the proximal distance of an
ion from the line connecting all the solute beads can be smaller than
$\sigma_{\rm bx}$ which is the closest distance of approach between a solute
bead and an ion.
\subsection{Effect of salt concentration}
\begin{figure}[h!]
\includegraphics[width=0.4\textwidth]{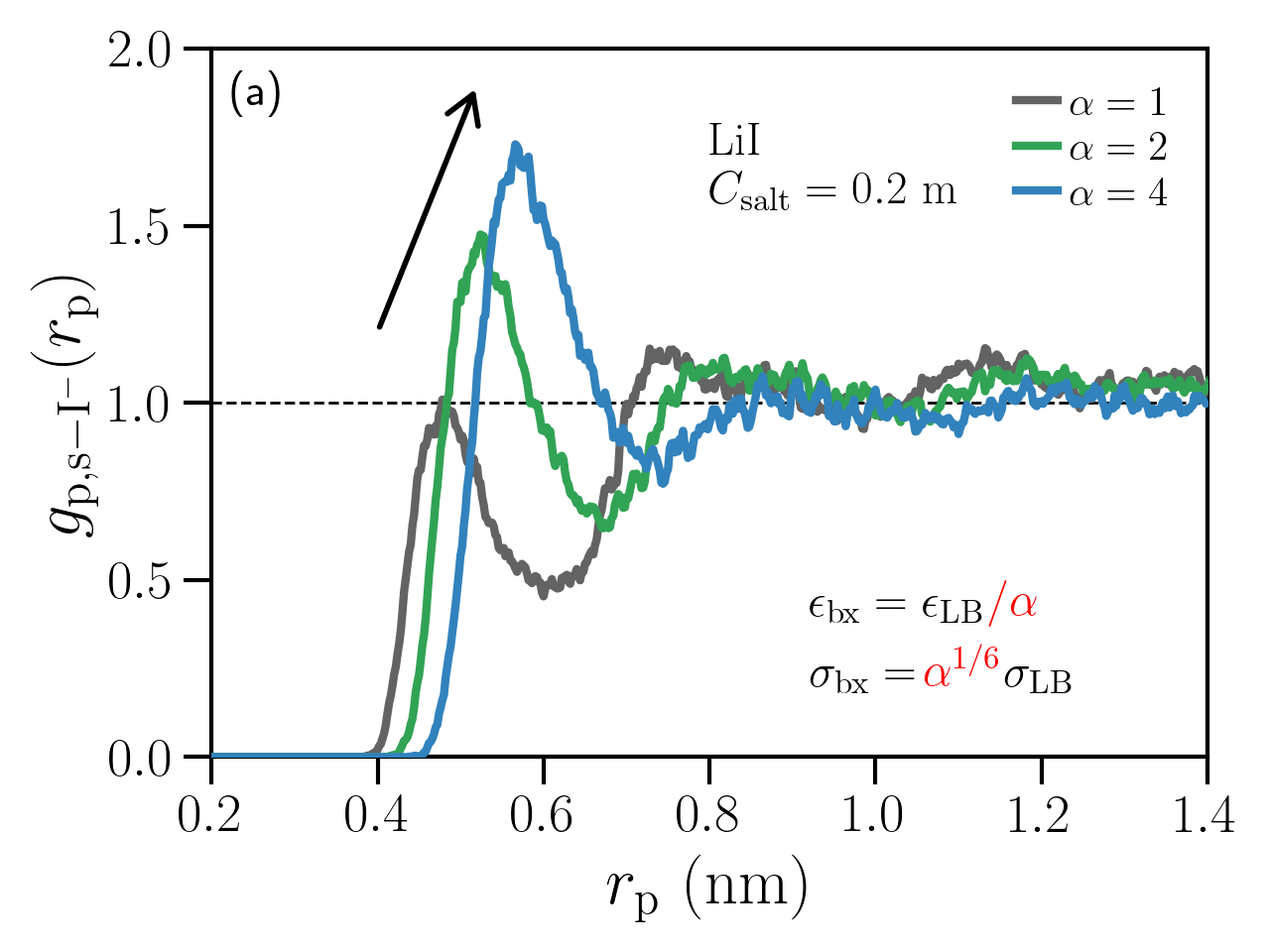}
\includegraphics[width=0.4\textwidth]{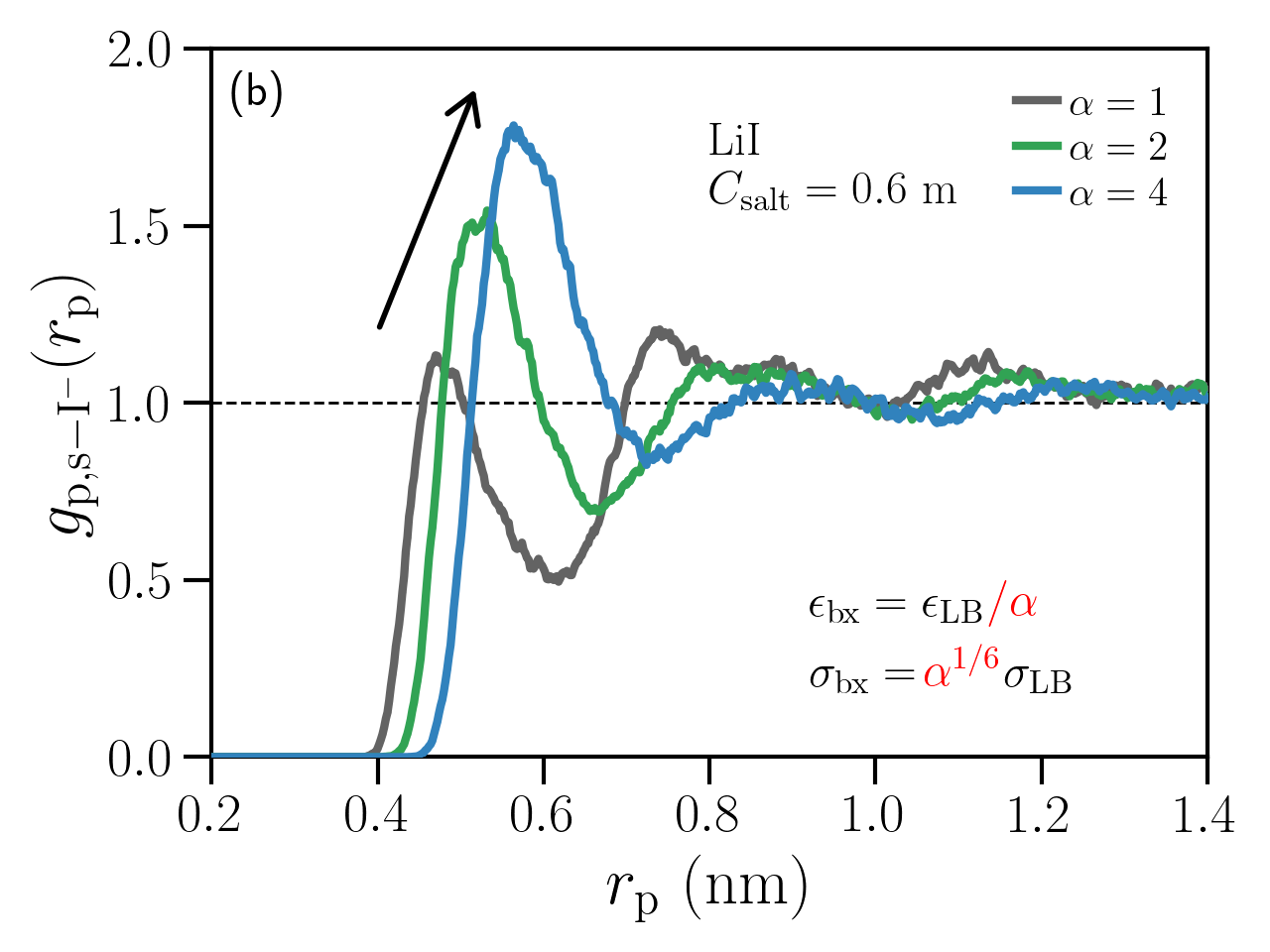}\\
\includegraphics[width=0.4\textwidth]{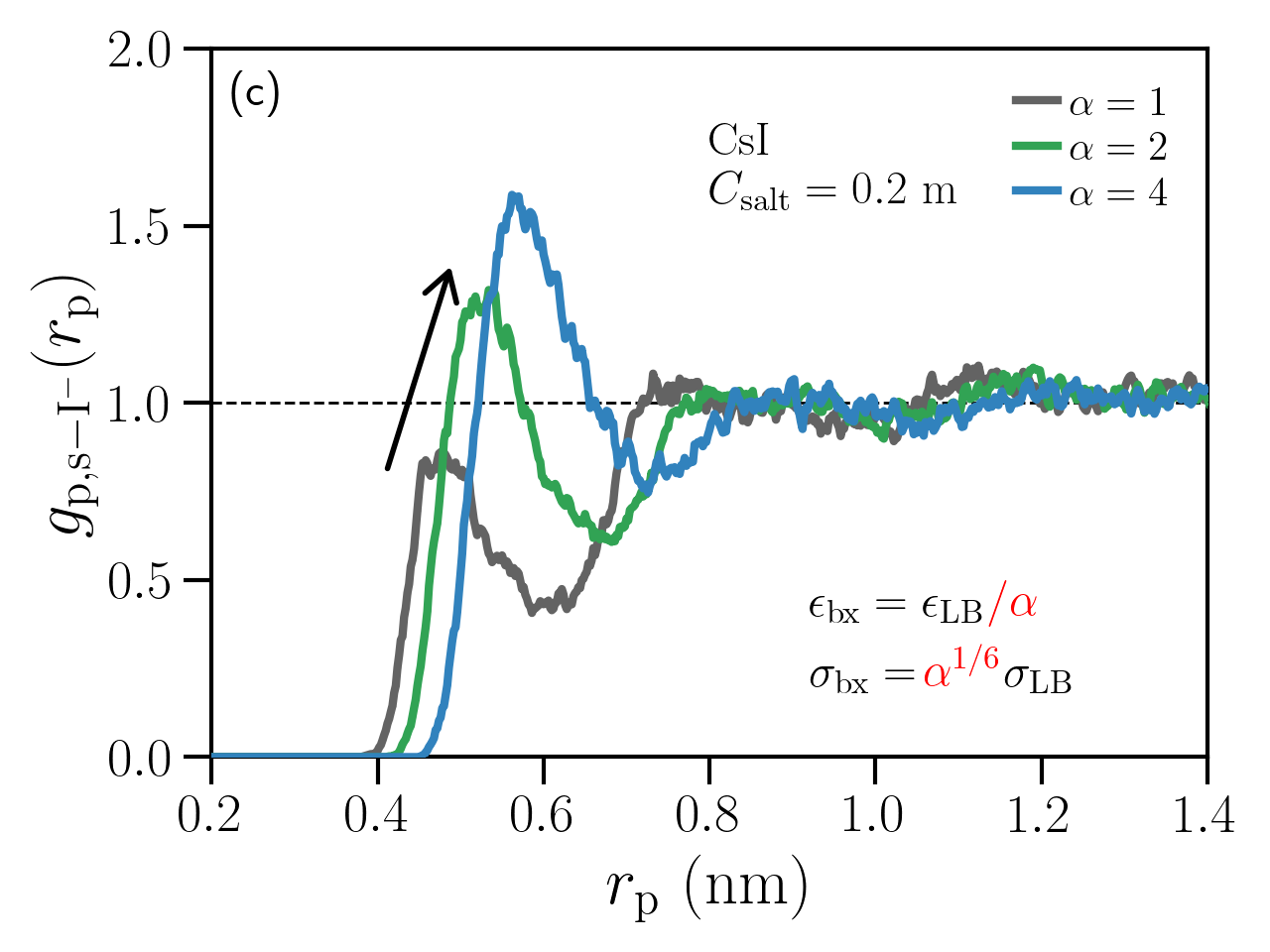}
\includegraphics[width=0.4\textwidth]{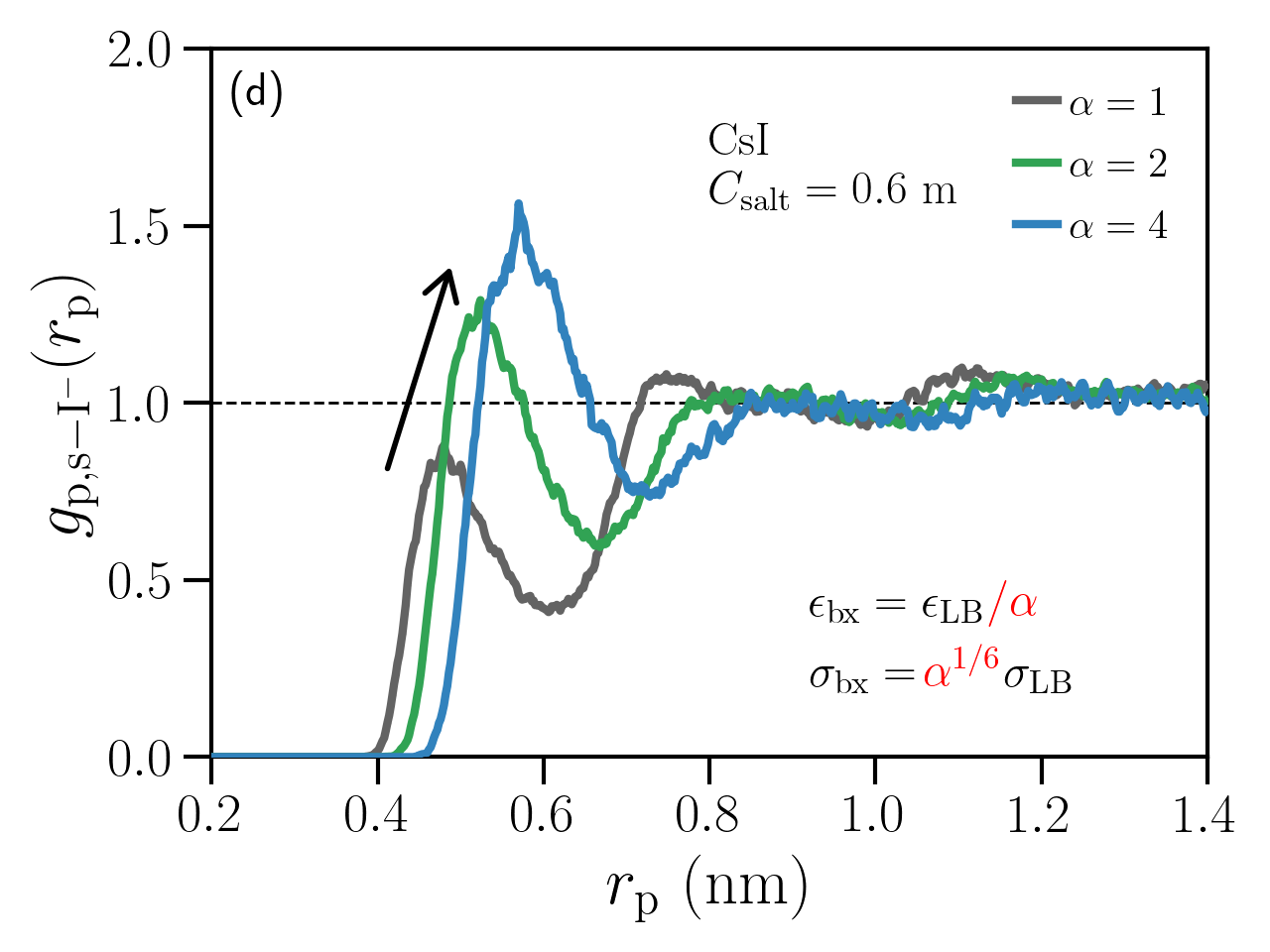}
\caption{Solute-iodide proximal RDF profiles, $g_{\rm p,s-I^{-}}(r_{\rm p})$, at
different values of $\alpha$ in aqueous LiI and CsI solutions for the  coupled
$\sigma_{\rm bx}-\epsilon_{\rm bx}$ variation. The left and right columns
correspond to profiles for 0.2 m and 0.6 m salt solutions, respectively.}
\label{fig:prdf_iodide_alpha_conc} \end{figure}

\begin{figure}[h!]
\includegraphics[width=0.4\textwidth]{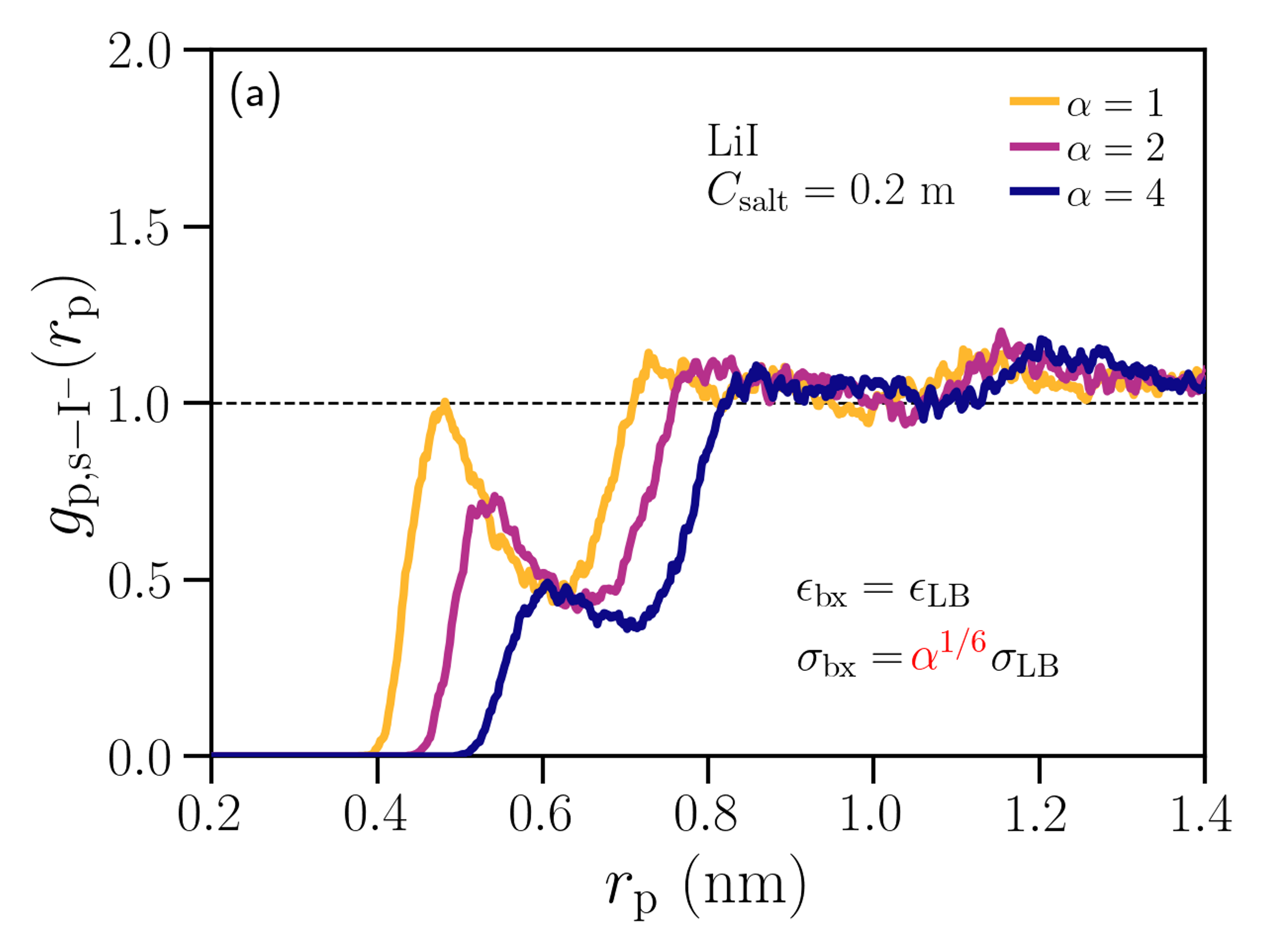}
\includegraphics[width=0.4\textwidth]{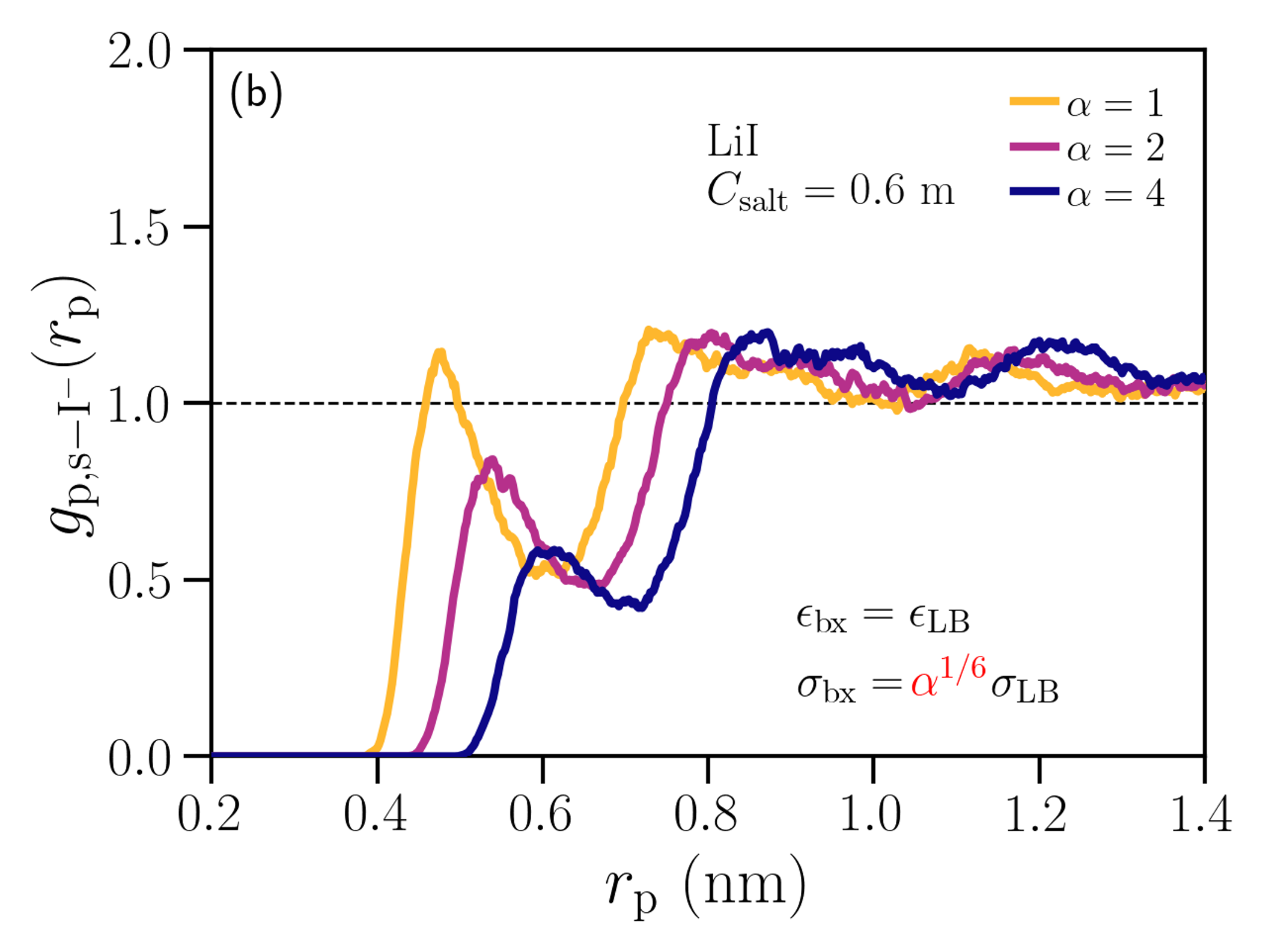}
\includegraphics[width=0.4\textwidth]{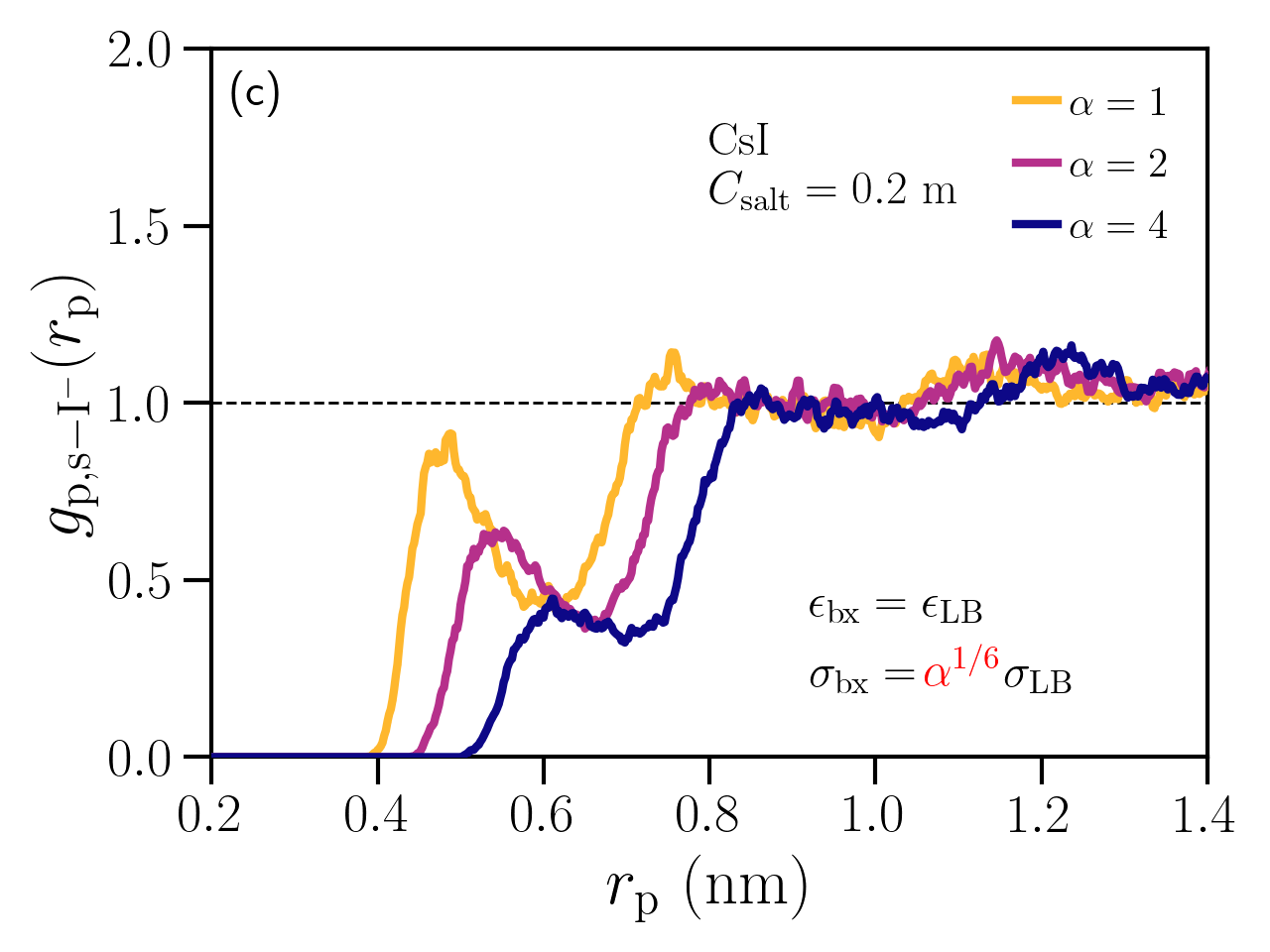}
\includegraphics[width=0.4\textwidth]{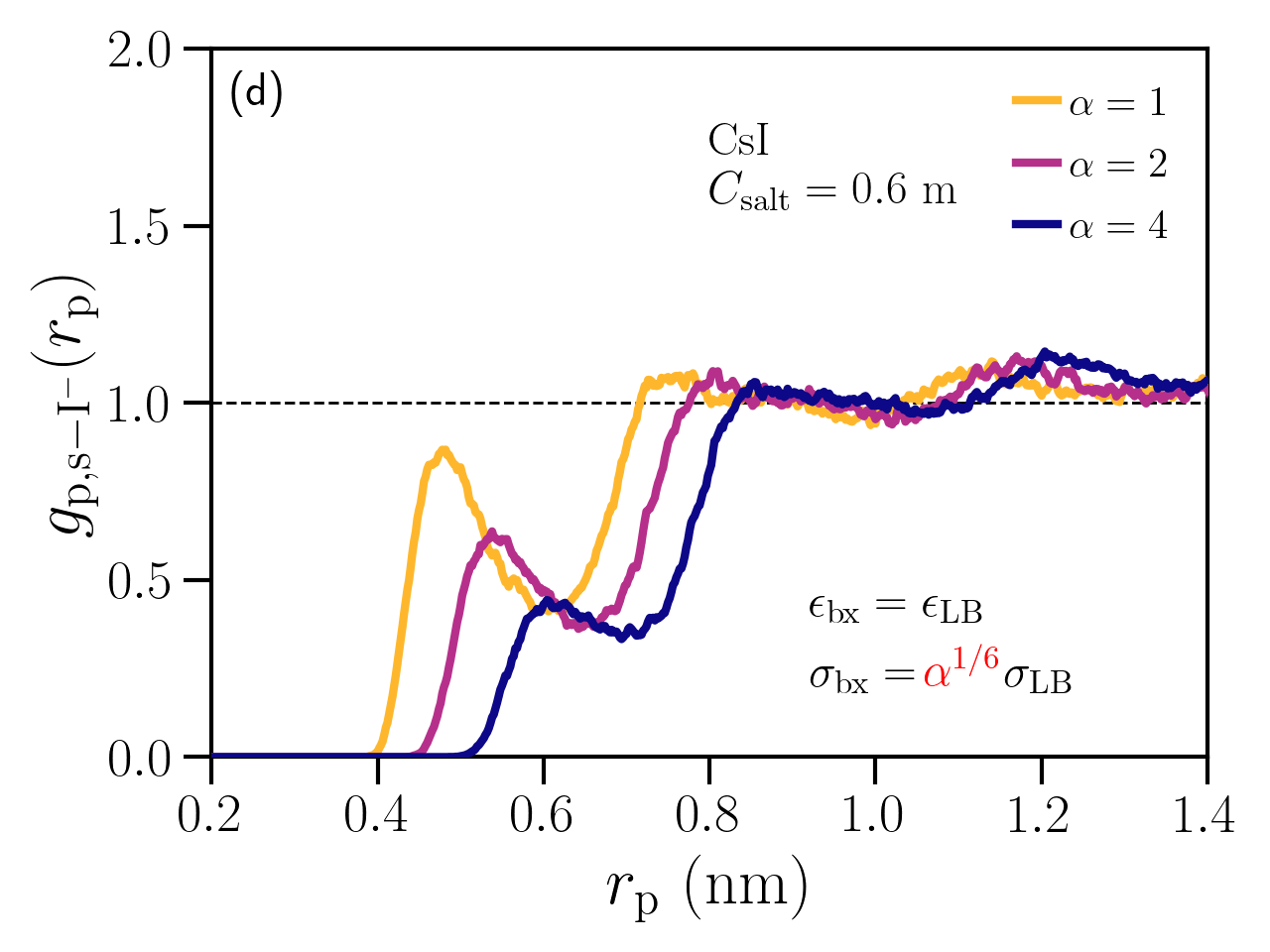}
\caption{Solute-iodide proximal RDF profiles, $g_{\rm p,s-I^{-}}(r_{\rm p})$, at
different values of $\alpha$ in aqueous LiI and CsI solutions for the
$\sigma_{\rm bx}$-only variation. The left and right columns correspond to
profiles for 0.2 m and 0.6 m salt solutions, respectively.}
\label{fig:prdf_iodide_sigma_conc}
\end{figure}

Figure~\ref{fig:prdf_iodide_alpha_conc} show the solute-iodide proximal RDFs at
different values of $\alpha$ for the coupled $\sigma_{\rm bx}-\epsilon_{\rm bx}$
variation in LiI and CsI solutions with salt concentrations of 0.2 m and 0.6 m.
Similar to the trends observed at a salt concentration of 1 m, the first peak of $g_{\rm
p,s-I^{-}}$ increases monotonically with increasing $\alpha$ for both CsI and LiI
solutions at both salt concentrations. Figure~\ref{fig:prdf_iodide_sigma_conc}
show the solute-iodide proximal RDFs at different values of $\alpha$ for the
$\sigma_{\rm bx}$-only variation in LiI and CsI solutions with salt
concentrations of 0.2 m and 0.6 m. The first peak of  $g_{\rm p,s-I^{-}}$ decreases
monotonically with increasing $\alpha$, and in agreement with the trends
observed in 1 m salt solutions.
 

\section{Extended solute in aqueous alkali thiocyanate solutions}
\subsection{Solute-water and solute-cation RDFs at 1 m salt concentration}
\begin{figure}[h!]
\includegraphics[width=0.4\textwidth]{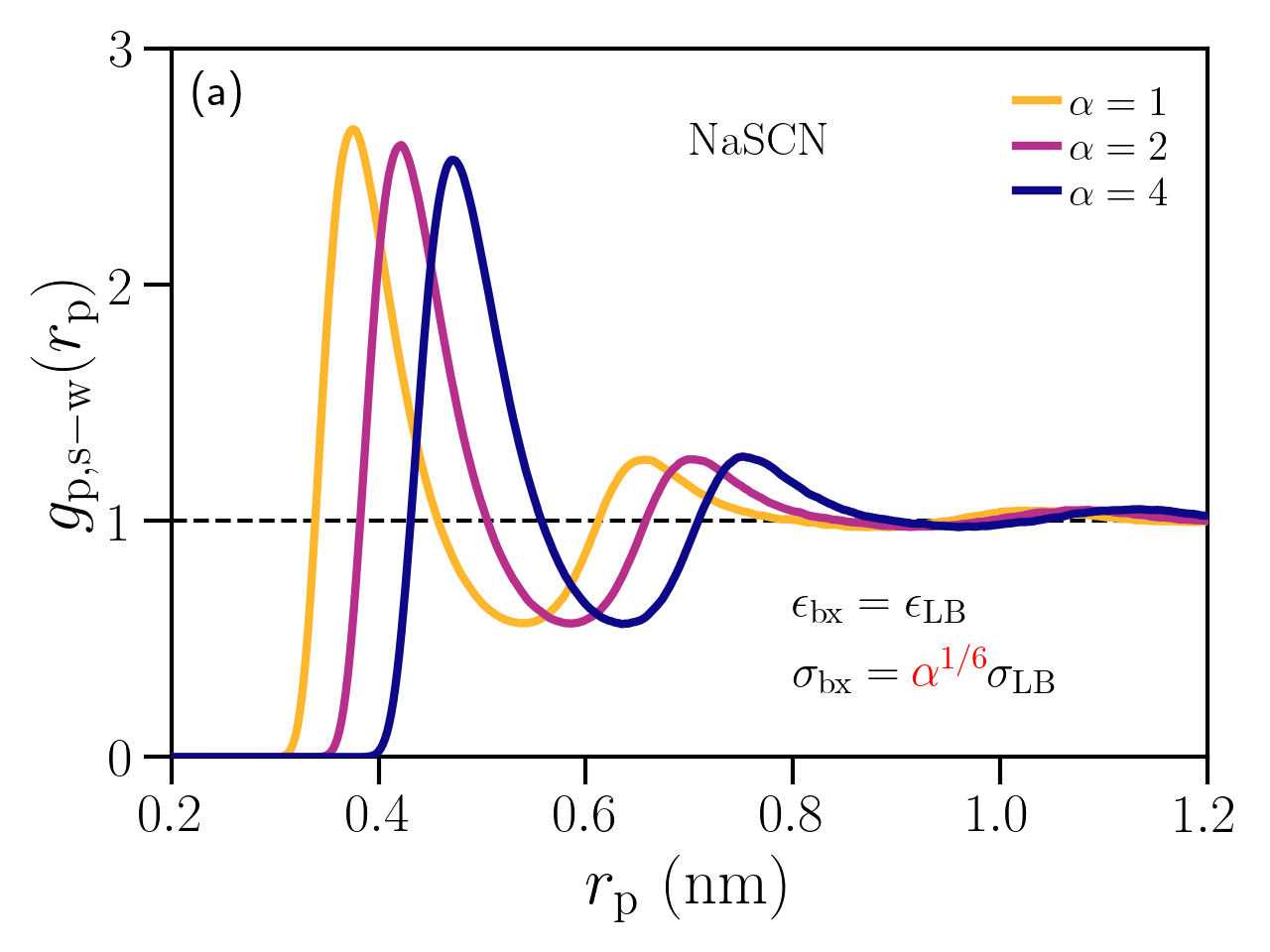}
\includegraphics[width=0.4\textwidth]{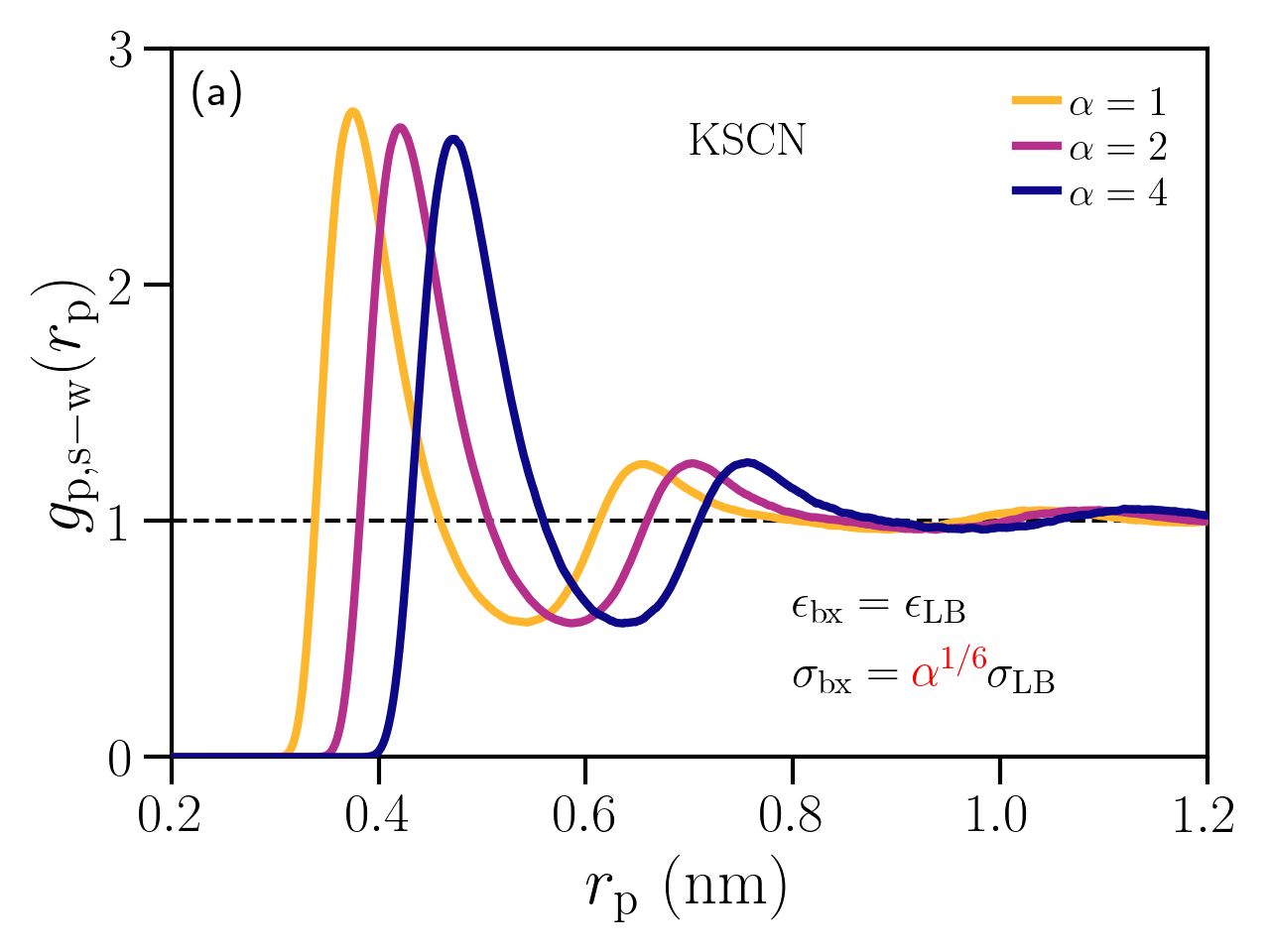}
\includegraphics[width=0.4\textwidth]{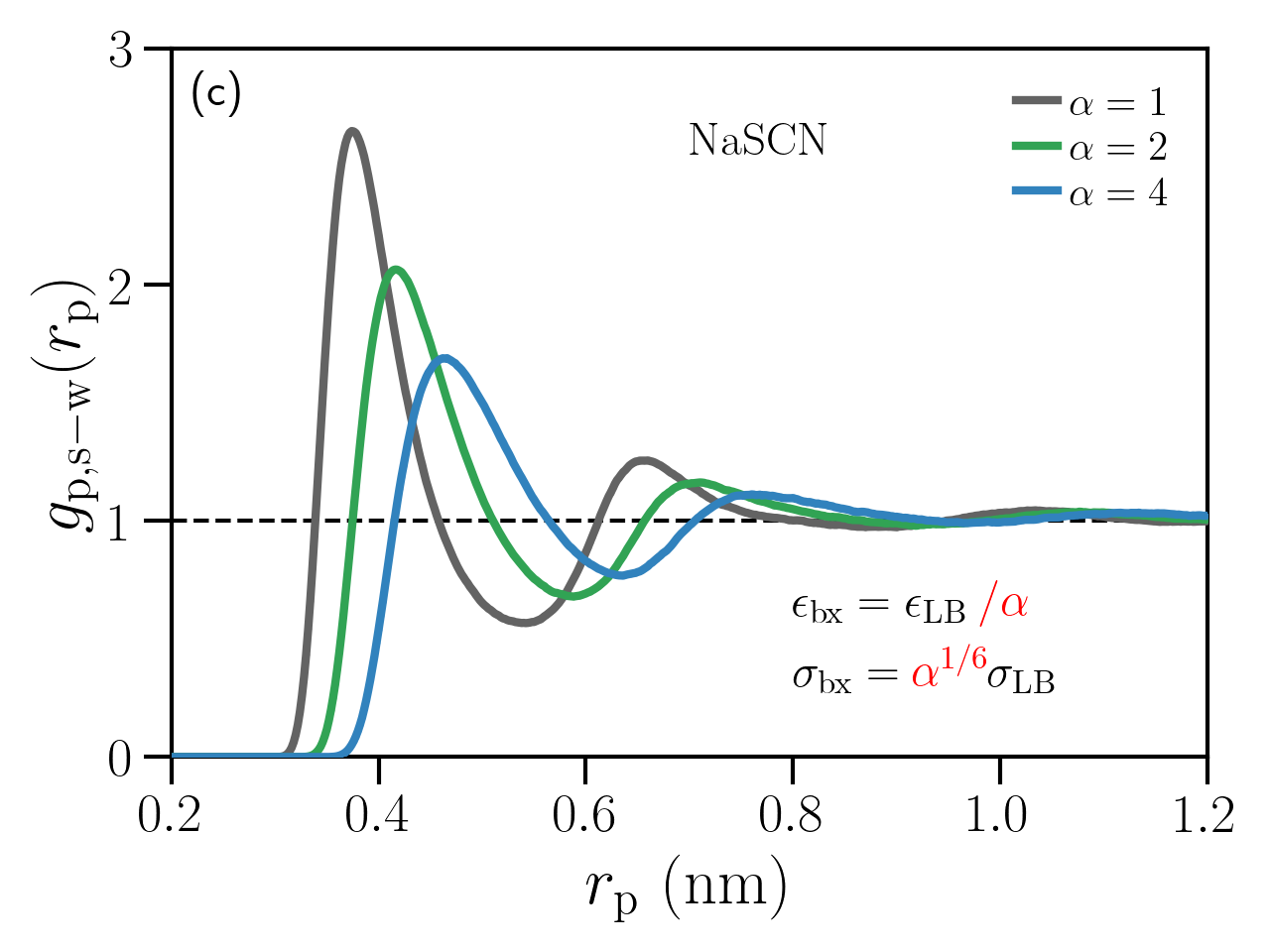}
\includegraphics[width=0.4\textwidth]{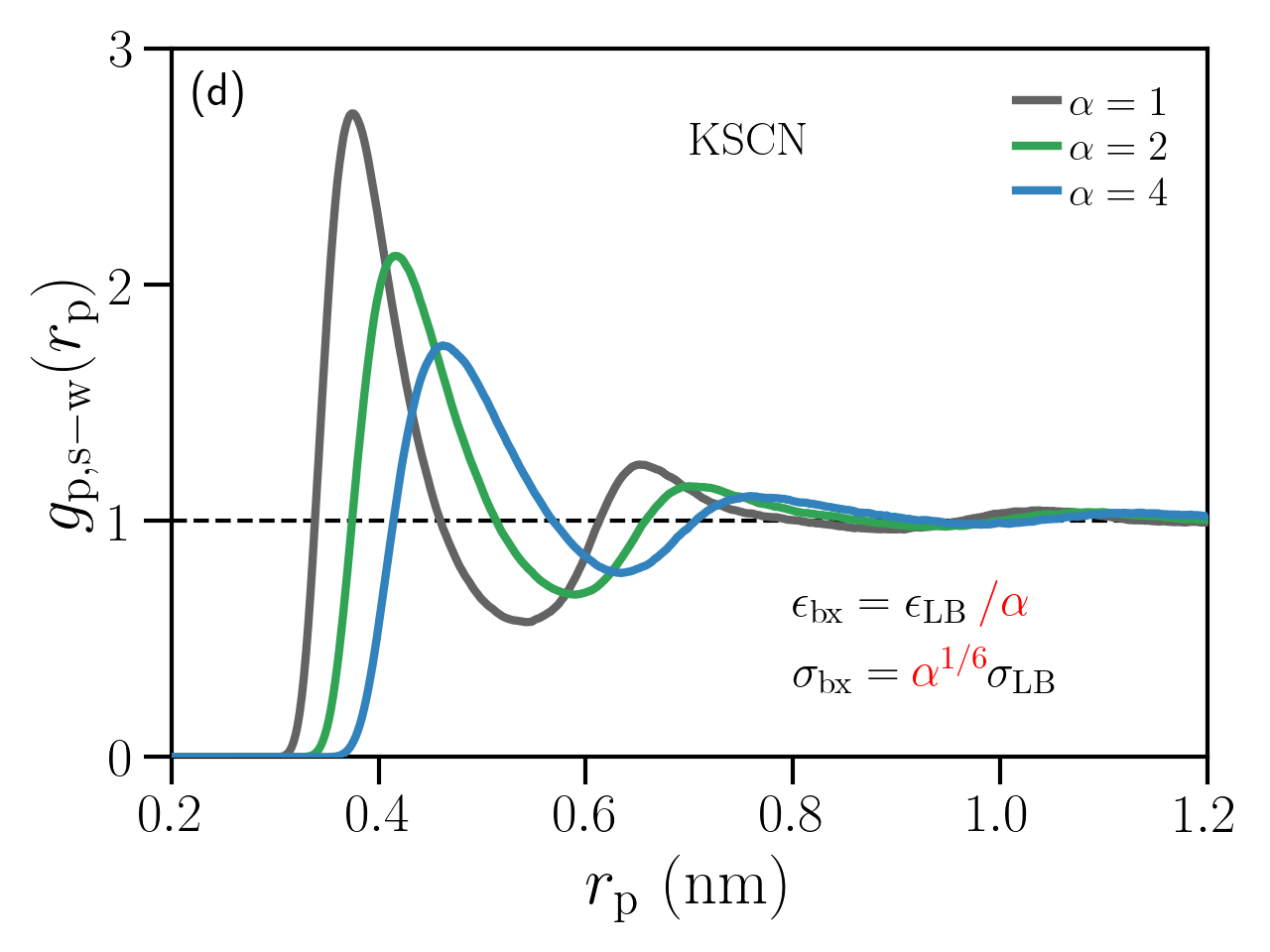}
\caption{Solute-water proximal RDF profiles, $g_{\rm p,s-w}(r_{\rm
p})$, at different values of $\alpha$ in aqueous NaSCN and KSCN solutions. The top
and bottom rows correspond to profiles for the $\sigma_{\rm bx}$-only  and coupled $\sigma_{\rm
sx}-\epsilon_{\rm bx}$ variations, respectively.}
\label{fig:prdf_thiocyanate_water}
\end{figure}

\begin{figure}[h!]
\includegraphics[width=0.4\textwidth]{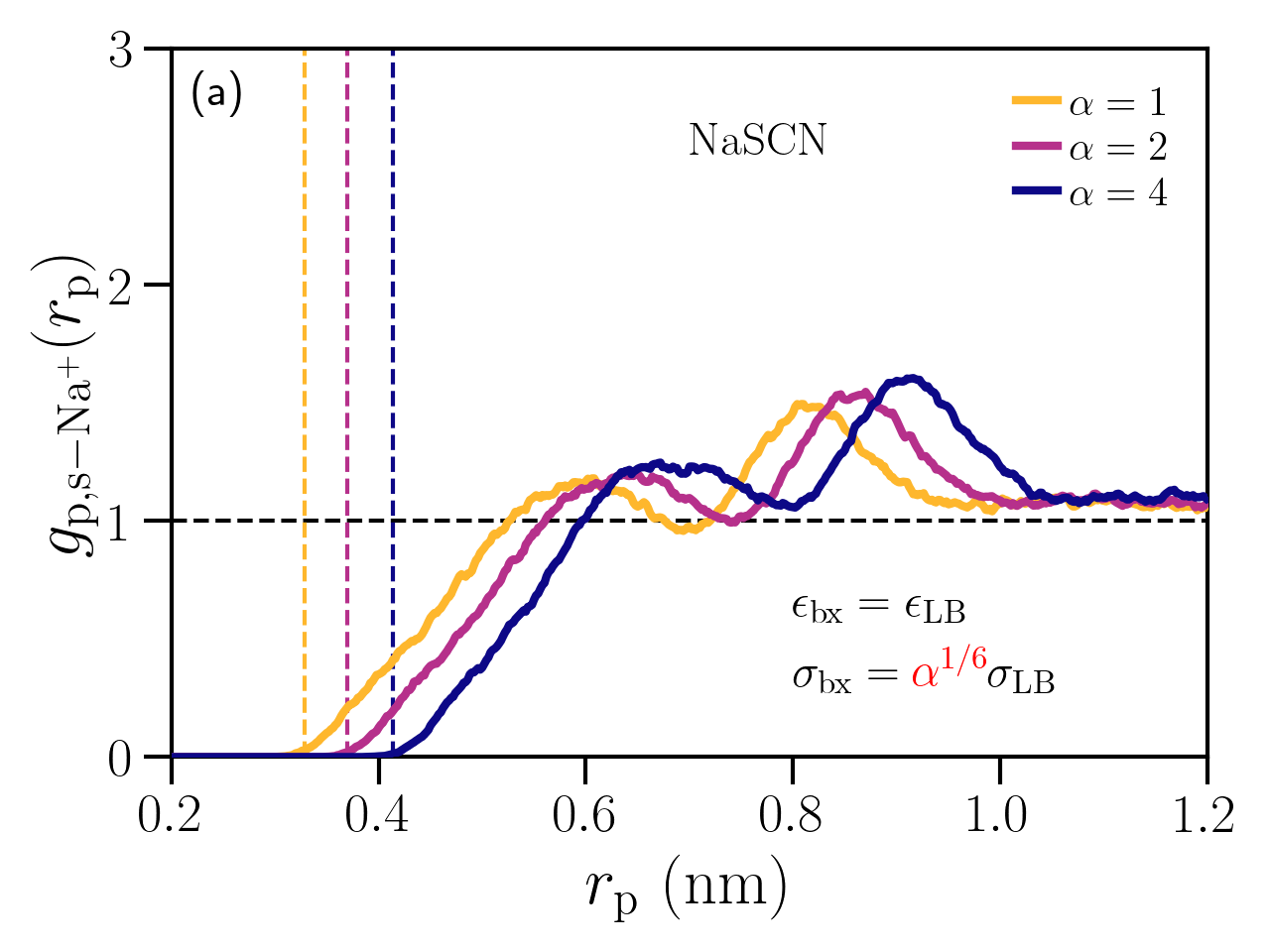}
\includegraphics[width=0.4\textwidth]{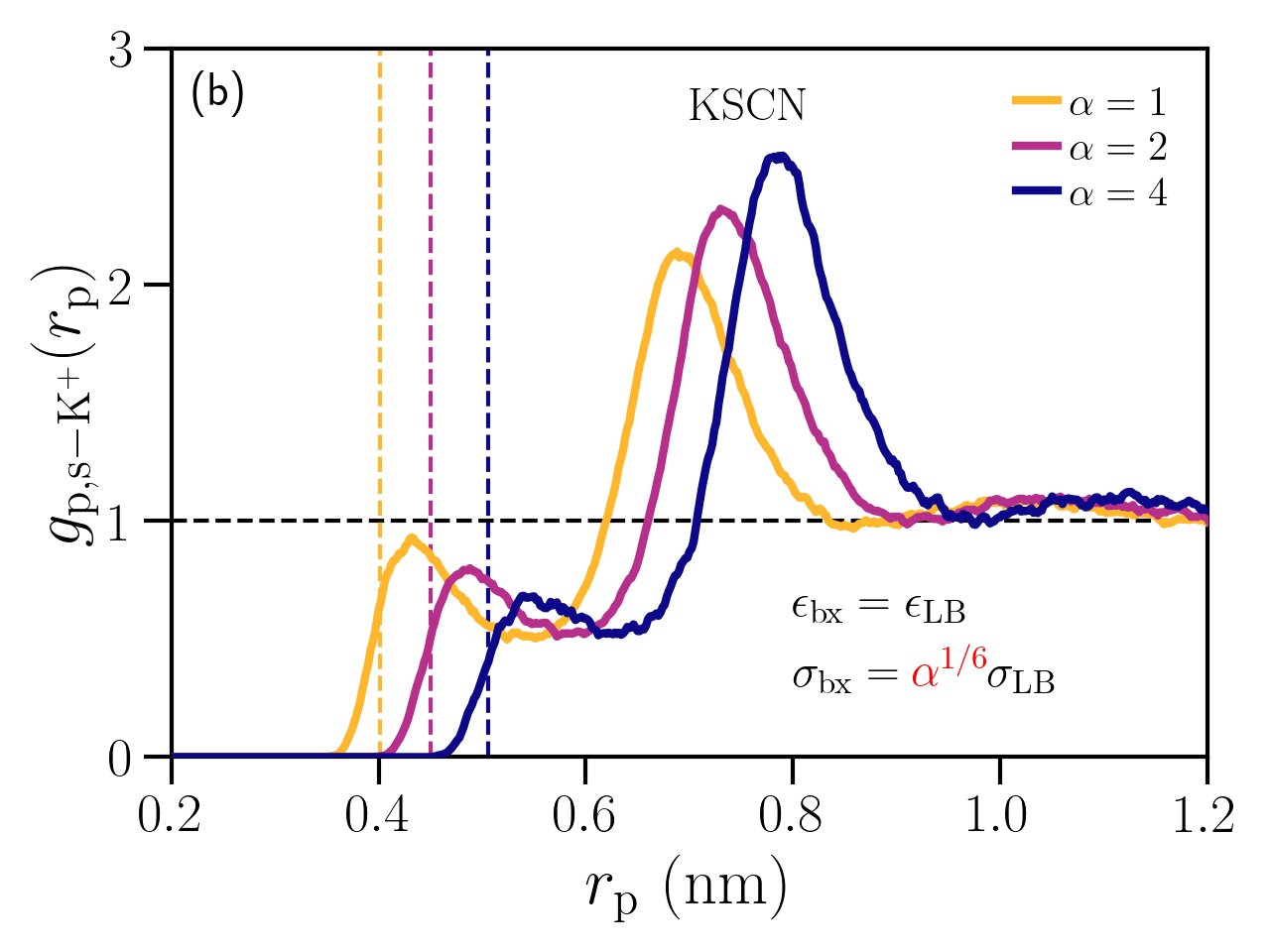}
\includegraphics[width=0.4\textwidth]{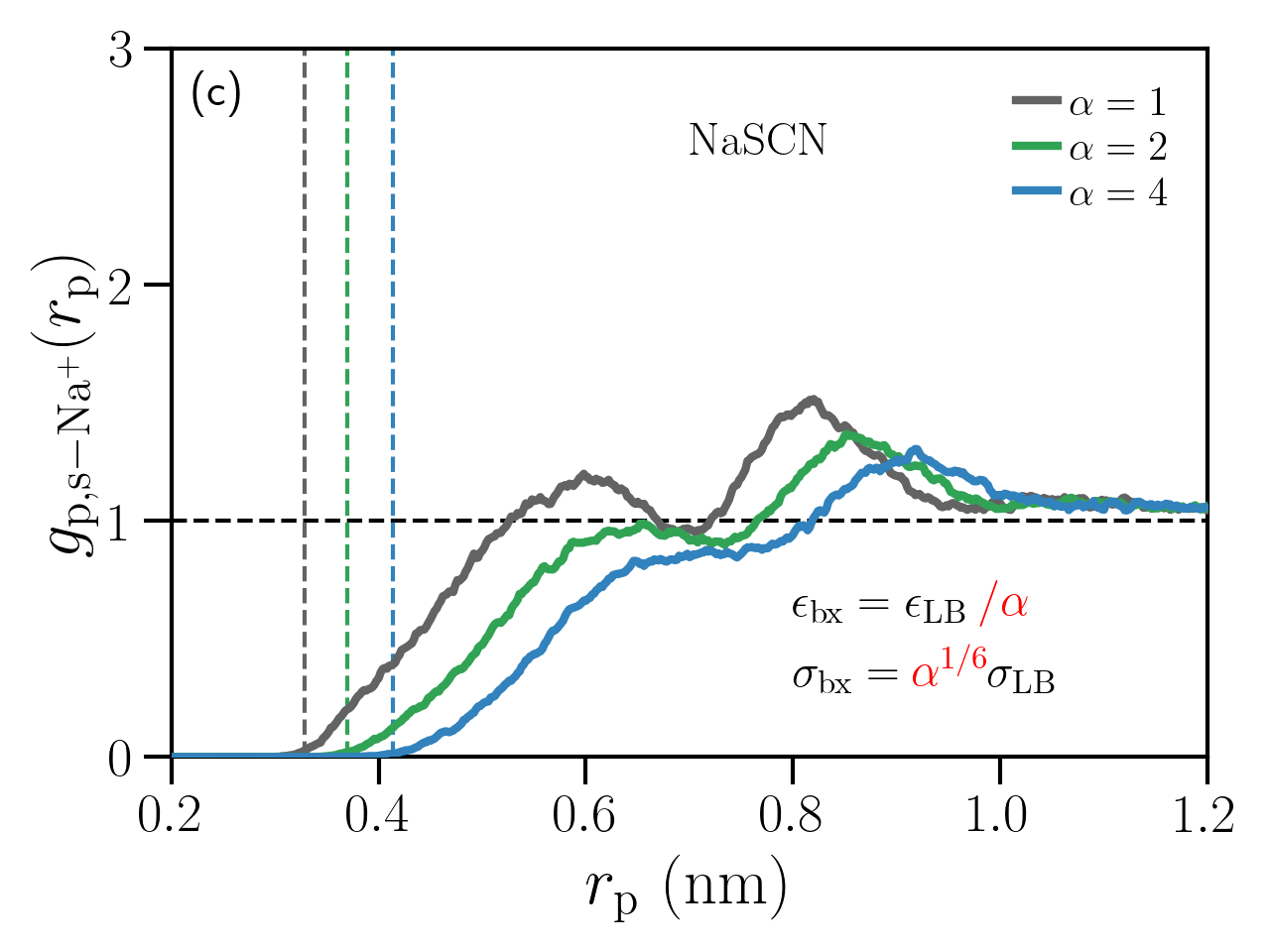}
\includegraphics[width=0.4\textwidth]{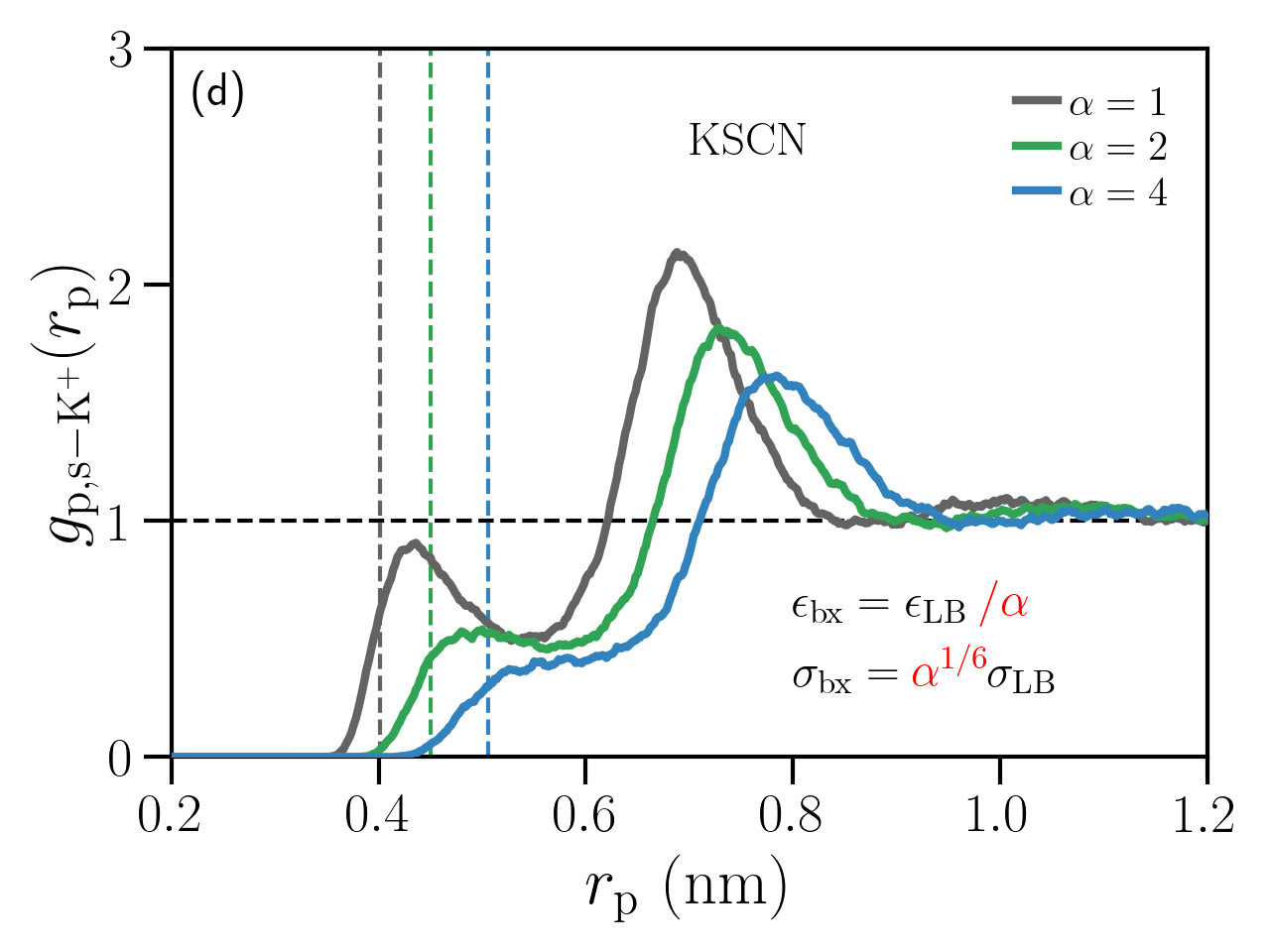}
\caption{Solute-cation proximal RDF profiles, $g_{\rm p,s-X^{+}}(r_{\rm
p})$, at different values of $\alpha$ in aqueous NaSCN and KSCN solutions. The top
and bottom rows correspond to profiles for the $\sigma_{\rm bx}$-only  and coupled $\sigma_{\rm
sx}-\epsilon_{\rm bx}$ variations, respectively.}
\label{fig:prdf_thiocyanate_cation}
\end{figure}

Figures~\ref{fig:prdf_thiocyanate_water} (a)-(b) show the solute-water proximal
RDF profiles at different values of $\alpha$ for the coupled $\sigma_{\rm
bx}-\epsilon_{\rm bx}$ variation in aqueous NaSCN and KSCN solutions with a salt
concentration of 1 m. Figure~\ref{fig:prdf_thiocyanate_water} (c)-(d) show the
solute-water proximal RDF profiles at different values of $\alpha$ for the
$\sigma_{\rm bx}$-only variation in aqueous NaSCN and KSCN solutions with a salt
concentration of 1 m. 
Figures~\ref{fig:prdf_thiocyanate_cation} (a)-(b) show the solute-cation proximal
RDF profiles at different values of $\alpha$ for the coupled $\sigma_{\rm
bx}-\epsilon_{\rm bx}$ variation in aqueous NaSCN and KSCN solutions with a salt
concentration of 1 m. Figure~\ref{fig:prdf_thiocyanate_cation} (c)-(d) show the
solute-cation proximal RDF profiles at different values of $\alpha$ for the
$\sigma_{\rm bx}$-only variation in aqueous NaSCN and KSCN solutions with a salt
concentration of 1 m. 

\end{document}